\documentclass{elsart}

\usepackage[square,comma]{natbib}
\usepackage{graphicx}
\usepackage{pxfonts}
\usepackage{amssymb}

\begin{document}

\thispagestyle{empty}
\begin{Large}
\textbf{DEUTSCHES ELEKTRONEN-SYNCHROTRON}

\textbf{\large{in der HELMHOLTZ-GEMEINSCHAFT}\\}
\end{Large}

DESY 06-127

August 2006

\begin{eqnarray}
\nonumber &&\cr \nonumber && \cr \nonumber &&\cr
\end{eqnarray}
\begin{eqnarray}
\nonumber
\end{eqnarray}
\begin{center}
\begin{Large}
\textbf{Fourier Optics treatment of Classical Relativistic
Electrodynamics}
\end{Large}
\begin{eqnarray}
\nonumber &&\cr \nonumber && \cr
\end{eqnarray}

\begin{large}
Gianluca Geloni, Evgeni Saldin, Evgeni Schneidmiller and Mikhail
Yurkov
\end{large}
\textsl{\\Deutsches Elektronen-Synchrotron DESY, Hamburg}
\begin{eqnarray}
\nonumber
\end{eqnarray}
\begin{eqnarray}
\nonumber
\end{eqnarray}
\begin{eqnarray}
\nonumber
\end{eqnarray}
ISSN 0418-9833
\begin{eqnarray}
\nonumber
\end{eqnarray}
\begin{large}
\textbf{NOTKESTRASSE 85 - 22607 HAMBURG}
\end{large}
\end{center}
%\end{widetext}
\clearpage
\newpage

\begin{frontmatter}



\title{Fourier Optics treatment of Classical Relativistic Electrodynamics}


\author[DESY]{Gianluca Geloni}
\author[DESY]{Evgeni Saldin}
\author[DESY]{Evgeni Schneidmiller}
\author[DESY]{Mikhail Yurkov}

\address[DESY]{Deutsches Elektronen-Synchrotron (DESY), Hamburg,
Germany}

\begin{abstract}
In this paper we couple Synchrotron Radiation (SR) theory with a
branch of physical optics, namely laser beam optics. We show that
the theory of laser beams is successful in characterizing
radiation fields associated with any SR source. Both radiation
beam generated by an ultra-relativistic electron in a magnetic
device and laser beam are solutions of the wave equation based on
paraxial approximation. It follows that they are similar in all
aspects. In the space-frequency domain SR beams appear as laser
beams whose transverse extents are large compared with the
wavelength. In practical situations (e.g. undulator, bending
magnet sources), radiation beams exhibit a virtual "waist" where
the wavefront is often plane. Remarkably, the field distribution
of a SR beam across the waist turns out to be strictly related
with the inverse Fourier transform of the far-field angle
distribution. Then, we take advantage of standard Fourier Optics
techniques and apply the Fresnel propagation formula to
characterize the SR beam. Altogether, we show that it is possible
to reconstruct the near-field distribution of the SR beam outside
the magnetic setup from the knowledge of the far-field pattern.
The general theory of SR in the near-zone developed in this paper
is illustrated for the special cases of undulator radiation, edge
radiation and transition undulator radiation. Using known
analytical formulas for the far-field pattern and its inverse
Fourier transform we find analytical expressions for near-field
distributions in terms of far-field distributions. Finally, we
compare these expressions with incorrect or incomplete literature.
\end{abstract}


\end{frontmatter}

% main text

\clearpage

\section{\label{sec:intro} Introduction}

In a series of previous works \cite{OUR1,OUR2,OUR3} we developed a
formalism which, we believe, is ideally suited for analysis of any
Synchrotron Radiation problem. In particular, in \cite{OUR3} we
took advantage of Fourier Optics ideas. Fourier Optics (see e.g.
\cite{GOOD}) provides an extremely successful approach which
revolutionized the treatment of wave optics problems and, in
particular, laser beam optics problems. It deals with paraxial
fields, propagating along a chosen direction (say the longitudinal
$z$-axis) and directions close to it. In any given plane
transverse to the $z$-axis, the field is represented in the
space-frequency domain by a complex scalar function.
Alternatively, one may deal with a monochromatic signal
oscillating as $\exp[-i\omega t]$, $\omega$ being the frequency
and $t$ being the time. Then, one can represent the slowly
varying\footnote{With respect to the wavelength $\lambda = 2 \pi
c/\omega$.} field envelope with a complex scalar function
$\widetilde{E}(\vec{r}_\bot, z)$. Here $z$ is the longitudinal
position of the observer, and $\vec{r}_\bot$ is a two-dimensional
vector identifying transverse coordinates $(x,y)$ of the observer.
$\widetilde{E}(\vec{r}_\bot, z)$ represents an electric field
amplitude polarized everywhere in the same, fixed direction. The
effects of free-space propagation are mathematically represented
by the Fresnel formula. Elementary systems, such as lenses and
filters placed along the $z$ axis are introduced as complex
scalar-amplitude transmission functions $T(\vec{r}_\bot)$, which
yield the output field after multiplication with an input field.
More complex systems are realized by stacking free-space sections
and amplitude transmission functions in series.

The use of Fourier Optics led us in a natural way to establish, in
this work, basic foundations for the treatment of SR fields in
terms of laser beam optics. In particular, radiation from an
ultra-relativistic particle can be interpreted as radiation from a
virtual source, which produces a laser-like beam. The virtual
source itself is the waist of the laser-like beam, and often
exhibits a plane wavefront. In this case it is completely
specified, for any given polarization component, by a real-valued
amplitude distribution of the field. As we will see, the
laser-like representation of SR is intimately connected with the
ultra-relativistic nature of the electron beam. In particular,
paraxial approximation always applies. Then, free space basically
acts as a spatial Fourier transformation. This means that the
field in the far zone is, aside for a phase factor, the Fourier
transform of the field at any position $z$ down the beamline. It
is also, aside for a phase factor, the Fourier transform of the
virtual source. Once the field at the virtual source is known, the
field at other longitudinal positions, both in the far and in the
near zone up to distances to the sources comparable with the
radiation wavelength, can be obtained with the help of the Fresnel
propagation formula. This is equivalent to state that the
near-zone field can be calculated from the knowledge of the
far-zone field, i.e. from the acceleration term of the
Lienard-Wiechert field only.  This is possible due to the paraxial
approximation. On the one hand, as said above, the Fresnel formula
can be used to recover the field up to longitudinal positions
which are a wavelength far away from the sources. On the other
hand, due to the paraxial approximation, the wavelength of
interest in SR emission is very short compared with the radiation
formation length and with the linear dimensions of the magnetic
setup. In this view, it is incorrect to consider the far-zone
field as an asymptotic expression only: its knowledge completely
specifies, through the Fresnel integral, the near-zone field as
well. In the case when the electron generating the field is not
ultra-relativistic, though, the paraxial approximation cannot be
applied. As a result, the radiation wavelength is comparable with
the radiation formation length. It is then impossible to
reconstruct the near-field distribution from the knowledge of the
far-field pattern\footnote{Our arguments hold in the case near
zone and formation zone coincide. For example, this is the case
for bending magnet radiation. However, depending on the parameter
choice, this may not be the case for complex magnetic setups. We
will treat this subject in a more extensive way in Section
\ref{sec:dis}}.

In this work we consider a SR source composed of an arbitrary
number of magnetic devices under the standard ultra-relativistic
approximation. We show that every such magnetic setup is
equivalent to several virtual sources and can be modelled as a
sequence of appropriately chosen plane sources inserted between
the edges of each magnetic structure. This equivalence provides
conceptual insight regarding SR sources and should facilitate
their design and analysis. In fact, great simplifications in the
solution for the problem of propagation of radiation through a
beamline result after adopting our viewpoint. Since the analysis
of SR sources can be reduced to that of laser-like sources, it
follows that any result, method of analysis or design and any
algorithm specifically developed for laser beam optics (e.g. the
code ZEMAX \cite{ZEMA}) are also applicable to SR sources.

To give an example of the applicability of our method we first
consider the case of undulator radiation. We work within the
applicability region of the resonance approximation. This means
that we consider an undulator with a large number $N_w$ of
periods, i.e. $N_w \gg 1$, and that we are interested in
frequencies near the fundamental. We find the field distribution
of the virtual source with the help of the far-zone field
distribution and we propagate to any distance of interest.

Subsequently, we treat the relatively simple case of edge
radiation, studying the emission from a setup composed by a
straight section and two bending magnets, upstream and downstream
of the straight section. Bright, long-wavelength edge radiation is
expected to be emitted by such setup for $\lambda \gg \lambda_c$,
$\lambda_c$ being the critical wavelength of SR produced by
bending magnets. In particular, we derived an expression for the
field from a straight section that is valid at arbitrary
observation position. This result is fundamental importance
because, due to the superposition principle, our expression may be
used as building block for more complicated setups. We begin
calculating an analytical expression for the edge radiation (see,
among others, \cite{BOSF,CHUN,BOS4,BOS7, PROY, BOS2, BOS3, WILL})
from a single electron in the far zone. Then, through the
virtual-source technique we find novel results in the near zone.
Two alternative description of the field propagation are given,
based on a single virtual source located in the middle of the
edge-radiation setup and based on two virtual sources located at
the edges of the setup.

With the help of the knowledge of the edge-radiation case, we turn
to analyze a more complicated setup consisting of an undulator
preceded and followed by two straight sections and two (upstream
and downstream) bending magnets. We restrict ourselves to the
long-wavelength region. Edge radiation from this kind of setup is
commonly known as Transition Undulator Radiation (TUR). The first
study on TUR appeared more than a decade ago \cite{KIM1}. In that
work it was pointed out for the first time that, since an electron
entering or leaving an undulator experiences a sudden change in
longitudinal velocity, radiation with broadband spectral features,
similar to transition radiation, had to be expected in the
low-frequency region in addition to the usual undulator radiation.
That paper constituted a theoretical basis for many other studies,
developed in the years that followed. We remind here a few
\cite{BOS4,KINC,CAST,BOS6,ROY2}, dealing both with theoretical and
experimental issues. More recently, TUR has been given
consideration in the framework of large X-Ray Free Electron Laser
(XFEL) projects. For example, a characterization of Coherent
Transition Radiation from the LCLS undulator has been given in
\cite{REI1}. A method to obtain intense infrared/visible light
pulses naturally synchronized to x-ray pulses from the LCLS XFEL
by means of Coherent TUR has been proposed in \cite{ADA1}, that
should be used for pump-probe experiments in the femtosecond scale
resolution. In view of these applications, there is a need to
extend the knowledge of TUR to the near zone and to the coherent
case\footnote{In this paper we limit our analysis to the
single-particle case, leaving the study of coherent radiation for
the future}.

Specification of what precedes and follows the undulator is of
fundamental importance. As has been recognized in particular for
TUR many years ago \cite{BOS4}, it does make sense to discuss
about the intensity distribution of TUR only when a detailed
knowledge is given, of what precedes and follows the undulator in
a specific setup. According to our novel approach, the two
straight sections and the undulator will be associated to virtual
sources with plane wavefronts. The field from the setup can then
be described, in the near as well as in the far field, as a
superposition of laser-like beams, radiating at the same
wavelength and separated by different phase shifts.

Our work is organized as follows. In Section \ref{sec:meth} we
discuss our method in full generality. In fact, the analogy with
laser beams can be extended to include SR from setups not
discussed in this paper.  In the following Section \ref{sec:dis}
we discuss our method in more detail and we focus on particular
issues like the importance of the paraxial approximation, the
accuracy of our calculations and a easily misleading relationship
between velocity and radiation field. In Section \ref{sec:exa} we
apply our algorithm to the case of usual undulator radiation.
Section \ref{sec:four} is dedicated to the study of the simplest
edge radiation setup, composed of a straight section and two
bending magnets. An undulator is then added to the setup in
Section \ref{sec:five}, where radiation is given as a
superposition of laser-like beam from different virtual sources.
Our findings are compared with results in literature in Section
\ref{sec:crit}. Finally, in Section \ref{sec:conc}, we come to
conclusions.

\section{\label{sec:meth} Method for obtaining the electromagnetic field generated by an ultra-relativistic electron}

In the present Section \ref{sec:meth} we discuss our method in all
generality. In the next paragraphs \ref{sub:der1} and
\ref{sub:der2} we show that radiation from an ultra-relativistic
particle can be represented as a laser-like beam with a given
virtual source. Our algorithm is based on two separately
well-known expressions, the Fresnel propagation formula and a
far-field presentation of the electromagnetic field.

\subsection{\label{sub:der1} Inverse problem with far-field data}

We represent the electric field in time domain $\vec{E}(\vec{r},
t)$ as a time-dependent function of an observation point located
at position $\vec{r}=\vec{r}_{\bot}+z\vec{z}=x \vec{x}+y\vec{y}
+z\vec{z}$, where $\vec{x}$, $\vec{y}$ and $\vec{z}$ are defined
as (adimensional) unit vectors along horizontal, vertical and
longitudinal direction. The field $\vec{E}(\vec{r}, t)$ satisfies,
in free space, the source-free wave equation:

\begin{eqnarray}
\left[\nabla^2 -\frac{1}{c^2} \frac{\partial^2}{\partial
t^2}\right] \vec{E} = 0~.\label{timedw}
\end{eqnarray}
Here $c$ indicates the speed of  light in vacuum. For
monochromatic waves of angular frequency $\omega$, the wave
amplitude has the form

\begin{eqnarray}
\vec{E}_\bot(z,\vec{r}_\bot,t) =
\vec{\bar{E}}_\bot(z,\vec{r}_\bot) \exp\left[-i\omega t\right] +
C.C.~, \label{wavemo}
\end{eqnarray}
where "C.C." indicates the complex conjugate of the preceding term
and $\vec{\bar{E}}_\bot$ describes the variation of the wave
amplitude. The vector $\vec{\bar{E}}_\bot$ actually represents the
amplitude of the electric field in the space-frequency domain.

We assume that the ultra-relativistic approximation is satisfied,
that is always the case for SR setups. In this case paraxial
approximation applies \cite{OUR1}. Paraxial approximation implies
a slowly varying envelope of the field with respect to the
wavelength. It is therefore convenient to introduce the slowly
varying envelope of the transverse field components as

\begin{equation}
\vec{\widetilde{E}}_\bot = \vec{\bar{E}}_\bot \exp{\left[-i\omega
z/c\right]}~, \label{vtilde}
\end{equation}
where the propagation constant $\omega/c$ is related to the
wavelength $\lambda$ by $\omega/c = 2\pi/\lambda$.  In paraxial
approximation and in free space, the following parabolic equation
holds for the complex envelope ${\widetilde{E}}$ of the Fourier
transform of the electric field along a fixed polarization
component:

\begin{equation}
\left({ {\nabla}_\bot}^2 + \frac{2 i \omega}{c}
\frac{\partial}{{\partial
 {z}}}\right) {\widetilde{E}}_\bot = 0 ~.\label{parabo}
\end{equation}
The derivatives in the Laplacian operator ${ {\nabla}_\bot}^2$ are
taken with respect to the transverse coordinates. One has to solve
Eq. (\ref{parabo}) with a given initial condition at ${z}$, which
defines a Cauchy problem. Note that Eq. (\ref{parabo}) is similar
to the time-dependent Schroedinger equation.  We obtain

\begin{equation}
{ \widetilde{E}_\bot}( {z}_o,\vec {r}_{o\bot}) = \frac{i \omega}{2
\pi c( {z}_o- {z})} \int d \vec{ {r}'}_{\bot}~
\widetilde{E}_\bot(z,\vec{r'}_{\bot}) \exp{\left[\frac{i \omega
\left|{\vec{ {r}}}_{o\bot}-\vec{ {r}'_{\bot}}\right|^2}{2 c (
{z}_o- {z})}\right]}~, \label{fieldpropback}
\end{equation}
where the integral is performed over the transverse plane. It is
important to note that since $\tilde{E}$ is a slowly-varying
function with respect to the wavelength, within the accuracy of
the paraxial approximation one cannot resolve the evolution of the
field on a longitudinal scale of order of the wavelength. In order
to do so, the paraxial equation Eq. (\ref{parabo}) should be
replaced by a more general Helmholtz equation.

Next to the propagation equation for the field in free space, Eq.
(\ref{fieldpropback}), we can discuss a propagation equation for
the spatial Fourier transform of the field, which can also be
derived from Eq. (\ref{parabo}) and will be useful later on. We
will indicate the spatial Fourier transform of
$\widetilde{E}_\bot(z,\vec{r'}_{\bot})$ with $\mathrm{F}(z,
\vec{u})$\footnote{For the sake of completeness we explicitly
write the definitions of the two-dimensional Fourier transform and
inverse transform of a function $g(\vec{r})$ in agreement with the
notations used in this paper. The Fourier transform and inverse
transform pair reads:

\begin{eqnarray}
\nonumber \tilde{g}(\vec{k}) = \int \d \vec{r} ~g(\vec{r})
\exp\left[i \vec{r} \cdot \vec{k} \right]~;~~{g}(\vec{r}) =
\frac{1}{4\pi^2} \int \d \vec{k} ~\tilde{g}(\vec{k}) \exp\left[-i
\vec{r} \cdot \vec{k} \right] , \label{spatgft}
\end{eqnarray}
the integration being understood over the entire plane. If $g$ is
circular symmetric we can introduce the Fourier-Bessel transform
and inverse transform pair:

\begin{eqnarray}
\nonumber \tilde{g}({k}) = 2\pi \int_0^{\infty} d{r}~ r g(r) J_o(k
r)~;~~ {g}({r}) = \frac{1}{2\pi} \int_0^{\infty} d{k}~ k
\tilde{g}(k) J_o(k r)~,\label{bessspatgft}
\end{eqnarray}
$r$ and $k$ indicating the modulus of the vectors $\vec{r}$ and
$\vec{k}$ respectively, and $J_o$ being the zero-th order Bessel
function of the first kind.}:

\begin{equation}
\mathrm{F}\left(z,\vec{u}\right) = \int d \vec{r'}_{\bot}
\widetilde{E}_\bot(z,\vec{r'}_{\bot}) \exp{\left[i
\vec{r'}_{\bot}\cdot\vec{u}\right]}~. \label{ftfield}
\end{equation}
Eq. (\ref{parabo}) can be rewritten in terms of $\mathrm{F}$ as

\begin{equation}
\left({ {\nabla}_\bot}^2 + \frac{2 i \omega}{c}
\frac{\partial}{{\partial {z}}}\right) \left\{\int d\vec{u}~
\mathrm{F}\left(z,\vec{u}\right) \exp{\left[-i
\vec{r}_{\bot}\cdot\vec{u}\right]}\right\} = 0 ~.\label{paraboft}
\end{equation}
Eq. (\ref{paraboft}) requires that

\begin{equation}
\left(-\left|\vec{u}\right|^2+ \frac{2 i \omega}{c}
\frac{\partial}{{\partial {z}}}\right)
\mathrm{F}\left(z,\vec{u}\right) = 0 ~.\label{paraboft2}
\end{equation}
Solution of Eq. (\ref{paraboft2}) can be presented as

\begin{eqnarray}
\mathrm{F}\left(z,\vec{u}\right) =
\mathrm{F}\left(0,\vec{u}\right) \exp{\left[-\frac{i c |\vec{
u}|^2 z}{2 \omega}\right]}~. \label{fieldpropback35}
\end{eqnarray}
It should be noted that the definition of $\mathrm{F}(0, \vec{u})$
is a matter of initial conditions. In many practical cases,
including the totality of the situation treated in this paper,
$\mathrm{F}$ (or $\tilde{E}$) may have no direct physical meaning
at $z=0$. For instance, in all cases considered in this paper,
$z=0$ is within the magnetic setup. However, $\mathrm{F}(0,
\vec{u})$ can be considered as the spatial Fourier transform of
the field produced by a \textit{virtual} source. Such a source is
defined by the fact that, supposedly placed at $z=0$, it would
produce, at any distance from the magnetic system under study, the
same field as the real source does. The result in Eq.
(\ref{fieldpropback35}) is very general. On the one hand, the
spatial Fourier transform of the electric field exhibits an almost
trivial behavior in $z$, since $|\mathrm{F}(z)|^2 =
\mathrm{const}$. On the other hand, the behavior of the electric
field itself is not trivial at all. These properties directly
follow from the propagation equation for the field and its Fourier
transform.

Let us discuss the physical meaning of Eq.
(\ref{fieldpropback35}). The spatial Fourier transform of the
field, $\mathrm{F}(z,\vec{u})$, may be interpreted as a
superposition of plane waves (the so-called \textit{angular
spectrum}).  Once the frequency $\omega$ is fixed, the wave number
$k=\omega/c$ is fixed as well, and a given value of the transverse
component of the wave vector $\vec{k}_\bot=\vec{u}$ corresponds to
a given angle of propagation of a plane wave. Different
propagation directions correspond to different distances travelled
to get to a certain observation point. Therefore, they also
correspond to different phase shifts, which depend on the position
along the $z$ axis (see, for example, reference \cite{GOOD}
Section 3.7). Free space basically acts as a Fourier
transformation. This means that the field in the far zone is,
phase factor and proportionality factor aside, the spatial Fourier
transform of the field at any position $z$. We now demonstrate
this fact. This proof will directly result in an operative method
to calculate the field at the virtual source, once the far field
is known.  We first recall that if we know the field at a given
position $(z,\vec{r'}_{\bot})$ we may use Eq.
(\ref{fieldpropback}) to calculate the field at another position
$(z_o,\vec{r}_{o\bot})$ . Let us now consider the limit $z_o
\longrightarrow \infty$, with finite ratio $\vec{r}_{o\bot}/z_o$.
In this case, the exponential function in Eq.
(\ref{fieldpropback}) can be expanded giving

\begin{eqnarray}
{ \widetilde{E}}_\bot( {z}_o,\vec{r}_{o\bot}) &=& \frac{i
\omega}{2 \pi c {z}_o} \int d \vec{ {r}'}_{\bot}~
\widetilde{E}_\bot(z,\vec{r'}_{\bot}) \cr &&\times
\exp{\left[\frac{i \omega}{{2 c {z}_o}}\left( |\vec{
{r}}_{o\bot}|^2-2 \vec{ {r}}_{o\bot}\cdot\vec{
{r}'}_{\bot}+\frac{z |\vec{
{r}}_{o\bot}|^2}{{z}_o}\right)\right]}~.\cr &&
\label{fieldpropback2}
\end{eqnarray}
Letting $\vec{\theta} = \vec{r}_{o\bot}/z_o$ we have

\begin{eqnarray}
{ \widetilde{E}}_\bot( \vec{\theta} )&=& \frac{i \omega}{2 \pi c
{z}_o} \exp{\left[\frac{i \omega |\vec{ \theta}|^2}{2
c}(z_o+z)\right]} \mathrm{F}\left(z,-\frac{\omega
\vec{\theta}}{c}\right)~. \label{fieldpropback3}
\end{eqnarray}
With the help of Eq. (\ref{fieldpropback35}), Eq.
(\ref{fieldpropback3}) may be presented as

\begin{eqnarray}
{ \widetilde{E}}_\bot(\vec{\theta}) &=& \frac{i \omega}{2 \pi c
{z}_o} \exp{\left[\frac{i \omega |\vec{ \theta}|^2}{2
c}z_o\right]} \mathrm{F}\left(0,-\frac{\omega
\vec{\theta}}{c}\right)~ . \label{fieldpropback3tris}
\end{eqnarray}
Eq. (\ref{fieldpropback3tris}) shows what we wanted to
demonstrate: free space basically acts as a Fourier
transformation. Also, from Eq. (\ref{fieldpropback3tris}) one
directly obtains

\begin{eqnarray}
{ \widetilde{E}}_\bot( 0,\vec{r}_{\bot} )&=& \frac{i \omega
{z}_o}{2 \pi c} \int d\vec{\theta}\exp{\left[-\frac{i \omega
|\vec{ \theta}|^2}{2 c}z_o\right]}{
\widetilde{E}}_\bot(\vec{\theta}) \exp\left[\frac {i \omega}{c}
\vec{r}_{\bot}\cdot \vec{\theta}\right] ~ , \label{virfie}
\end{eqnarray}
where the transverse vector $\vec{r}_{\bot}$ defines a transverse
position on the virtual source plane at $z=0$. Eq. (\ref{virfie})
allows to calculate the field at the virtual source once the field
in the far zone is known. Identification of the position $z=0$
with the virtual source position is always possible, but not
always convenient (although often it is). In general, if the
virtual source is at position $z=z_s$, one obtains, from Eq.
(\ref{fieldpropback3}) that Eq. (\ref{virfie}) should be modified
to

\begin{eqnarray}
{ \widetilde{E}}_\bot( z_s,\vec{r}_{\bot} )&=& \frac{i \omega
{z}_o}{2 \pi c} \int d\vec{\theta}\exp{\left[-\frac{i \omega
|\vec{ \theta}|^2}{2 c}(z_o+z_s)\right]}{
\widetilde{E}}_\bot(\vec{\theta}) \exp\left[\frac {i \omega}{c}
\vec{r}_{\bot}\cdot \vec{\theta}\right] ~ . \label{virfiemody}
\end{eqnarray}

\subsection{\label{sub:der2} Far-field pattern calculations}

All is left to do in order to solve the propagation problem in
paraxial approximation is to calculate the field in the far zone.
We indicate the electron velocity in units of $c$ with
$\vec{\beta}(t')$, the Lorentz factor (that will be considered
fixed throughout this paper) with $\gamma$, the (negative) charge
of the electron with $(-e)$ and the electron trajectory as
$\vec{r'}(t')$. Finally, the direction from the retarded position
of the electron to the observer is fixed by the unit vector
$\vec{n}$,

\begin{equation}
\vec{n} =
\frac{\vec{r}_o-\vec{r'}(t')}{|\vec{r}_o-\vec{r'}(t')|}~.
\label{enne}
\end{equation}
Condition $\vec{n} = \mathrm{constant}$ defines the far-field
zone. In this zone, a widely-used result presented in textbooks
(see \cite{JACK}) holds:

\begin{eqnarray}
\vec{\bar{E}}(\vec{r_o},\omega) &=& -{i\omega e\over{c
r_o}}\exp\left[\frac{i \omega}{c}\vec{n}\cdot\vec{r}_o\right]
\int_{-\infty}^{\infty}
dt'~{\vec{n}\times(\vec{n}\times{\vec{\beta}})}\exp
\left[i\omega\left(t'-{1\over{c}}{\vec{n}\cdot
\vec{r'}}\right)\right] .\cr && \label{revwied}
\end{eqnarray}
The transverse components of the envelope of the field in Eq.
(\ref{revwied}) can be written as (see \cite{OUR1})

\begin{eqnarray}
\vec{\widetilde{{E}}}_\bot(z_o, \vec{r}_{\bot o},\omega) &=& -{i
\omega e\over{c^2}z_o} \int_{-\infty}^{\infty} dz' {\exp{\left[i
\Phi_T\right]}}  \left[\left({v_x(z')\over{c}} \right.\right. \cr
&&\left.\left. -{x_o-x'(z')\over{z_o}}\right){\vec{x}}
+\left({v_y(z')\over{c}}-{y_o-y'(z')\over{z_o}}\right){\vec{y}}\right]
~, \label{generalfin}
\end{eqnarray}
where the total phase $\Phi_T$ is

\begin{equation}
\Phi_T = \omega \left[{s(z')\over{v}}-{z'\over{c}}\right]+
\omega\left(\frac{1}{z_o}+\frac{z'}{z_o^2}\right)
{\left[x_o-x'(z')\right]^2+\left[y_o-y'(z')\right]^2 \over{2c}}~
.\label{totph}
\end{equation}
Eq. (\ref{generalfin}) can be obtained starting directly with
Maxwell's equations in the space-frequency domain. Textbooks (see
for example \cite{JACK}) usually follow a not so direct
derivation. They start with the solution of Maxwell's equation in
the space-time domain, the well-known Lienard-Wiechert expression,
and they subsequently apply a Fourier transformation. Eq.
(\ref{generalfin}) is automatically subject to the paraxial
approximation. Here $v_x(z')$ and $v_y(z')$ are the horizontal and
the vertical components of the transverse velocity of the
electron, while $x'(z')$ and $y'(z')$ specify the transverse
position of the electron as a function of the longitudinal
position. Finally, we define the curvilinear abscissa $s(z')=v
t'(z')$, $v$ being the modulus of the velocity of the electron.

Eq. (\ref{generalfin}) can be used to characterize the far field
from an electron moving on any trajectory as long as the
ultra-relativistic approximation is satisfied. Then, once the far
field is known, Eq. (\ref{virfie}) (or Eq. (\ref{virfiemody})) can
be used to calculate the field distribution at the virtual source.
Finally, Eq. (\ref{fieldpropback}) solves the propagation problem
at any observation position $z_o$. Note that part of the phase in
Eq. (\ref{totph}) compensates with the phase in $|\vec{\theta}|^2$
in Eq. (\ref{virfie}). If Eq. (\ref{generalfin}) describes a field
with a spherical wavefront with center at $z=0$, such compensation
is complete. The centrum of the spherical wavefront is a
privileged point, and the plane at $z=0$ exhibits a plane
wavefront. This explains why the point $z=0$ is often privileged
with respect to others.

Let us conclude this Section with a short summary. We considered
the electric field represented in the space-frequency domain.
Since the system is ultra-relativistic, paraxial approximation
applies. We saw that within the applicability region of the
paraxial approximation, all information about the field, both in
the near and in the far zone, is encoded in the far zone field
alone. An explicit expression for the near field can in fact be
recovered with the help of the paraxial Green's function for
Maxwell's equation. Such Green's function is the same Green's
function considered in Fourier Optics. This consideration allows
to describe simple ultra-relativistic systems like, for instance,
an electron moving through a bending magnet, an undulator or a
straight section in terms of laser-like beams. Each beam has a
virtual source, which often exhibits a plane wavefront and is
completely specified, for any given polarization component, by a
real-valued amplitude distribution for the field. Note that the
square modulus of such distribution can be physically associated
with an intensity, and can be detected as the image of a properly
placed, ideal converging lens. The Fourier transform of such
distribution instead is linked to the electric field in the far
zone. Then, once the electric field in the far zone is calculated,
a virtual source for the laser-like beam can be specified. Once
the field at the virtual source is known, the field at any
position can be found by applying usual propagation equations
based on Fourier Optics.

\section{\label{sec:dis} Discussion}

In the present Section \ref{sec:dis} we discuss the algorithm
proposed in Section \ref{sec:meth}. First we present a thorough
derivation of Eq. (\ref{generalfin}) based on paraxial Green's
function techniques. Then we discuss the region of applicability
and the accuracy of this derivation. Finally, we deal with some
delicate question related to the nature of the so-called "velocity
field".

\subsection{\label{sub:par} A derivation of the electric field
based on paraxial Green's function}

Eq. (\ref{generalfin}) is an expression that can be used to
characterize the far field from of an electron moving on any
trajectory in paraxial approximation. For the sake of completeness
we present, here, a derivation of Eq. (\ref{generalfin}) based on
\cite{OUR1}.

Accounting for electromagnetic sources, i.e. in a region of space
where current and charge densities are present, the following
equation for the field in the space-frequency domain holds in all
generality:

\begin{equation}
c^2 \nabla^2 \vec{\bar{E}} + \omega^2 \vec{\bar{E}} = 4 \pi c^2
\vec{\nabla} \bar{\rho} - 4 \pi i \omega \vec{\bar{j}}~,
\label{trdisoo}
\end{equation}
where $\bar{\rho}(\vec{r},\omega)$ and
$\vec{\bar{j}}(\vec{r},\omega)$ are the Fourier transforms of the
charge density $\rho(\vec{r},t)$ and of the current density
$\vec{j}(\vec{r},t)$. Eq. (\ref{trdisoo}) is the well-known
Helmholtz equation, that has elliptic characteristic. With the
help of Eq. (\ref{vtilde}), Eq. (\ref{trdisoo}) can be written as

\begin{equation}
c^2 \exp\left[{i\omega z/c}\right] \left( \nabla^2 +{2 i
\omega\over{c}}{\partial\over{\partial z}}\right)
\vec{\widetilde{E}} = 4 \pi c^2 \vec{\nabla} \bar{\rho} - 4 \pi i
\omega \vec{\bar{j}}~. \label{trdiso2}
\end{equation}
A system of electromagnetic sources in the space-time can be
conveniently described by $\rho(\vec{r},t)$ and
$\vec{j}(\vec{r},t)$.  In this paper  we will be concerned about a
single electron. Using the Dirac delta distribution, we can write,
following \cite{OUR1}

\begin{equation}
\rho(\vec{r},t) = -e \delta(\vec{r}-\vec{r'}(t)) =
-\frac{e}{v_z(z)}
\delta(\vec{r}_\bot-\vec{r'}_\bot(z))\delta\left(\frac{s(z)}{v}-t\right)\label{charge}
\end{equation}
and

\begin{eqnarray}
\vec{j}(\vec{r},t) &=& -{e} \vec{v}(t)
\delta(\vec{r}-\vec{r'}(t))= -\frac{e}{v_z(z)} \vec{v}(z)
\delta(\vec{r}_\bot-\vec{r'}_\bot(z))\delta\left(\frac{s(z)}{v}-t\right)
, \cr&& \label{curr}
\end{eqnarray}
where $\vec{r'}(t)$ and $\vec{v}(t)$ are, respectively, the
position and the velocity of the particle at a given time $t$ in a
fixed reference frame, and $v_z$ is the longitudinal velocity of
the electron.

In the space-frequency domain, Eq. (\ref{charge}) and Eq.
(\ref{curr}) transform to:

\begin{equation}
\bar{\rho}(\vec{r_\bot},z,\omega) = -{e\over{v_z(z)}}
\delta\left(\vec{r_\bot}-\vec{r'_\bot}(z)\right) \exp\left[{i
\omega s(z)/v}\right] \label{charge2tr}
\end{equation}
and

\begin{equation}
\vec{\bar{j}}(\vec{r_\bot},z,\omega) = -{e\over{v_z(z)}}
\vec{v}(z) \delta\left(\vec{r_\bot}-\vec{r'_\bot}(z)\right)
\exp\left[{i \omega s(z)/v}\right]~. \label{curr2tr}
\end{equation}
By substitution in Eq. (\ref{trdiso2}) we obtain

\begin{eqnarray}
\left({\nabla}^2 + {2 i \omega \over{c}} {\partial\over{\partial
z}}\right) \vec{\widetilde{E}} &=& {4 \pi e\over{v_z(z)}}
\exp\left\{{i \omega
\left({s(z)\over{v}}-{z\over{c}}\right)}\right\}  \cr && \times
\left[{i\omega\over{c^2}}\vec{v}(z)
\delta\left(\vec{r_\bot}-\vec{r'_\bot}(z)\right)-\vec{\nabla}
\delta\left(\vec{r_\bot}-\vec{r'_\bot}(z)\right)
\right]~.\label{incipit}
\end{eqnarray}
Eq. (\ref{incipit}) is still fully general and may be solved in
any fixed reference system $(x,y,z)$ of choice with the help of an
appropriate Green's function.

When the longitudinal velocity of the electron, $v_z$, is close to
the speed of light $c$ (i.e. $\gamma_z^2 \gg 1$), the Fourier
components of the source are almost synchronized with the
electromagnetic wave travelling at the speed of light. In this
case the phase $\omega ({s(z)/{v}}-{z/{c}})$ is a slow function of
$z$ compared to the wavelength. For example, in the particular
case of motion on a straight section, one has $s(z) = z/v_z$, so
that $\omega ({s(z)/{v}}-{z/{c}}) = \omega z/(2\gamma_z^2 c)$, and
if $\gamma_z^2 \gg 1$ such phase grows slowly in $z$ with respect
to the wavelength. For a more generic motion, one has

\begin{equation}
\omega \left({s(z_2)-s(z_1)\over{v}}-{z_2-z_1\over{c}}\right) =
\int_{z_1}^{z_2} d \bar{z} \frac{\omega}{2 \gamma_z^2(\bar{z})
c}~. \label{moregen}
\end{equation}
Mathematically, the phase in Eq. (\ref{moregen}) enters in the
Green's function solution of Eq. (\ref{incipit}) as a factor in
the integrand. As we integrate along $z'$, the factor
$\omega(s(z')/v - z'/c)$ leads to an oscillatory behavior of the
integrand over a certain integration range in $z'$. Such range can
be identified with the value of $z_2-z_1$ for which the right hand
side of Eq. (\ref{moregen}) is of order unity, and it is naturally
defined as the radiation formation length $L_f$ of the system at
frequency $\omega$. It is easy to see by inspection of Eq.
(\ref{moregen}) that if $v$ is sensibly smaller than $c$ (but
still of order $c$), i.e. $v\sim c$ but $1/\gamma_z^2 \sim 1$,
then $L_f \sim \lambda$. On the contrary, when $v$ is very close
to $c$, i.e. $1/\gamma_z^2 \ll 1$, the right hand side of Eq.
(\ref{moregen}) is of order unity for $L_f = z_2-z_1 \gg \lambda$.
When the radiation formation length is much longer than the
wavelength, $\vec{\widetilde{E}}$ does not vary much along $z$ on
the scale of $\lambda$, that is $\mid
\partial_z \widetilde{E}_{x,y}\mid \ll \omega/c \mid
\widetilde{E}_{x,y}\mid$. Therefore, the second order derivative
with respect to $z$ in the $\nabla^2$ operator on the left hand
side of Eq. (\ref{incipit}) is negligible with respect to the
first order derivative. Eq. (\ref{incipit}) can then be simplified
to

\begin{eqnarray}
\left({\nabla_\bot}^2 + {2 i \omega \over{c}}
{\partial\over{\partial z}}\right) \vec{\widetilde{E}}_\bot &=& {4
\pi e\over{c}} \exp\left\{{i \omega
\left({s(z)\over{v}}-{z\over{c}}\right)}\right\} \cr&& \times
\left[{i\omega\over{c^2}}\vec{v}_\bot(z)
\delta\left(\vec{r}_\bot-\vec{r'}_\bot(z)\right)-\vec{\nabla}_\bot
\delta\left(\vec{r}_\bot-\vec{r'}_\bot(z)\right) \right],\cr &&
\label{incipit2}
\end{eqnarray}
where we consider transverse components of $\vec{\widetilde{E}}$
only and we substituted $v_z(z)$ with $c$, having used the fact
that $1/\gamma_z^2 \ll 1$. Eq. (\ref{incipit2}) is Maxwell's
equation in paraxial approximation. In this way we transformed Eq.
(\ref{incipit}), which is an elliptic partial differential
equation, into Eq. (\ref{incipit2}), that is of parabolic type.
Note that the applicability of the paraxial approximation depends
on the ultra-relativistic assumption $\gamma^2 \gg 1$ but not on
the choice of the $z$ axis. In fact, if the longitudinal $z$
direction is chosen in such a way that $\gamma_z^2 \sim 1$, than
the formation length is very short $L_f \sim \lambda$, and the
radiated field is practically zero. As a result, Eq.
(\ref{incipit2}) can always be applied, i.e. the paraxial
approximation can always be applied, whenever $\gamma^2 \gg 1$.
Formally, this statement is in contradiction with the
approximation $1/\gamma_z^2 \ll 1$, explicitly declared to obtain
Maxwell's equation in paraxial approximation, i.e. Eq.
(\ref{incipit2}). However, suppose that one is using a detector
not accurate enough to distinguish, on an axis where $\gamma_z^2
\gg 1$, between solution of Maxwell's equation with paraxial
approximation and solution without it. Let us fix any observation
wavelength of interest $\lambda$. Then, with the accuracy of the
paraxial approximation, one would detect no photon flux along an
axis where $\gamma_z^2 \sim 1$. Thus, from a practical viewpoint,
the paraxial approximation can always be applied independently of
$\gamma_z$. When it is not formally applicable, the error
resulting from its application cannot be resolved by our detector,
within the accuracy of the paraxial approximation, because we are
working with ultra-relativistic particles, and radiation is highly
collimated. In this regard, it should be remarked here that the
status of the paraxial equation Eq. (\ref{incipit2}) in
Synchrotron Radiation theory is different from that of the
paraxial equation in Physical Optics. In the latter case, the
paraxial approximation is satisfied only by small observation
angles. For example, one may think of a setup where a thermal
source is studied by an observer positioned at a long distance
from the source and behind a limiting aperture. Only if a
small-angle acceptance is considered the paraxial approximation
can be applied. On the contrary, due to the ultra-relativistic
nature of the electrons that generate radiation, Synchrotron
Radiation is highly collimated. As a result the paraxial equation
can always be applied in practice, because it practically returns
zero field at angles where it should not be applied.

The Green's function for Eq. (\ref{incipit2}), namely the solution
corresponding to the unit point source, satisfies the equation

\begin{eqnarray}
\left({\nabla_\bot}^2 + {2 i \omega \over{c}}
{\partial\over{\partial z}}\right) G(z_o-z;\vec{r_{\bot
o}}-\vec{r_\bot}) =  \delta(\vec{r_{\bot
o}}-\vec{r_\bot})\delta(z_o-z)~, \label{greeneq}
\end{eqnarray}
and, in an unbounded region, can be explicitly written  as

\begin{eqnarray}
G(z_o-z';\vec{r}_{\bot o}-\vec{r'}_\bot) = -{1\over{4\pi
(z_o-z')}} \exp\left\{ i\omega{\mid \vec{r}_{\bot o}
-\vec{r'}_\bot\mid^2\over{2c (z_o-z')}}\right\} \label{green}~,
\end{eqnarray}
assuming $z_o-z' > 0$. When $z_o-z' < 0$ the paraxial
approximation does not hold, and the paraxial wave equation Eq.
(\ref{incipit2}) should be substituted, in the space-frequency
domain, by the more general Helmholtz equation. However, the
radiation formation length for $z_o - z'<0$ is very short with
respect to the case $z_o - z' >0$, i.e. there is no radiation for
observer positions $z_o-z' <0$. As a result, in this paper we will
consider only $z_o - z'> 0$. It follows that the observer is
located downstream of the sources.

This leads to the solution

\begin{eqnarray}
\vec{\widetilde{E}}_\bot(z_o, \vec{r}_{\bot o},\omega) &=&
-{e\over{c}} \int_{-\infty}^{\infty} dz' {1\over{z_o-z'}} \int d
\vec{r'}_{\bot} \cr && \times
\left[{i\omega\over{c^2}}\vec{v_\bot}(z')
\delta\left(\vec{r'}_{\bot}-\vec{r'}_\bot(z')\right)
-\vec{\nabla'}_\bot \delta\left(\vec{r'}_{\bot}
-\vec{r'}_\bot(z')\right)\right] \cr && \times
\exp\left\{i\omega\left[{\mid \vec{r}_{\bot o}-\vec{r'}_\bot
\mid^2\over{2c (z_o-z')}}+
\left({s(z')\over{v}}-{z'\over{c}}\right)\right] \right\} ~,
\label{blob}
\end{eqnarray}
where $\vec{\nabla'}_\bot$ represents the gradient operator with
respect to the source point. The integration over transverse
coordinates can be carried out leading to the final result:

\begin{eqnarray}
\vec{\widetilde{E}}_\bot(z_o, \vec{r}_{\bot o},\omega) &=& -{i
\omega e\over{c^2}} \int_{-\infty}^{\infty} dz' \frac{\exp\left[{i
\Phi_T}\right]}{z_o-z'} \left[\left({v_x(z')\over{c}}
\right.\right. \cr &&  \left.\left.
-{x_o-x'(z')\over{z_o-z'}}\right)\vec{x}
+\left({v_y(z')\over{c}}-{y_o-y'(z')\over{z_o-z'}}\right)\vec{y}\right]
~, \label{generalfin2}
\end{eqnarray}
where the total phase $\Phi_T$ is given by

\begin{equation}
\Phi_T = \omega \left[{s(z')\over{v}}-{z'\over{c}}\right]+ \omega
\left[
{\left(x_o-x'(z')\right)^2+\left(y_o-y'(z')\right)^2\over{2c
(z_o-z')}}\right]~. \label{totph2}
\end{equation}
As we will see in Section \ref{sub:hel}, Eq. (\ref{generalfin2})
is valid at any observation position $z_o$ such that the paraxial
approximation is valid, i.e. up to distances between the observer
and the electromagnetic sources comparable with the radiation
wavelength. In the far zone Eq. (\ref{generalfin2}) reduces to Eq.
(\ref{generalfin}).

\subsection{\label{sub:hel} Exact solution of Helmholtz equation}

Up to now we presented an algorithm based on the paraxial
approximation. It makes sense to ask what is the range of observer
positions where the paraxial approximation applies, and what is
the accuracy of our result in this range. We will demonstrate that
paraxial approximation holds with good accuracy up to positions of
the observer such that its distance from the electromagnetic
sources in the space-frequency domain, denoted with $d$, is of
order of $\lambda$.

In order to investigate the accuracy of our findings we have to
compare results from the paraxial treatment with results obtained
within a more general framework without the help of the paraxial
approximation. Therefore we go back to the most general Eq.
(\ref{trdisoo}), that can be solved by direct application of the
Green's function for the Helmholtz equation

\begin{eqnarray}
G(\vec{r}_{o}-\vec{r'}) &=& -{1\over{4\pi
\left|\vec{r}_{o}-\vec{r'}\right| }} \exp\left\{
i{\omega\over{c}}{\mid \vec{r}_{o} -\vec{r'}\mid}\right\}
\label{greenhyp}~,
\end{eqnarray}
that automatically includes the Sommerfeld radiation condition

\begin{equation}
|\vec{r_{o}}-\vec{r'}|\left(\frac{\partial G}{\partial \vec{n}}-i
\frac{c G}{\omega} \right) \longrightarrow 0 ~~~~\mathrm{as}~~
|\vec{r_{o}}-\vec{r'}| \longrightarrow \infty~.\label{reso}
\end{equation}
Integrating by parts the term in $\vec{\nabla}\bar{\rho}$ we have

\begin{eqnarray}
\vec{\bar{E}} &=& -\int d\vec{r'} \left[{i\omega \over{c
\left|\vec{r_{o}}-\vec{r'}\right| }} \left(\bar{\rho} \vec{n} -
{\vec{\bar{j}}\over{c}}\right)+{\bar{\rho}
\vec{n}\over{\left|\vec{r_{o}}-\vec{r'}\right|^2}}\right]
\exp\left\{ i{\omega\over{c}}{\mid \vec{r_{o}}
-\vec{r'}\mid}\right\} \label{hypE}~.
\end{eqnarray}
Use of explicit expressions for $\bar{\rho}$, i.e. Eq.
(\ref{charge2tr}), and for $\vec{\bar{j}}$, i.e. Eq.
(\ref{curr2tr}), leads straightforwardly to the following
expression for the field envelope:

\begin{eqnarray}
\vec{\widetilde{E}}(\vec{r_o},\omega) &=& -{i\omega e\over{c}}
\int_{-\infty}^{\infty} dz'
{1\over{v_z(z')}}\left[{\vec{\beta}-\vec{n}\over{|\vec{r_o}-\vec{r'}(z')|}}-{ic\over{\omega}}{\vec{n}\over{
|\vec{r_o}-\vec{r'}(z')|^2}}\right]\cr && \times
\exp\left\{i\omega\left[\left({s(z')\over{v}}-{z'\over{c}}\right)+
\left({|\vec{r_o}-\vec{r'}(z')|\over{c}}-{z_o-z'\over{c}}\right)\right]\right\}~.
\label{rev2zeta}
\end{eqnarray}
It should be emphasized that in usual Physical Optics cases one is
interested in solving Maxwell's equations with particular boundary
conditions. Typically, boundary conditions are artificial and not
consistent with Maxwell's equation. Thus, it is not possible to
control approximations applied. On the contrary, in our case, Eq.
(\ref{rev2zeta}) is an exact solution of Maxwell's equations with
boundary conditions at infinity. In this case the scalar Green's
function in Eq. (\ref{greenhyp}) can be used without introducing
any extra assumption. These situations are completely different.
The exactness of the solution in Eq. (\ref{rev2zeta}) allows us to
control the overall accuracy of the paraxial approximation by
comparing results from  Eq. (\ref{generalfin2}) with results from
the exact solution of Maxwell's equations, Eq. (\ref{rev2zeta}).
We can present here a conservative estimate. Let us pose

\begin{equation}
\lambdabar = \frac{\lambda}{2\pi}~. \label{lbar}
\end{equation}
When $d \gtrsim L_f$ this accuracy can be estimated to amount to
$\lambdabar/L_f$, but quickly decreases as $L_f \gg d \gg
\lambdabar$ remaining, \textit{at least}, of order of
$\lambdabar/d$. The words "at least" emphasize the
before-mentioned fact that this is a conservative estimate. %There
%are situations where the accuracy of the paraxial approximation is
%of order $1/\gamma^2$ and, thus, completely independent of the
%distance $d$ between observer and electromagnetic sources. In
%Section \ref{sec:four} we will discuss an example of this
%situation, constituted by ultra-relativistic electrons moving on a
%straight section preceded and followed by switchers (e.g. bending
%magnets) whose function is that of effectively switch on and off
%the electromagnetic source at the harmonic of interest. However,
%the fact remains, that comparison of Eq. (\ref{generalfin2}) with
%Eq. (\ref{rev2zeta}) allows full control of the accuracy of the
%paraxial approximation.

The most general conclusion that can be drawn from Eq.
(\ref{rev2zeta}) is that, independently of any approximation, the
use of the space-frequency domain automatically presents a very
natural scale, the wavelength $\lambda$. As it can be verified by
inspecting Eq. (\ref{rev2zeta}), the integrand term scaling as
$|\vec{r_o}-\vec{r'}(z')|^{-2}$ can be dropped whenever $d \gg
\lambda$. This is to be taken as a sufficient condition, but not
necessary in general. Dropping this term is automatically enforced
when the paraxial approximation is applied, but it is not
sufficient for the paraxial approximation to hold. In order for
the paraxial approximation to hold, as seen before, the system
must be ultra-relativistic.

While the present discussion about Helmholtz equation is important
to understand the region of applicability of Eq.
(\ref{generalfin2}) we should stress the fact that the inverse
field problem, which relies on far-field data only, cannot be
solved without application of the paraxial approximation. In fact,
let us suppose that the paraxial approximation was not applicable.
Then, instead of Eq. (\ref{parabo}) we should solve the
homogeneous Helmholtz equation

\begin{equation}
c^2 \nabla^2 \vec{\bar{E}} + \omega^2 \vec{\bar{E}} = 0~.
\label{Hho}
\end{equation}
Boundary conditions are now constituted by the knowledge of the
field on a open surface (for example, a transverse plane) and
additionally, Rayleigh-Sommerfeld radiation condition (separately
for all polarization components) at infinity, i.e.

\begin{equation}
|\vec{r_{o}}-\vec{r'}|\left(\frac{\partial \bar{E}_{x,y}}{\partial
\vec{n}}-i \frac{c \bar{E}_{x,y}}{\omega} \right) \longrightarrow
0 ~~~~\mathrm{as}~~ |\vec{r_{o}}-\vec{r'}| \longrightarrow
\infty~.\label{reso2}
\end{equation}
However, this is not enough to reconstruct the field at any
position in space. In order to do so, we would need to specify the
sources. In this case though, Eq. (\ref{hypE}) can be directly
applied. The boundary conditions specified above allow one to
solve the direct transmission problem, but not the inverse one.
However, as we have seen, if the paraxial approximation is
applicable, the inverse field problem has unique and stable
solution.

Based on this discussion we can formulate a definition of several
observation zones of interest (together with their complementary
zones in square brackets).

\begin{itemize}
\item \textbf{Far zone [Near zone].}  Defined by $d$ such that $\vec{n} = \mathrm{const.}$ [for the near zone read: $\vec{n} \ne \mathrm{const.}$]
throughout the integration in Eq. (\ref{rev2zeta}).
\item \textbf{Formation zone [Non-formation zone].} Defined by $d \lesssim L_f$\footnote{In this region one may say that the part of the electron trajectory actively contributing to the field at the observer position is the nearest to the observer and of length of order $L_f$.} [for the non-formation zone read: $d \gg L_f$] .
\item \textbf{Radiation zone [Non-radiation zone] .} Defined by $d$ such that the electromagnetic
field is \textit{disentangled} [for the non-radiation zone read:
\textit{entangled}] from the particle field and can be interpreted
as radiation\footnote{This means that in the radiation zone the
flux of the Poynting vector through any transverse plane is
independent of its longitudinal position.}.
\item \textbf{$\mathbf{1/R-}$zone [$\mathbf{1/R^2-}$zone] .} Defined by $d$ such that the term
in $1/R^2=|\vec{r_o}-\vec{r'}(z')|^{-2}$ \textit{can be neglected}
[for the ${1/R^2-}$zone read: \textit{cannot be neglected}] in the
integrand of Eq. (\ref{rev2zeta}).
\item \textbf{Reconstruction zone [Non-reconstruction zone].} Defined by $d$ such that a solution
of the inverse problem based on far-field data \textit{can be
given} [for the non-reconstruction zone read: \textit{cannot be
given}].
\end{itemize}
In all generality, from Eq. (\ref{rev2zeta}) follows that the
$1/R-$zone always coincides with the reconstruction zone, because
field can be reconstructed from the far-zone data if and only if
the term containing $1/R^2$ in the integrand of Eq.
(\ref{rev2zeta}) can be neglected. This is clear because the
far-zone data do not include the term in Eq. (\ref{rev2zeta}).
Moreover if an observer is in the $1/R-$zone (or in the
reconstruction zone), it necessarily belongs to the radiation zone
as well. However we cannot prove, in general, that if an observer
is in the radiation zone, it necessarily belongs to the $1/R-$zone
(or to the reconstruction zone).

%On the contrary, coincidence of these two zones with the radiation
%zone is not automatically satisfied in all generality. As a
%counter-example, one may think of the similarity between the field
%from a ultra-relativistic charge and the fields from a pulse of
%radiation. The method of virtual quanta (see \cite{JACK}) relies
%upon this similarity.

Let us now consider a slightly less general standpoint and
discuss, independently of the paraxial approximation, both
non-ultra-relativistic systems\footnote{Note that there is a
certain abuse of language in this definition, as it does not
include the case $\beta \ll 1$. }, such that $\beta\simeq 1$ and
$\gamma^2 \simeq 1$, and ultra-relativistic systems such that
$\gamma^2 \gg 1$ (and obviously $\beta \simeq 1$).

In the case of non-ultra-relativistic systems the characteristic
size of the system $a$, the formation length $L_f$ and the
radiation wavelength $\lambda$ are of the same
magnitude\footnote{This is almost always the case. It is possible,
however, to organize particular situations where contributions
from many structure sum up coherently within a limited angular and
spectral range. For instance, one may think of contributions from
a row of bends spaced ad hoc, as in a wiggler, behaving as an
array of antennas.}. As a result, there are only two observation
zones of interest. The first, such that $d \gg \lambdabar$,
incorporates far zone, non-formation zone, radiation zone,
$1/R-$zone and reconstruction zone. The second, such that
$d\lesssim \lambdabar$, incorporates near zone, formation zone,
non-radiation zone, $1/R^2-$zone and non-reconstruction zone.

In the case of ultra-relativistic systems instead, we can say that
$a \gtrsim L_f$ and $L_f \gg \lambda$. As a result the near zone
is now defined by $d \lesssim a$, the formation zone by $d
\lesssim L_f$, the radiation zone by $d \gg \lambdabar$. As for
the reconstruction zone and the $1/R-$zone we can say that a
necessary and sufficient condition to be in these zones is that $d
\gg \lambdabar$. %However, this last condition is not necessary. In
%particular, in Section \ref{sec:four} we will see that there are
%situations where the accuracy of the paraxial approximation is
%completely independent of the distance $d$ between observer and
%electromagnetic sources. Then, it is a known fact that the
%paraxial approximation can be applied only in the $1/R-$zone. It
%follows that condition  $d \gg \lambdabar$ is sufficient but not
%always necessary to be in the $1/R-$zone (and in the
%reconstruction zone).

\subsection{\label{sub:vel} Radiation contribution from the
"velocity field"}

As one can see, the case of ultra-relativistic system presents an
increased level of complexity with respect to the
non-ultra-relativistic system. This complexity is at the origin of
several misconceptions, that we here discuss.

We presented, in the space-frequency domain, an exact solution of
Maxwell's equation, Eq. (\ref{rev2zeta}) (and its paraxial form
Eq.(\ref{generalfin2})), starting directly with Maxwell's
equations in the space-frequency domain. As said before, textbooks
(see for example \cite{JACK}) usually follow a not so direct
derivation. They start with the solution of Maxwell's equation in
the space-time domain, the well-known Lienard-Wiechert expression,
and they subsequently apply a Fourier transformation. The
Lienard-Wiechert expression for the electric field of a point
charge $(-e)$ reads (see, e.g. \cite{JACK}):

\begin{eqnarray}
\vec{{E}}(\vec{r}_o,t) &=& -e {\vec{n}-\vec{\beta}\over{\gamma^2
(1-\vec{n}\cdot\vec{\beta})^2 |\vec{r}_o-\vec{r'}|^2}} -{e
\over{c}} {\vec{n}\times[(\vec{n}-\vec{\beta})\times
\dot{\vec{\beta}}]\over{(1-\vec{n}\cdot\vec{\beta})^2|\vec{r}_o-\vec{r'}|}}
. \label{LWt}
\end{eqnarray}
As before, $R=|\vec{r}_o-\vec{r'}(t')|$ denotes the displacement
vector from the retarded position of the charge to the point where
the fields are calculated. Moreover, $\vec{\beta}=\vec{v}/c$,
$\vec{\dot{\beta}}=\vec{\dot{v}}/c$, while $\vec{v}$ and
$\vec{\dot{v}}$ denote the retarded velocity and acceleration of
the electron. Finally, the observation time $t$ is linked with the
retarded time $t'$ by the retardation condition $R =  c(t-t')$. As
is well-known, Eq. (\ref{LWt}) serves as a basis for the
decomposition of the electric field into a sum of two quantities.
The first term on the right-hand side of Eq. (\ref{LWt}) is
independent of acceleration, while the second term linearly
depends on it. For this reason, the first term is called "velocity
field", and the second "acceleration field" \cite{JACK}. Velocity
fields differ from acceleration fields in several respects, one of
which is the behavior in the limit for a very large distance from
the electron. There one finds that the velocity fields decrease
like $R^{-2}$, while acceleration fields only decrease as
$R^{-1}$. Let us apply a Fourier transformation:

\begin{eqnarray}
\vec{\bar{E}}(\vec{r}_o,\omega) &=& -e\int_{-\infty}^{\infty} dt'
{\vec{n}-\vec{\beta}\over{\gamma^2 (1-\vec{n}\cdot\vec{\beta})^2
|\vec{r}_o-\vec{r'}|^2}} \exp
\left[i\omega\left(t'+{|\vec{r}_o-\vec{r'}(t')|\over{c}}\right)\right]
\cr &&-{e \over{c}}\int_{-\infty}^{\infty}dt'
{\vec{n}\times[(\vec{n}-\vec{\beta})\times
\dot{\vec{\beta}}]\over{(1-\vec{n}\cdot\vec{\beta})^2|\vec{r}_o-\vec{r'}|}}
\exp
\left[i\omega\left(t'+{|\vec{r}_o-\vec{r'}(t')|\over{c}}\right)\right]
. \label{LW}
\end{eqnarray}
As in Eq. (\ref{LWt}) one may formally recognize a velocity and an
acceleration term in Eq. (\ref{LW}) as well. Since Eq. (\ref{LW})
follows directly from Eq. (\ref{LWt}), that is valid in the time
domain, the magnitude of the velocity and acceleration parts in
Eq. (\ref{LW}), that include terms in $1/R^2$ and $1/R$
respectively, do not depend on the wavelength $\lambda$. On the
contrary, as we have seen, the terms including $1/R$ and $1/R^2$
in Eq. (\ref{rev2zeta}) do. It is instructive to take advantage of
integration by parts to show that Eq. (\ref{rev2zeta}) and Eq.
(\ref{LWt}) are equivalent. With the help of

\begin{eqnarray}
{1\over{c}}{d\over{dt'}}|\vec{r_o}-\vec{r'}(t')| =
-\vec{n}\cdot\vec{\beta}~. \label{usefulrel}
\end{eqnarray}
and

\begin{eqnarray}
{d\vec{n}\over{dt'}} = {c\over{|\vec{r}_o-\vec{r'}(t')|}}
\left[-\vec{\beta}+\vec{n}\left(\vec{n}\cdot\vec{\beta}\right)\right]~,
\label{usefulrel0}
\end{eqnarray}
Eq. (\ref{LW}) can be written as

\begin{eqnarray}
\vec{\bar{E}}&(&\vec{r}_o, \omega) = -{ e} \int_{-\infty}^{\infty}
dt'{\vec{n}\over{|\vec{r}_o-\vec{r'}(t')|^2}}  \exp
\left[i\omega\left(t'+{|\vec{r}_o-\vec{r'}(t')|\over{c}}\right)\right]
\cr &&+ {e\over{c}} \int_{-\infty}^{\infty} dt'{d\over{dt'}}\left[
{\vec{\beta}-\vec{n}\over{({1-\vec{n}\cdot\vec{\beta}})|\vec{r}_o-\vec{r'}(t')|}}\right]
\exp
\left[i\omega\left(t'+{|\vec{r}_o-\vec{r'}(t')|\over{c}}\right)\right].\cr
&& \label{revtrasfbiss}
\end{eqnarray}
Eq. (\ref{revtrasfbiss}) may now be integrated by parts. When edge
terms can be dropped\footnote{The only assumption made going from
Eq. (\ref{LW}) to Eq. (\ref{revtrasf}) is that edge terms in the
integration by parts can be dropped. This assumption can be
justified by means of physical arguments in the most general
situation accounting for the fact that the integral in $dt'$ has
to be performed over the entire history of the particle and that
at $t'=-\infty$ and $t'=+\infty$ the electron does not contribute
to the field anymore. Let us give a concrete example for an
ultra-relativistic electron. In this case, switchers may be
obtained placing bending magnets at the beginning and at the end
of the setup, and requiring that they deflect the electron
trajectory of an angle much larger than the maximal observation
angle of interest for radiation from a bending magnet.
Equivalently, this means that the magnets would be longer than the
formation length associated with the bends, i.e. $L_\mathrm{fb}
\sim (c \rho^2/\omega)^{1/3}$, where $\rho$ is the bending radius.
In this way, intuitively, the magnets act like switches: the first
magnet switches the radiation on, the second switches it off.
Then, what precedes the upstream bend and what follows the
downstream bend does not contribute to the field detected at the
screen position. With these \textit{caveat} Eq. (\ref{revtrasf})
is completely equivalent to Eq. (\ref{LW}).} one obtains

\begin{eqnarray}
\vec{\bar{E}}(\vec{r_o},\omega) &=& -{i\omega e\over{c}}
\int_{-\infty}^{\infty} dt'
\left[{\vec{\beta}-\vec{n}\over{|\vec{r_o}-\vec{r'}(t')|}}-{ic\over{\omega}}{\vec{n}\over{
|\vec{r_o}-\vec{r'}(t')|^2}}\right]\cr &&\times
\exp\left\{i\omega\left(t'+{|\vec{r_o}-\vec{r'}(t')|\over{c}}\right)\right\}~.
\label{revtrasf}
\end{eqnarray}
Recalling $t' = z'/v_z(z') = s(z')/v$ and the definition in Eq.
(\ref{vtilde}) one sees that Eq. (\ref{revtrasf}) is equivalent to
Eq. (\ref{rev2zeta}). Eq. (\ref{revtrasf}) is the starting point
for numerical codes based on direct calculation of the near field
(the most well-known example is perhaps SRW \cite{CHU2}).

The derivation of Eq. (\ref{revtrasf}) is particularly instructive
because shows that each term in Eq. (\ref{revtrasf}) is due to a
combination of velocity and acceleration terms in Eq. (\ref{LW}).
In other words the terms in $1/R$ and in $1/R^2$ in Eq.
(\ref{revtrasf}) appear as a combination of the terms in $1/R$
(acceleration term) and $1/R^2$ (velocity term) in Eq. (\ref{LW}).
As a result, one can say that there exist contributions to the
radiation from the velocity part in Eq. (\ref{LW}). The
presentation in Eq. (\ref{revtrasf}) (or Eq. (\ref{rev2zeta})) is
more interesting with respect to that in Eq. (\ref{LW}) (although
equivalent to it) because the magnitude of the $1/R^2$-term in Eq.
(\ref{revtrasf}) can be directly compared with the magnitude of
the $1/R$-term inside the sign of integral, and is related, as we
have seen, to the $1/R$-zone and to the reconstruction zone, where
the field can be reconstructed from far-field data. It must be
clear that the $1/R^2$-term in Eq. (\ref{revtrasf}) does not
coincide with the $1/R^2$-term in Eq. (\ref{LW}), that is the
velocity field. If one forgets about this fact he would come to
the paradoxical conclusion that far-field data can be used to
reconstruct the field in the far-zone only (on the contrary, we
have seen that when $d \gg \lambdabar$ we are in the
reconstruction zone). Note that while this conclusion is, in
general, a misconception, it does give incorrect results in the
case of ultra-relativistic systems only, because in the case of
non-ultra-relativistic systems the reconstruction zone and the far
zone coincide.

The bottom line is that only to the integral in Eq. (\ref{LW}) or
Eq. (\ref{revtrasf}) can be ascribed physical sense. The integrand
is, in fact, an artificial construction. In this regard, it is
interesting to note that the integration by parts giving Eq.
(\ref{revtrasf}) is not unique. An expression equivalent to Eq.
(\ref{revtrasf}) for the Fourier transform of the electric fields
can be found in the very interesting, but perhaps little known
reference \cite{LUCC}. After starting with Eq. (\ref{LW}), the
authors of \cite{LUCC} organized integration by part in a
different way compared with what has been done in Eq.
(\ref{revtrasfbiss}). First they found that

\begin{eqnarray}
{\vec{n}\times[(\vec{n}-\vec{\beta})\times\vec{\dot{\beta}}]
\over{|\vec{r_o}-\vec{r'}|(1-\vec{n}\cdot\vec{\beta})^2}} &=&
{1\over{|\vec{r_o}-\vec{r'}|}} {d\over{dt'}}
\left[{\vec{n}\times(\vec{n}\times\vec{\beta})\over{(1-\vec{n}\cdot\vec{\beta})}}\right]
\cr &&
-\left[{\vec{\dot{n}}(\vec{n}\cdot\vec{\beta})+\vec{n}(\vec{\dot{n}}\cdot\vec{\beta})
-
\vec{\dot{n}}(\vec{n}\cdot\vec{\beta})^2-\vec{\beta}(\vec{\dot{n}}\cdot\vec{\beta})
\over{|\vec{r_o}-\vec{r'}|(1-\vec{n}\cdot\vec{\beta})^2}}\right].
\label{luccioparti}
\end{eqnarray}
Note that Eq. (\ref{luccioparti}) accounts for the fact that
$\vec{n} = (\vec{r_o}-\vec{r'}(t'))/|\vec{r_o}-\vec{r'}(t')|$ is
not a constant in time. Using Eq. (\ref{luccioparti}) in the
integration by parts, authors of \cite{LUCC} obtained

\begin{eqnarray}
\vec{\bar{E}}(\vec{r_o},\omega) &=& -{i \omega e\over{c}}
\int_{-\infty}^{\infty} dt'
\left[-{\vec{n}\times(\vec{n}\times{\vec{\beta}})\over{|\vec{r_o}-\vec{r'}(t')|}}
+ {i {c}\over{\omega}}
{\vec{\beta}-\vec{n}-2\vec{n}(\vec{n}\cdot\vec{\beta})\over{|\vec{r_o}-\vec{r'}(t')|^2}}\right]
\cr && \times
\exp\left\{i\omega\left(t'+{|\vec{r_o}-\vec{r'}(t')|\over{c}}\right)\right\}
~. \label{revluccio}
\end{eqnarray}
Similarly as before,  the edge terms have been dropped. Eq.
(\ref{LW}), Eq. (\ref{revtrasf}) and Eq. (\ref{revluccio}) are
equivalent but include different integrands. This is no mistake,
as different integrands can lead to the same integral.

Another way to state this fact, in a more general way, is that Eq.
(\ref{LW}), Eq. (\ref{revtrasf}) and Eq. (\ref{revluccio}) can be
obtained one from the other by means of the addition of a full
derivative to the respective integrals. For example, the
integration by parts used  to transform Eq. (\ref{LW}) into Eq.
(\ref{revtrasf}) is equivalent to add to the integrand of Eq.
(\ref{LW}) the full derivative $d\Omega_1/dt'$ of the quantity

\begin{equation}
\Omega_1 = -
{\vec{\beta}-\vec{n}\over{({1-\vec{n}\cdot\vec{\beta}})|\vec{r}_o-\vec{r'}(t')|}}
\exp
\left[i\omega\left(t'+{|\vec{r}_o-\vec{r'}(t')|\over{c}}\right)\right]~,
\label{unop}
\end{equation}
while the integration by parts used  to transform Eq. (\ref{LW})
into Eq. (\ref{revluccio}) is equivalent to add to the integrand
of Eq. (\ref{LW}) the full derivative $d\Omega_2/dt'$ of the
quantity

\begin{equation}
\Omega_2=-
{\vec{n}\times(\vec{n}\times\vec{\beta})\over{(1-\vec{n}\cdot\vec{\beta})|\vec{r}_o-\vec{r'}(t')|}}
\exp
\left[i\omega\left(t'+{|\vec{r}_o-\vec{r'}(t')|\over{c}}\right)\right]~.
\label{duep}
\end{equation}
There are infinite full derivatives of functions $\Omega$ that can
be summed to the integrand of the field-integral, giving infinite
representations of the same physical quantity, the electric field.
From this viewpoint, summing a full derivative to the integrand of
the field-integral is analogous to a gauge transformation, that
leaves invariant the observable quantity (the field), but
transforms the electromagnetic potential in ways that may be
useful in particular situations. Of all possible choice of
$\Omega$, the function $\Omega_1$, that gives Eq.
(\ref{revtrasf}), is particularly useful because, as we have seen,
the magnitude of the $1/R^2$-term in Eq. (\ref{revtrasf}) can be
directly compared with the magnitude of the $1/R$-term inside the
sign of integral. However, the field obtained by summing full
derivatives to the integrand is invariant for any choice of
$\Omega$.

Let us summarize the result of this discussion. We have seen how
different zones scale with the other characteristic lengths of the
problem, the wavelength $\lambda$, the formation length $L_f$ and
the typical dimension of the system $a$. Complexities arise in the
ultra-relativistic case, when $\lambda$, $a$ and $L_f$ are very
different quantities. In particular far zone, reconstruction zone
and radiation zone are, in general, different concepts. Due to
this fact, paradoxes may arise when ascribing separate meaning to
velocity and acceleration part of the Lienard-Wiechert fields
after Fourier transformation, i.e. in Eq. (\ref{LWt}). In
particular, we have shown that the integrand of the field-integral
has no physical meaning. However, the representation of the field
integral in Eq. (\ref{revtrasf}) is particularly important, as it
allows a better physical insight, very much as a particular choice
of electromagnetic gauge can do in particular situations.

\section{\label{sec:exa} Application 1. Undulator radiation as a
laser-like beam}

An important exemplification of our algorithm is given here for
the particular case of undulator radiation. We assume, for
simplicity, that the resonance condition with the fundamental
harmonic is satisfied, that is

\begin{eqnarray}
\frac{\omega}{2 \gamma^2 c} \left(1+\frac{K^2}{2}\right) =
\frac{2\pi}{\lambda_w}~, \label{rsfirsth}
\end{eqnarray}
where $\lambda_w$ is the undulator period, while $K$ is the
undulator parameter

\begin{equation}
K=\frac{\lambda_w e H_w}{2 \pi m_\mathrm{e} c^2}~, \label{Kpara}
\end{equation}
$m_\mathrm{e}$ being the electron mass and $H_w$ being the maximum
of the magnetic field produced by the undulator on the $z$ axis.

We also specify that position $z=0$ is taken in the middle of the
undulator. We consider a planar undulator, so that the transverse
velocity of an electron can be written as

\begin{equation}
\vec{v}_\bot(z') = - {c K\over{\gamma}} \sin{\left(k_w z'\right)}
\vec{x}~. \label{vuzo}
\end{equation}
A well-known expression for the angular distribution of the first
harmonic field in the far-zone can be obtained from Eq.
(\ref{generalfin}) or Eq. (\ref{revwied}). Such expression is
axis-symmetric, and can therefore be presented as a function of a
single observation angle $\theta$, where

\begin{equation}
\theta^2 = \theta_x^2+\theta_y^2~, \label{thsq}
\end{equation}
$\theta_x$ and $\theta_y$ being angles measured from the undulator
$z$-axis in the horizontal and in the vertical direction. One
obtains the following distribution:

\begin{eqnarray}
\widetilde{{E}}_{\bot}(z_o, \theta)&=& -\frac{K \omega e L_w} {c^2
z_o \gamma} A_{JJ}\exp\left[i\frac{\omega z_o}{2 c}
\theta^2\right] \mathrm{sinc}\left[\frac{\omega L_w\theta^{2}}{4
c}\right] ~,\label{undurad4bis}
\end{eqnarray}
where the field is polarized in the horizontal direction.  Here
$L_w = \lambda_w N_w$ is the undulator length and  $N_w$ the
number of undulator periods. Finally, $A_{JJ}$ is defined as

\begin{equation}
A_{JJ} = J_o\left(\frac{K^2}{4+2K^2}\right)
-J_1\left(\frac{K^2}{4+2K^2}\right)~, \label{AJJdef}
\end{equation}
$J_n$ being the n-th order Bessel function of the first kind.
Eq.(\ref{undurad4bis}) describes a field with spherical wavefront
centered in the middle of the undulator. Eq. (\ref{virfie}) can
now be used to calculate the field distribution at the virtual
source yielding

\begin{eqnarray}
\widetilde{{E}}_{\bot}(0, r_{\bot}) &=& i \frac{K \omega e} {c^2
\gamma} A_{JJ}\left[\pi - 2\mathrm{Si} \left(\frac{\omega {r}_{
\bot}^2}{L_w c}\right)\right]~, \label{undurad5}
\end{eqnarray}
where $\mathrm{Si}(\cdot)$ indicates the sin integral function and
$r_{\bot} = |\vec{r}_{\bot}|$ is the distance from the $z$ axis on
the virtual-source plane. Note that $\widetilde{{E}}_{\bot}(0,
r_{\bot})$ is axis-symmetric. Eq. (\ref{undurad5}), that has been
already presented in \cite{OUR3}, describes a virtual field with a
plane wavefront. Let us compare this virtual field with a
laser-beam waist. In laser physics, the waist  is located in the
center of the optical cavity. In analogy with this, in our case
the virtual source is located in the center of the undulator. Both
in laser physics and in our situation the waist has a plane phase
front and the transverse dimension of the waist is much longer
than the wavelength. Note that the phase of the wavefront in Eq.
(\ref{undurad5}) is shifted of $-\pi/2$ with respect to the
spherical wavefront in the far zone. Such phase shift is analogous
to the Guoy phase shift in laser physics. Finally, in laser
physics, the Rayleigh range for a laser beam is presented in the
form $z_R = (\omega/c) w_o^2$, $w_o$ being the radius of the beam
at the location of the waist (i.e. at that position along $z$
where the phase front is flat). This is defined, for example, by
requiring that the intensity on the edge of an aperture of radius
$w_o$ be one fourth of the intensity at the center of the
radiation spot. In the undulator source case, the definition given
above amounts to $w_o = 0.9 (c L_w/\omega)^{1/2}$ and $z_R=0.8 L_w
\simeq L_w$.  In the case of a laser beam the Rayleigh length is
related to the resonator geometrical factor. In analogy with this,
in the case of a undulator source the Rayleigh length is related,
instead, to the undulator geometrical factor.

Finally, use of Eq. (\ref{fieldpropback}) gives the field
distribution at arbitrary observation position $z_o$ outside the
magnetic setup:

\begin{equation}
\widetilde{E}_{\bot}\left({z}_o,r_\bot\right) = \frac{K \omega e
A_{JJ}}{c^2 \gamma} \left[ \mathrm{Ei} \left(\frac{i \omega
r_\bot^2
  }{2{z}_o c - L_w c}\right)- \mathrm{Ei} \left(\frac{i \omega
r_\bot^2   }{2{z}_o c + L_w c}\right) \right]~, \label{Esum}
\end{equation}
where $\mathrm{Ei}(\cdot)$ indicates the exponential integral
function\footnote{Note that the field is singular in the point
$z_o = L_w/2$ and ${r}_\bot = 0$. This feature is related with the
use of the resonance approximation. As has been seen before, the
field distributions in the virtual source and in the far zone are
linked by a Fourier transformation. This fact justifies the
reciprocal relation linking small features in the near zone to
large feature in the far zone (and viceversa large features in the
near zone to small features in the far zone). Thus we may say that
the singularity at $z_o = L_w/2$ and $r_\bot = 0$ is another way
of stating the well known fact that resonance approximation fails,
in the far zone, for angles comparable with $1/\gamma_z$ (or
larger). From this viewpoint, singularity in Eq. (\ref{Esum}) is
not specific of our approach, but is intrinsically related with
the use of resonance approximation. Note that, while for the first
harmonic the far zone field does not exhibit singular behaviors at
large angles, for the second harmonic one has a logarithmic
divergence of the spectral flux integrated over angles of
observation, due to a different behavior of the field distribution
as a function of angles (see \cite{HAR2}).}. Taking square modulus
of Eq. (\ref{undurad5}) and Eq. (\ref{Esum}) one obtains,
respectively, the intensity profile for the virtual source and the
evolution of the intensity profile for undulator radiation both in
the near and in the far zone. Namely, introducing normalized units

\begin{figure}
\begin{center}
\includegraphics*[width=140mm]{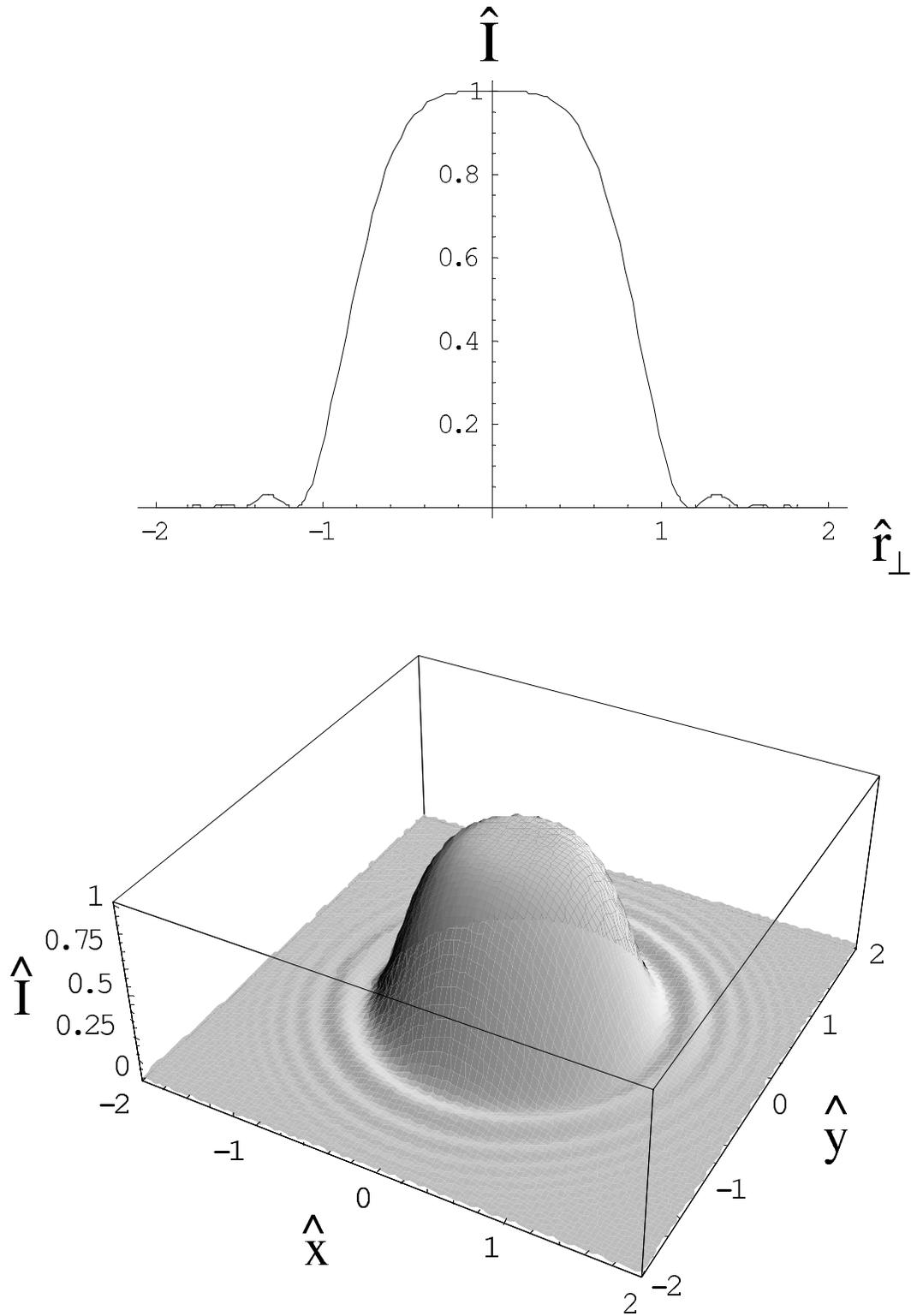}% Here is how to import EPS art
\caption{\label{virundu} Normalized intensity distribution at the
beam waist location, $\hat{I}$ as a function of
$|\vec{\hat{r}}_\bot|$ (upper plot) and 3D view as a function of
$\hat{x}$ and $\hat{y}$.}
\end{center}
\end{figure}

\begin{equation}
\vec{\hat{r}}_\bot = \sqrt{\frac{\omega }{L_w c}} \vec{r}_{\bot}~,
\label{rsnoe2}
\end{equation}
\begin{equation}
\vec{\hat{\theta}} = \sqrt{\frac{\omega L_w}{c}} \vec{\theta}~,
\label{thnoe2}
\end{equation}
and

\begin{equation}
{\hat{z}} =  \frac{z}{L_w}~ \label{zat}
\end{equation}

\begin{figure}
\begin{center}
\includegraphics*[width=140mm]{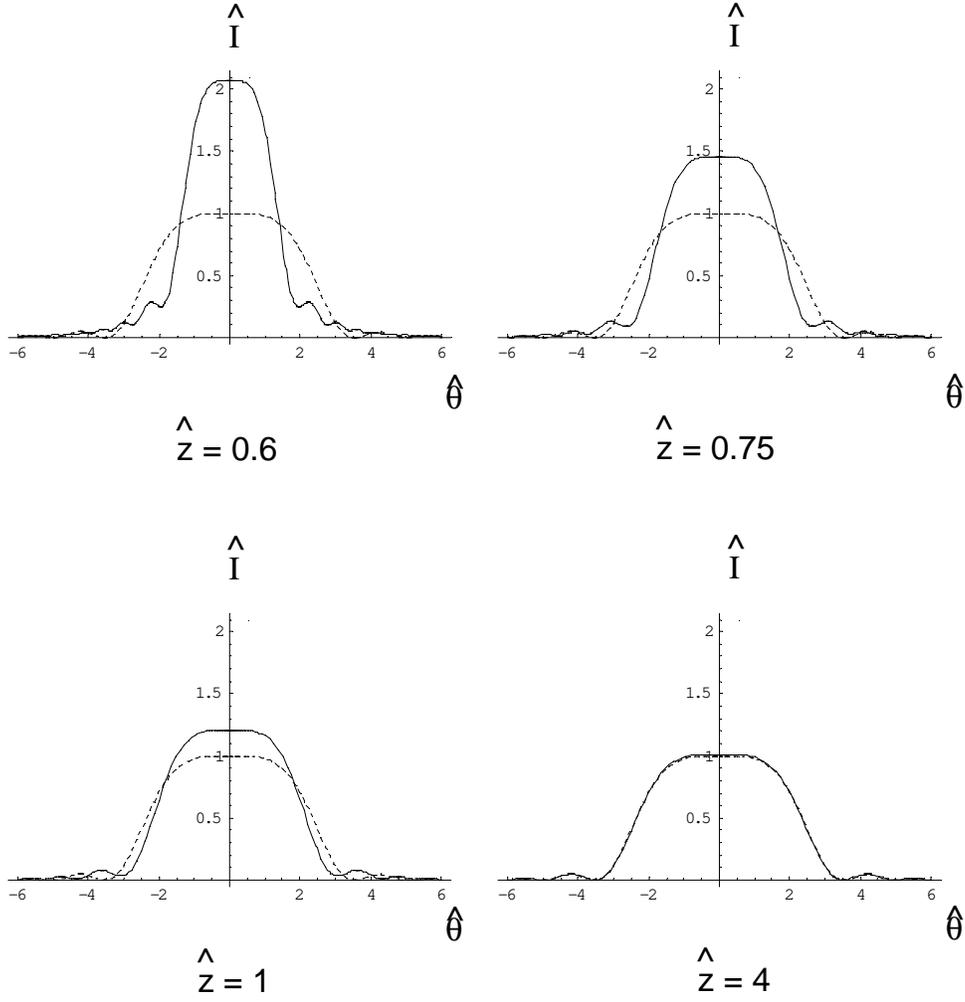}% Here is how to import EPS art
\caption{\label{anydundu} Evolution of the intensity profile for
undulator radiation according to Eq. (\ref{intsuper}) (solid
lines) and comparison with the far field asymptotic
$\mathrm{sinc}^2(\hat{\theta}^2/4)$ (dashed lines). The profiles
according to Eq. (\ref{intsuper}) are shown as a function of
angles at different observation distances $\hat{z}_o=0.6$,
$\hat{z}_o=0.75$, $\hat{z}_o=1.0$ and $\hat{z}_o=4.0$. }
\end{center}
\end{figure}

we obtain the relative intensity at the virtual source

\begin{eqnarray}
\hat{I}(0, \hat{r}_\bot) = \frac{1}{\pi^2} \left[\pi -
2\mathrm{Si} \left({\hat{r}_\bot^2}\right)\right]^2~,
\label{intsou}
\end{eqnarray}
and at any distance $\hat{z}_o$, both in the near and in the far
zone:

\begin{eqnarray}
\hat{I}\left(\hat{z}_o,\hat{\theta}\right) = \hat{z}_o^2 \left|
\mathrm{Ei} \left(\frac{i  \hat{z}_o^2 \hat{\theta}^2 }{2\hat{z}_o
- 1}\right)- \mathrm{Ei} \left(\frac{i \hat{z}_o^2
\hat{\theta}^2}{2\hat{z}_o  + 1}\right) \right|^2~,
\label{intsuper}
\end{eqnarray}
where $\hat{I}$ is defined as a normalized version of
$|\tilde{E}_\bot|^2$ in normalized units. Note that we use the
notation $\hat{\theta} = \hat{r}_\bot/\hat{z}_o$ in the near zone
as well, because it is convenient, for future discussions, to
present the intensity as in Eq. (\ref{intsuper}). Aside for
scaling factor, the intensity profile in Eq. (\ref{intsou}) can be
detected by imaging the virtual plane with an ideal lens.

The relative intensity at the virtual source is plotted in Fig.
\ref{virundu}. The evolution of the intensity profile at different
positions after the exit of the undulator according to Eq.
(\ref{intsuper}) is plotted, instead in Fig. \ref{anydundu}.

To conclude, a single electron produces a laser-like radiation
beam with a virtual source located in the center of the undulator
whose transverse size is much larger than the radiation
wavelength. Following \cite{OUR1}, Eq. (\ref{undurad4bis}) and Eq.
(\ref{undurad5}) can be generalized to the case of a particle with
a given offset $\vec{l}$ and deflection angle $\vec{\eta}$ with
respect to the longitudinal axis. The far-zone field, Eq.
(\ref{undurad4bis}), can be generalized to

\begin{eqnarray}
\widetilde{{E}}_{\bot}\left(z_o, \vec{\eta}, \vec{l},
\vec{\theta}\right)&=& -\frac{K \omega e L_w} {c^2 z_o \gamma}
A_{JJ}\exp\left[i\frac{\omega z_o}{2 c} \theta^2\right]
\exp\left[-i\frac{\omega}{c} \vec{\theta}\cdot\vec{l}\right] \cr
&&\times \mathrm{sinc}\left[\frac{\omega L_w
\left|\vec{\theta}-\vec{\eta}\right|^2}{4 c}\right]
~,\label{undurad4bisgg}
\end{eqnarray}
while the expression for the field at virtual source, Eq.
(\ref{undurad5}), is transformed to:

\begin{eqnarray}
\widetilde{{E}}_{\bot}\left(0,\vec{\eta}, \vec{l}, r_{\bot}\right)
&=& i \frac{K \omega e} {c^2 \gamma} A_{JJ}\exp\left[i
\frac{\omega}{c} \vec{\eta} \cdot
\left(\vec{r}_\bot-\vec{l}\right) \right]\left[\pi - 2\mathrm{Si}
\left(\frac{\omega \left|\vec{r}_{ \bot}-\vec{l}\right|^2}{L_w
c}\right)\right].\cr && \label{undurad5gg}
\end{eqnarray}
The meaning of Eq. (\ref{undurad4bisgg}) and Eq.
(\ref{undurad5gg}) is that offset and deflection of the single
electron motion with respect to the longitudinal axis of the
system result in offset and deflection of the waist plane.
Computer codes (e.g. ZEMAX \cite{ZEMA}, PHASE \cite{PHAS}), often
referred to as wavefront propagation codes have been developed and
are now standard tools used to carry out Fourier Optics-based
calculations. For example ZEMAX can be used for single-particle
field calculations. Once the virtual source Eq. (\ref{undurad5gg})
is specified, it can be used as an input for computer codes. This
allows propagation of the virtual wavefront in the presence of a
complicated setup, very much like it has been used as an input to
Eq. (\ref{fieldpropback}) in the particularly simple case of
free-space propagation. One may also deal with SR from a beam of
ultrarelativistic electrons. For example, in the case of
spontaneous (incoherent) SR, radiation produced by an electron
beam in an undulator can be treated as an incoherent collection of
laser beams with different offsets and deflections (i.e. by
summing up the intensities of laser-like beams with different
offsets and deflections)\footnote{Note that if the electron beam
is distributed coherently, radiation can be described as a
coherent collection of laser beams.}.

\section{\label{sec:four} Application 2. Edge radiation}

\begin{figure}
\begin{center}
\includegraphics*[width=140mm]{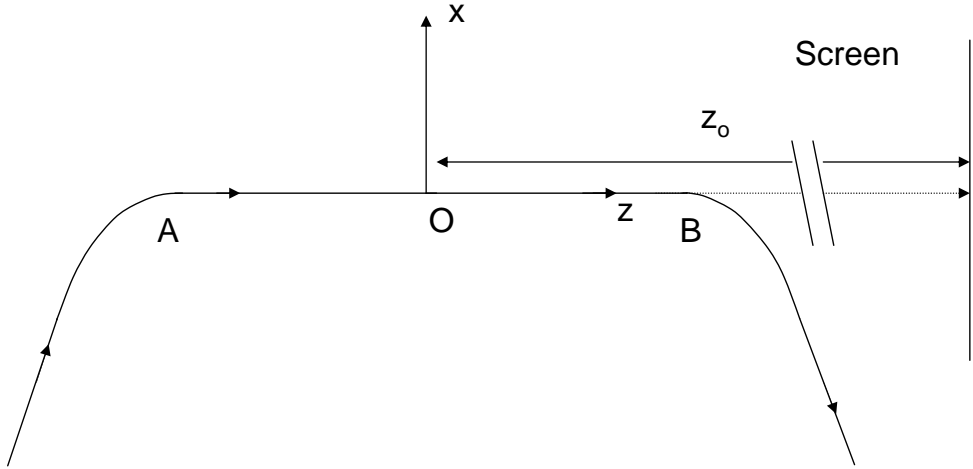}% Here is how to import EPS art
\caption{\label{geome} Edge radiation geometry. The beam enters
the system through a bending magnet, passes a straight section and
leaves the setup through another bend. Radiation is collected at a
distance $z_o$ from the origin of the reference system, located in
the middle of the straight section. }
\end{center}
\end{figure}
In this Section we apply the method described in Section
\ref{sec:meth} to the particular case of two dipole magnet edges.

We restrict ourselves to the system depicted in Fig. \ref{geome}
and we consider a single particle moving along the system. The
electron enters the setup via a bending magnet, passes through a
straight section (segment $AB$) and exits the setup via another
bend. Radiation is collected at a distance $z_o$ from the center
of the reference system, located in the middle of the straight
section. For wavelengths $\lambda \gg \lambda_c$, $\lambda_c$
being the critical wavelength of SR produced by bending magnets,
the passage of the electron through the setup results in the
emission of edge radiation (see, among the many references some
early works and more recent developments in
\cite{BOSF,CHUN,BOS4,BOS7, PROY, BOS2, BOS3, WILL} and references
therein).

In our case of study the trajectory and, therefore, the space
integration in Eq. (\ref{generalfin}) can be split in three parts:
the two bends, that will be indicated with $b_1$ and $b_2$, and
the straight sections $AB$.  One may write

\begin{eqnarray}
\vec{\widetilde{{E}}}_\bot(z_o, \vec{r}_{\bot o},\omega) &=&
\vec{\widetilde{{E}}}_{b1}(z_o, \vec{r}_{\bot o},\omega)+
\vec{\widetilde{{E}}}_{AB}(z_o, \vec{r}_{\bot o},\omega)+
\vec{\widetilde{{E}}}_{b2}(z_o, \vec{r}_{\bot o},\omega)~,
\label{splite}
\end{eqnarray}
with obvious meaning of notation.

We will denote the length of the segment $AB$ with $L$. This means
that points $A$ and $B$ are located at longitudinal coordinates
$z_A= -L/2$ and $z_B= L/2$.

First, with the help of Eq. (\ref{generalfin}) we will derive an
expression for the field in the far zone. The intensity
distribution in this case, result to be in agreement with
expression in references cited above. Then, we will calculate the
field distribution at the virtual source with the help of Eq.
(\ref{virfie}). Finally, Eq. (\ref{fieldpropback}) will allow us
to find an expression for the field both in the near and in the
far zone.

In the following we will only deal with a contribution of the
electric field, i.e. that from the straight section.

According to the superposition principle, one should sum the
contribution due to the straight section to that from the bends.
In some cases one can ignore the presence of the bending magnets
with good accuracy, and treat them as if they had zero length. In
other cases one cannot do that. When one can neglect the field
contribution due to the switchers, one can work in what can be
called the "zero-length switchers approximation". While we direct
the reader to Section \ref{sub:zerle} for details, we can mention
here that for long enough straight sections $L \gg \gamma^2
\lambdabar$, a condition for the zero-length switchers
approximation to apply is  $\lambda \gg \lambda_c$, $\lambda_c$
being the critical wavelength for SR from bends. Intuitively,
magnets act like switchers: the first magnet switches the
radiation harmonic on, the second switches it off. In the case
depicted in Fig. \ref{geome} the switchers are bending magnets,
but other setups can be considered where they have different
physical realizations. For example, one may think of a setup where
the phenomenon at study is the emission of bremsstrahlung in a
collision between a ultra-relativistic electron and a nucleus. In
this case, the switch-off process is taken care of by the
collision itself. The nucleus plays the role of the switcher and
the impact parameter, i.e. the minimal distance of the nucleus to
the electron, characterizes the switcher itself. Note that there
is no principle nor practical limitation to the length of the
switcher. In the bremsstrahlung case we can assume that such
length is very short, even with respect to the radiation
wavelength. In the case of bends it can be much longer than all
characteristic scales of interest, depending on the magnetic field
strength, and all kind of intermediate situations can be realized.
As one can see, the case of a straight line preceded and followed
by switchers has a number of physical realizations. The only
feature that different realizations must have in common by
definition of switcher is that the switching process depends
exponentially on the distance from the beginning of the process.
Then, a characteristic length $d_s$ is associated to any switcher.
Since electrodynamics is a linear theory, in the case when the
contribution from the finite straight-line cannot be considered a
good approximation of the total field, it can be considered as a
building block for a more complicated setup. An example of a more
sophisticated setup is studied in the next Section \ref{sec:five},
where the problem of Transition Undulator Radiation is addressed.

In general, one can say that the problem of studying the field
from a finite straight-line is of fundamental importance,
independently of the type of switchers considered. This justifies
our attention to the straight-section contribution to the field.

\subsection{\label{sub:fare} Far field pattern of edge radiation}

\subsubsection{\label{sub:stre} Field contribution calculated along
the straight section}

Accounting for the geometry in Fig. \ref{geome} we have

\begin{equation}
s(z') =  z' ~~~~~\mathrm{for}~z_A<z'<z_B ~\label{zssse}
\end{equation}
With the help of Eq. (\ref{generalfin}) we write the contribution
from the straight line $AB$ as

\begin{equation}
\vec{\widetilde{E}}_{AB}={i \omega e\over{c^2 z_o}}
\int_{-L/2}^{L/2} dz' \exp{\left[i\Phi_{AB}\right]} \left(\theta_x
{\vec{x}}+\theta_y {\vec{y}}\right) \label{ABcontre}
\end{equation}
where $\Phi_{AB}$ in Eq. (\ref{ABcontre}) is given by

\begin{equation}
\Phi_{AB} = \omega \left[ {\theta_x^2+\theta_y^2\over{2c}} z_o  +
{z'\over{2c}}\left({1\over{\gamma^2}}+\theta_x^2+\theta_y^2\right)\right]~,
\label{phiabe}
\end{equation}
$\theta_x = x_o/z_o$ and $\theta_y = y_o/z_o$ being the
observation angles in the horizontal and vertical direction. From
Eq. (\ref{ABcontre}) one obtains

\begin{eqnarray}
\vec{\widetilde{E}}_{AB}&&=\frac{ i \omega e L}{c^2 z_o}
\exp\left[\frac{i\omega \theta^2 z_o}{2c}\right] \vec{\theta} ~
\mathrm{sinc}\left[\frac{\omega L}{4
c}\left(\theta^2+\frac{1}{\gamma^2}\right)\right]
\label{ABcontrint4e}
\end{eqnarray}
where, as before

\begin{equation}
\theta^2 = \theta_x^2+\theta_y^2~. \label{thsq}
\end{equation}
Eq. (\ref{ABcontrint4e}) is an exact expression for the electric
field from the straight section $AB$. Note that Eq.
(\ref{ABcontrint4e}) describes a spherical wave. Moreover, it
explicitly depends on $L$ (this last remark will be useful later
on).

The formation length $L_\mathrm{fs}$  for the straight section
$AB$ can be written as

\begin{equation}
L_\mathrm{fs} \sim \min\left[\gamma^2 \lambdabar ,L\right] ~.
\label{lstfee}
\end{equation}
Depending on the wavelength of interest then, $L_\mathrm{fs} \sim
\gamma^2 \lambdabar$ or $L_\mathrm{fs} \sim L$. In both cases,
with the help of Eq. (\ref{phiabe})  we can give on a purely
mathematical basis an upper limit to the value of the observation
angle of interest related to the straight line:

\begin{equation}
\theta_{x,y}^2 \lesssim  \frac{\lambdabar}{L_\mathrm{fs}}
~.\label{maxinterlinee}
\end{equation}
Note that, if $L_\mathrm{fs} \sim  \gamma^2 \lambdabar$, the
maximal angle of interest is independent of the frequency.

Finally it should be remarked that the the far-zone asymptotic in
Eq. (\ref{ABcontre}) is valid at observation positions $z_o \gg
L$. This is a necessary and sufficient condition for the vector
$\vec{n}$ pointing from the retarded position of the source to the
observer, to be considered constant. This result is independent of
the formation length given by Eq. (\ref{lstfee}). Therefore, when
$L \lesssim \gamma^2\lambdabar$ we can say that an observer is the
far zone if and only if it is located many formation lengths away
from the origin. This is no more correct when $L \gg
\gamma^2\lambdabar$. In this case the observer can be located at a
distance $z_o \gg \gamma^2 \lambdabar$, i.e. many formation
lengths away from the origin of the reference system, but still at
$z_o \sim L$, i.e. in the near zone. As we see here, the formation
length $L_f$ is often, but not always related with the definition
of the far (or near) zone. This is the case for bending magnet
radiation, but not for edge radiation. As has been remarked before
in Section \ref{sec:dis}, the far (or near) is related with the
characteristic size of the system $a$ (in our case $a = L$). In
its turn $L_f \lesssim a$, which includes, as in the edge
radiation case when $\gamma^2 \lambdabar \ll L$, the situation
$L_f \ll a$.

\subsubsection{\label{sub:longe} Energy spectrum of radiation}

\begin{figure}
\begin{center}
\includegraphics*[width=140mm]{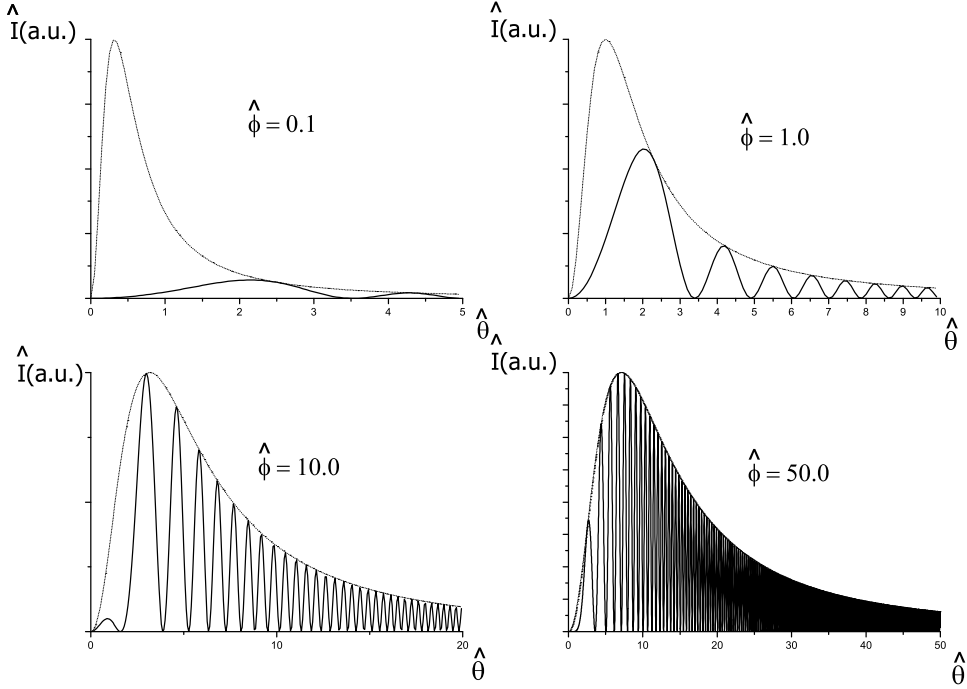}% Here is how to import EPS art
\caption{\label{Dirfar2} Directivity diagram (solid lines) of the
radiation from the setup in Fig. \ref{geome} for different values
of $\hat{\phi}$. Dotted lines show the envelope of the directivity
diagram.}
\end{center}
\end{figure}

The radiation energy density as a function of angles and
frequencies can be written as

\begin{eqnarray}
\frac{d W}{d\omega d \Omega} = \frac{c z_o^2}{4 \pi^2}
\left|\vec{\widetilde{E}}\right|^2~, \label{endene}
\end{eqnarray}
$d \Omega$ being the differential of the solid angle $\Omega$.
Here we have used Parseval theorem and the fact that the
irradiance (i.e. the energy per unit time per unit area, averaged
over a period of the carrier frequency) is given by

\begin{eqnarray}
P =\frac{c}{4\pi} \left<\left|{\vec{E}}\right|^2\right> =
\frac{c}{2\pi} \left|\vec{\widetilde{E}}\right|^2~,\label{powe}
\end{eqnarray}
where brackets $<...>$ denote averaging over a cycle of
oscillation of the carrier wave. Substituting Eq.
(\ref{ABcontrint4e}) in Eq. (\ref{endene}) it follows that

\begin{eqnarray}
&&\frac{d W}{d\omega d \Omega} = \frac{ \omega^2 e^2 L^2}{4 \pi^2
c^3} {\theta^2} ~ \mathrm{sinc}^2\left[\frac{\omega L}{4
c}\left(\theta^2+\frac{1}{\gamma^2}\right)\right]~.
\label{enden2e}
\end{eqnarray}
It is convenient to introduce normalized quantities

\begin{equation}
\vec{\hat{\theta}} = \sqrt{\frac{\omega L}{c}} \vec{\theta}~,
\label{thnoe}
\end{equation}
\begin{equation}
\hat{\phi} = \frac{\omega L}{\gamma^2 c}~, \label{phie}
\end{equation}
With Eq. (\ref{enden2e}) in mind and using normalized units, we
may write the directivity diagram $\hat{I}$ of the radiation as

\begin{eqnarray}
\hat{I} = \mathrm{const.} \times {\hat{\theta}^2} ~
\mathrm{sinc}^2\left[\frac{1}{4
}\left(\hat{\theta}^2+\hat{\phi}\right)\right] \label{normI1}
\end{eqnarray}
The directivity diagram in Eq. (\ref{normI1}) is plotted in Fig.
\ref{Dirfar2} for several values of $\hat{\phi}$ as a function of
the normalized angle $\hat{\theta}$. The natural angular unit is
evidently $(2\pi L/\lambda)^{-1/2}$.

There are two asymptotic cases for the problem parameter
$\hat{\phi}$: $\hat{\phi} \ll 1$ and $\hat{\phi} \gg 1$. When
$\hat{\phi} \gg 1$ the oscillating structures are fine with
respect to the envelope of the directivity diagram. The intensity
peaks where the envelope of the directivity diagram peaks, i.e. at
$\hat{\theta} \sim \sqrt{\hat{\phi}}$. This feature for the
far-field emission is shown in Fig. \ref{Dirfar2} for the case
$\hat{\phi} = 50$ and are typical for $\hat{\phi} \gg 1$. When the
length of the straight section becomes smaller and $\hat{\phi} \ll
1$, we reach the other asymptotic limit. In this case, the period
of the oscillations due to the $\sin(\cdot)$ function becomes much
larger than $\sqrt{\hat{\phi}}$, as it can also be seen in Fig.
\ref{Dirfar2} for the case $\hat{\phi} = 0.1$. Then, the maximum
intensity does not coincide anymore with the peak of the envelope
of the directivity diagram, but it is found at $\hat{\theta}
=2.2$.

The behavior of the far-field emission described here is
well-known in literature, and has been first reported long ago in
\cite{BOS4}. We take this as the starting point for further
investigations based on Fourier Optics.

\subsection{\label{sub:vire} Method of virtual sources}

\subsubsection{Edge radiation as a field from a single virtual source }

Eq. (\ref{virfie}) and Eq. (\ref{ABcontrint4e}) allow one to
characterize the virtual source through

\begin{eqnarray}
\vec{\widetilde{E}}(0,\vec{r}_{\bot}) =- \frac{ \omega^2 e L}{2\pi
c^3 } \int d\vec{\theta} ~\vec{\theta}~
\mathrm{sinc}\left[\frac{\omega L}{4 c}
\left(\theta^2+\frac{1}{\gamma^2}\right) \right]\exp\left[\frac{i
\omega}{c}\vec{r}_{\bot}\cdot \vec{\theta} \right] ~.\label{vir1e}
\end{eqnarray}
The Fourier transform in Eq. (\ref{vir1e}) is difficult to
calculate analytically in full generality. However, simple
analytical results can be found in the asymptotic case for
$\hat{\phi} \ll 1$, i.e. for $2\pi L/(\gamma^2 \lambda) \ll 1$. In
this limit, the right hand side of Eq. (\ref{vir1e}) can be
calculated with the help of polar coordinates. An analytic
expression for the field amplitude at the virtual source can then
be found and reads:

\begin{eqnarray}
\vec{\widetilde{E}}(0,\vec{r}_\bot) = \frac{4 \omega e}{c^2 L}
\vec{r}_\bot \mathrm{sinc}\left(\frac{\omega}{L c}
\left|\vec{r}_\bot\right|^2\right) ~.\label{vir3e}
\end{eqnarray}
It is useful to remark, for future use, that similarly to the
far-field emission Eq. (\ref{ABcontrint4e}), also the
non-normalized version of the field in Eq. (\ref{vir3e})
explicitly depend on $L$. This is true in general, for any value
of $\hat{\phi}$. After definition of the normalized transverse
position

\begin{equation}
\vec{\hat{r}}_\bot= \sqrt{\frac{\omega }{Lc}} \vec{r}_{\bot}~
\label{rsnoe}
\end{equation}
the intensity distribution of the virtual source is given by

\begin{figure}
\begin{center}
\includegraphics*[width=140mm]{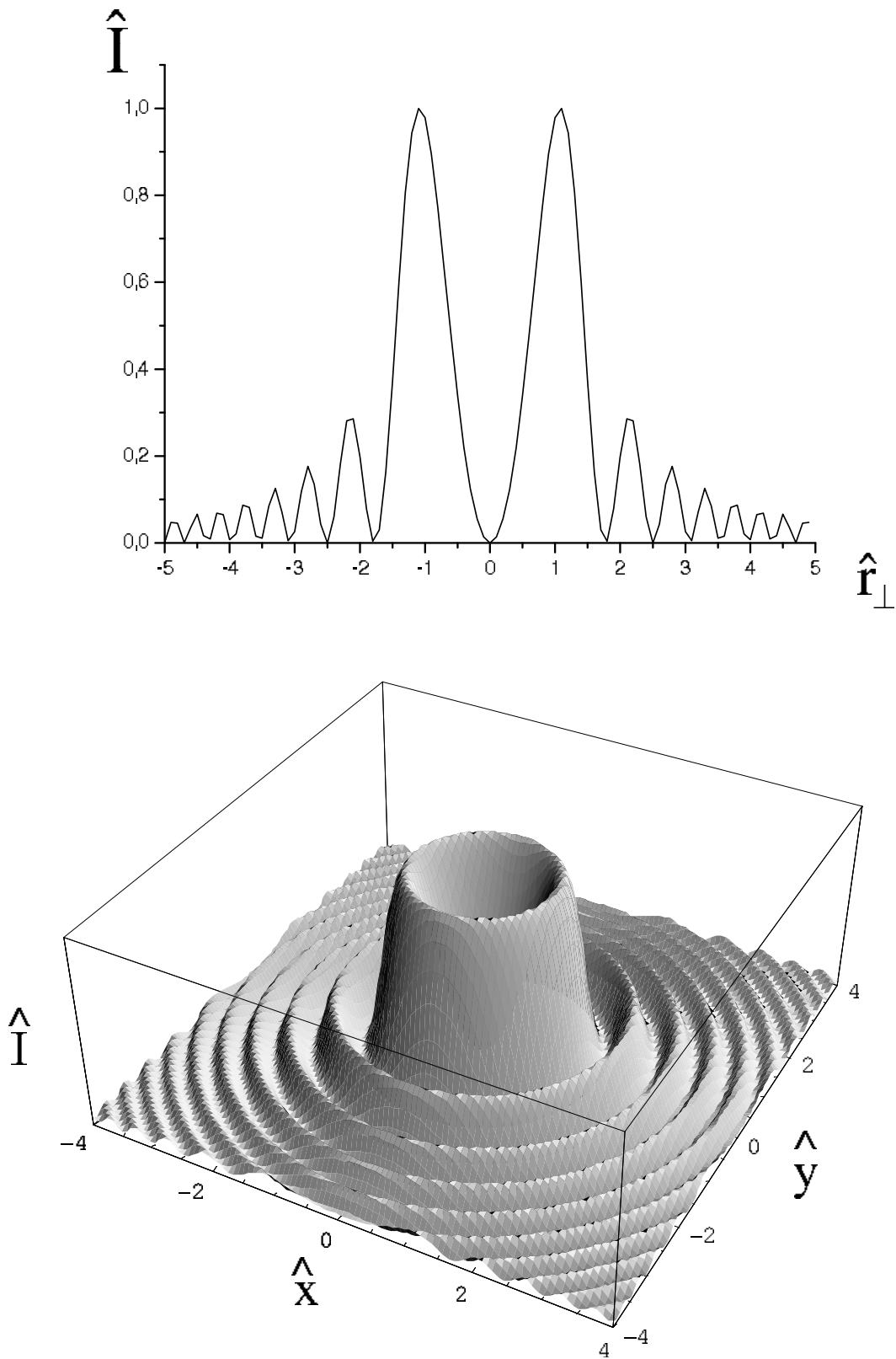}% Here is how to import EPS art
\caption{\label{viredge} Intensity distribution at the virtual
source, $\hat{I}$, as a function of $|\vec{\hat{r}_\bot}|$ (upper
plot) and 3D view as a function of $\hat{x}$ and $\hat{y}$.}
\end{center}
\end{figure}

\begin{eqnarray}
\hat{I}(\hat{r}_\bot) = \mathrm{const.} \times {\hat{r}_\bot}^2
\mathrm{sinc}^2\left({\hat{r}_\bot}^2\right) ~,\label{vir4ei}
\end{eqnarray}
it can be detected (aside for scaling factors) by imaging the
virtual plane with an ideal lens, and is plotted in Fig.
\ref{viredge}.

Note that Eq. (\ref{vir3e}) describes a virtual source
characterized by a plane wavefront. Application of the Fresnel
propagation formula, Eq. (\ref{fieldpropback}) to Eq.
(\ref{vir3e}) allows one to reconstruct the field both in the near
and in the far region. We obtain the following result:

\begin{eqnarray}
\vec{\widetilde{E}}({z}_o,\vec{r}_\bot) &&= -\frac{2 e}{c} \frac{
\vec{{r}}_\bot}{{r}_\bot^2}\exp\left[i \frac{\omega {r}_\bot^2}{2
c{z}_o}\right] \cr&&\times\left[ \exp\left(-\frac{i \omega
{r}_\bot^2 }{2  c {z}_o (1+2{z}_o/L)} \right)-\exp\left(\frac{i
\omega {r}_\bot^2 }{2  c {z}_o (-1+2{z}_o/L)} \right) \right]
~.\label{vir4e}
\end{eqnarray}
Eq. (\ref{vir4e}) solves the field propagation problem for both
the near and the far field in the limit for $\hat{\phi} \ll
1$\footnote{Eq. (\ref{vir4e}) is singular as $\vec{r}_\bot =0$ and
${z}_o = L/2$. Moreover, the spectral photon flux, integrated in
angles, is logarithmically divergent. In the undulator case, as we
have seen in footnote 10, singular features have been interpreted
as limits of applicability of the resonance approximation.
Similarly we can state here that the paraxial approximation is
valid, in the far zone, for angles of observation much smaller
than unity. The logarithmic singularity of the flux and the
singular behavior at $\vec{r}_\bot =0$ and ${z}_o = L/2$
correspond to a logarithmic divergence of the flux in the far zone
at large angles. These divergences, in the near as well as in the
far zone are outside the region of applicability of our
approximation, because we discuss the far-zone field within angles
much smaller than unity as well as near-zone field for transverse
displacements $r_\bot \gg \lambdabar$, where we have no
singularity at all.}.

The intensity profile associated with Eq. (\ref{vir4e}) is given
by

\begin{eqnarray}
{\hat{I}}(\hat{\theta}) &=& \frac{1}{\hat{\theta}^2} \left| \left[
\exp\left(-\frac{i \hat{\theta}^2 \hat{z}_o}{2 (1+2\hat{z}_o)}
\right)-\exp\left(\frac{i \hat{\theta}^2 \hat{z}_o }{2
(-1+2\hat{z}_o)} \right) \right]\right|^2 ~,\label{vir5e}
\end{eqnarray}
where $\hat{\theta} = \hat{r}_\bot/\hat{z}_o$ and
$\hat{z}_o=z_o/L$.

\begin{figure}
\begin{center}
\includegraphics*[width=140mm]{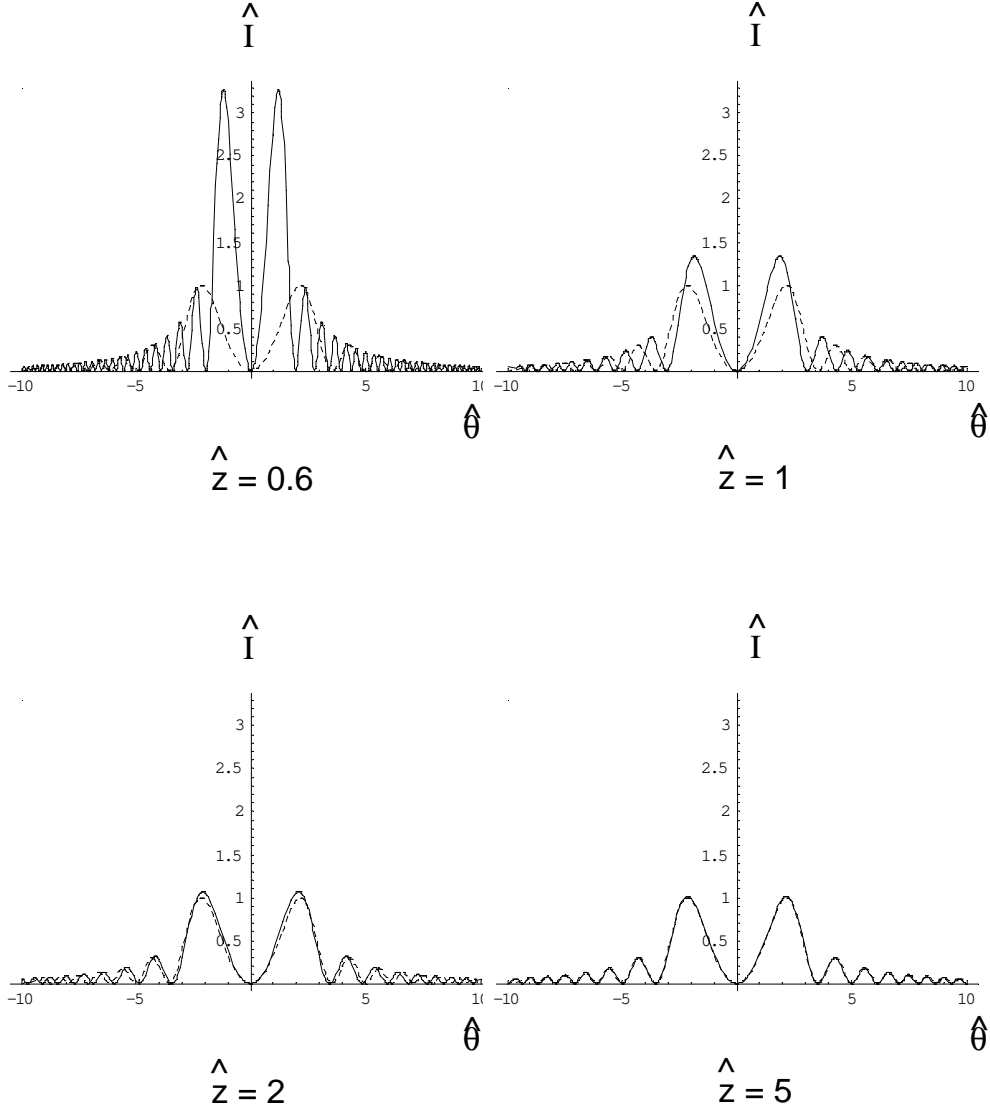}% Here is how to import EPS art
\caption{\label{Anyd} Evolution of the intensity profile for edge
radiation in the limit for $\hat{\phi} \ll 1$. These profiles are
shown as a function of angles at different observation distances
$\hat{z}_o=0.6$, $\hat{z}_o=1.0$, $\hat{z}_o=2.0$ and
$\hat{z}_o=5.0$ (solid lines). The dashed line always refers to
the far-zone intensity. }
\end{center}
\end{figure}

In the limit for $\hat{z}_o \longrightarrow \infty$ Eq.
(\ref{vir4e}) and Eq. (\ref{vir5e}) respectively transform to

\begin{eqnarray}
\vec{\widetilde{E}}&&=\frac{ i \omega e L}{c^2 z_o}
\exp\left[\frac{i\omega \theta^2 z_o}{2c}\right] \vec{\theta} ~
\mathrm{sinc}\left[\frac{\omega L \theta^2}{4 c}\right]
\label{lllim}
\end{eqnarray}
and

\begin{eqnarray}
\hat{I} = \mathrm{const.} \times {\hat{\theta}^2} ~
\mathrm{sinc}^2\left[\frac{\hat{\theta}^2}{4}\right]~,
\label{lllim2}
\end{eqnarray}
corresponding to Eq. (\ref{ABcontrint4e}) and Eq. (\ref{normI1})
in the asymptotic case for for $\hat{\phi} \ll 1$.

Note that when $\hat{\phi} \ll 1$ we have only two asymptotic
regions with respect to $\hat{z}_o$: the far zone for $\hat{z}_o
\gg 1$ and the near zone for $\hat{z}_o \ll 1$. Of course, it
should be stressed that in the case $\hat{z}_o \ll 1$ we still
hold the assumption that the approximation of zero-length
switchers is satisfied. It is interesting to study the evolution
of the intensity profile for edge radiation along the longitudinal
axis. This gives an idea of how good the far field approximation
($\hat{z}_o \gg 1$) is. A comparison between  intensity profiles
at different observation points is plotted in Fig. \ref{Anyd}.

%It is relevant here to suggest a possible application of Eq.
%(\ref{vir4e}) for a practical case of interest. Authors of
%reference \cite{STUP} study numerically the electric field from a
%bunch length monitor that will be

The case $\hat{\phi} \ll 1$ studied until now corresponds to a
short straight section $L \ll \gamma^2 \lambdabar$. When this is
condition is not satisfied, we find that the integral on the right
hand side of Eq. (\ref{vir1e}) is difficult to calculate
analytically. However, one can use numerical techniques. With the
help of polar coordinates, the right hand side of Eq.
(\ref{vir1e}) can be transformed in a one-dimensional integral,
namely

\begin{figure}
\begin{center}
\includegraphics*[width=140mm]{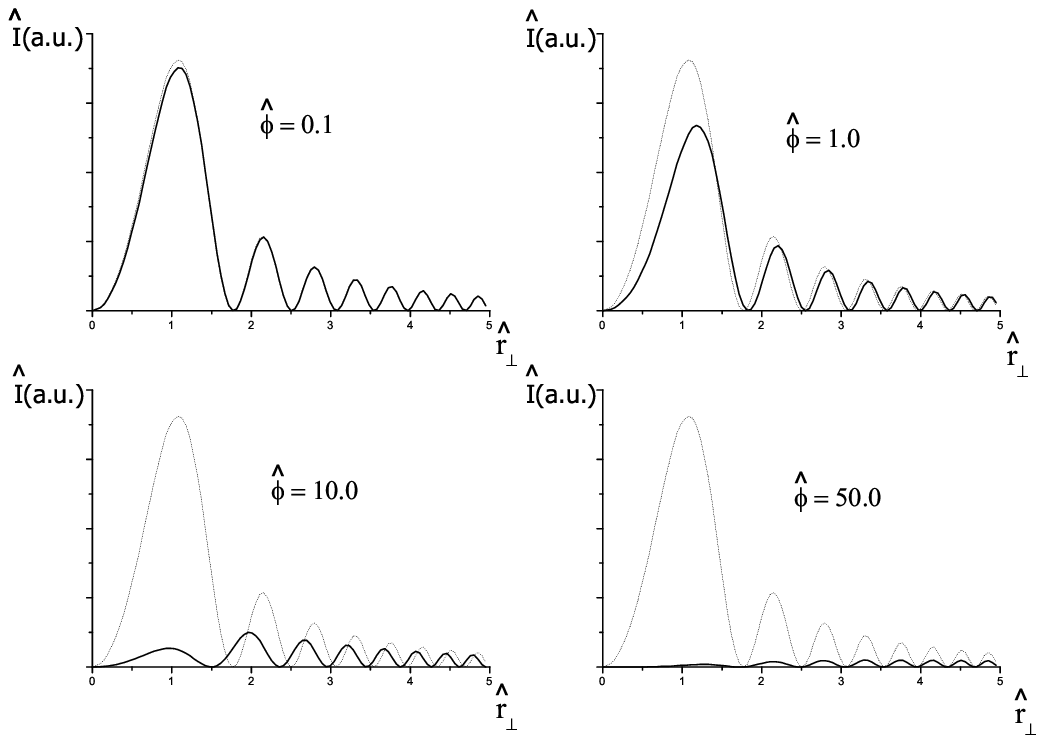}% Here is how to import EPS art
\caption{\label{Intvirphi} Intensity profiles of the virtual
source for the setup in Fig. \ref{geome}. These profiles are shown
for $\hat{\phi}=0.1$, $\hat{\phi}=1.0$, $\hat{\phi}=10.0$ and
$\hat{\phi}=50.0$ (solid lines). Solid curves are calculated with
the help of Eq. (\ref{vir2e1d}). The dotted lines show comparison
with the asymptotic limit for $\hat{\phi}\ll 1$, shown in Fig.
\ref{viredge} and calculated using Eq. (\ref{vir4ei}).}
\end{center}
\end{figure}

\begin{figure}
\begin{center}
\includegraphics*[width=140mm]{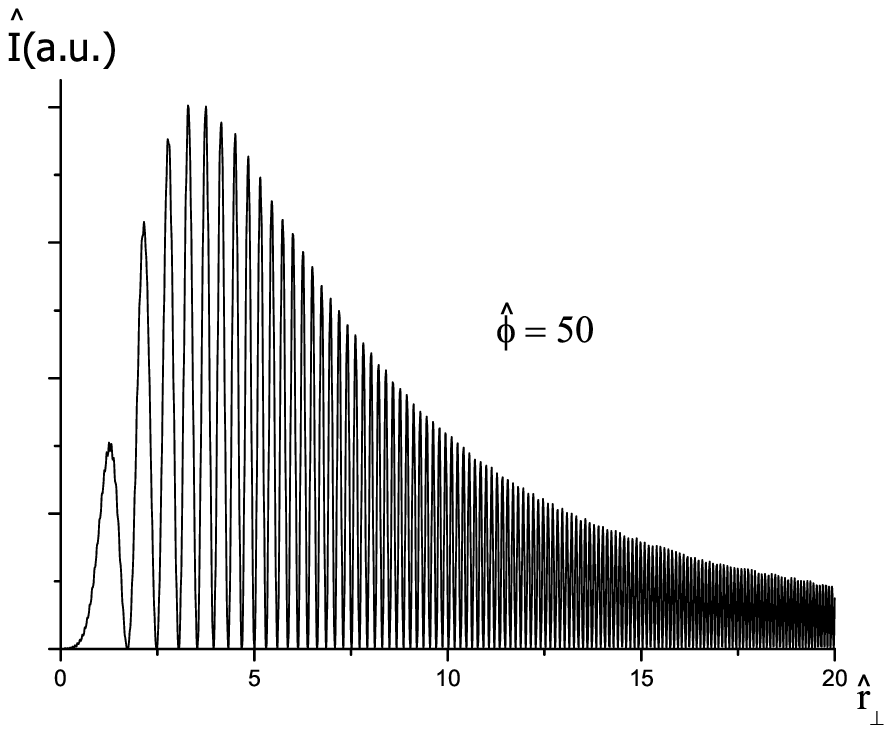}% Here is how to import EPS art
\caption{\label{phil} Intensity distribution at the virtual source
for the setup in Fig. \ref{geome} for $\hat{\phi}=50$ (enlargement
of the bottom right graph in Fig. \ref{Intvirphi}).}
\end{center}
\end{figure}

\begin{eqnarray}
\vec{\widetilde{E}}(0,\vec{r}_\bot) = - \frac{4 \omega e } {c^2}
\frac{\vec{r}_\bot}{r_\bot} \int_0^\infty
\frac{{\theta}^2}{{\theta}^2+1/\gamma^2} \sin\left[\frac{\omega
L}{4 c} \left(\theta^2 + \frac{1}{\gamma^2}\right)\right]
J_1\left(\frac{\omega {\theta}{r}_\bot}{c}\right)
d{\theta}~.\label{vir2e1d}
\end{eqnarray}
We calculated the intensity distribution associated with the
virtual source for values $\hat{\phi}=0.1$, $\hat{\phi}=1$,
$\hat{\phi}=10$ and $\hat{\phi}=50$, corresponding to directivity
diagrams in the far zone in Fig. \ref{Dirfar2}. We plot these
distributions in Fig. \ref{Intvirphi}. In particular, it is
instructive to make a separate, enlarged plot of the case
$\hat{\phi} = 50$, that is in the asymptotic case for $\hat{\phi}
\gg 1$. This is given in Fig. \ref{phil}. Fine structures are now
evident, and are consistent with the presence of fine structures
in Fig. \ref{Dirfar2} for the far zone. In fact, as we have seen,
the field in the far region and the virtual source are linked,
basically, by a Fourier transform. In principle, this allows to
qualitatively describe the situation at the virtual source through
the reciprocal relation. However, as we have said before, when
$\hat{\phi} \gg 1$, in the far zone we have fine structures with
variable width $\delta \theta$ (see also Fig. \ref{Dirfar2}). As a
result, in the limiting case $\hat{\phi} \gg 1$, use of the
reciprocal relation to describe the properties of the virtual
source is problematic. For example, a typical width $\delta
\hat{\theta} \sim 1$ should correspond to typical dimension of the
virtual source of order unity, that is in obvious disagreement
with Fig. \ref{phil}. Nonetheless, we managed to specified the
field at the virtual source by means of numerical techniques, even
in the case $\hat{\phi} \gg 1$ (see Fig. \ref{phil}). Once the
field at the virtual source is specified for any value of
$\hat{\phi}$, Fourier Optics can be used to propagate it. In free
space, the Fresnel formula must be used. However, we prefer to
proceed in another way. There is, in fact, an alternative way to
obtain the solution to the field propagation problem valid for any
value of $\hat{\phi}$ and capable of giving a better physical
insight for large values of $\hat{\phi}$.

\subsubsection{Edge radiation as a superposition of the field from two virtual sources}

Let us begin considering the far field in Eq.
(\ref{ABcontrint4e}). This can also be written as

\begin{eqnarray}
\vec{\widetilde{E}}\left({z}_o,\vec{\theta}\right) =
\vec{\widetilde{E}}_1\left({z}_o,\vec{\theta}\right)
+\vec{\widetilde{E}}_2\left({z}_o,\vec{\theta}\right) ~
\label{Efarsum2a}
\end{eqnarray}
where

\begin{eqnarray}
\vec{\widetilde{E}}_{1,2}\left({z}_o,\vec{\theta}\right) = \pm
\frac{2 e\vec{{\theta}}}{c {z}_o (\theta^2 + 1/\gamma^2)}
\exp\left[\pm\frac{i \omega L}{4 c \gamma^2}\right]
\exp\left[\frac{i \omega L \theta^2}{2 c
}\left(\frac{z_o}{L}\pm\frac{1}{2}\right)\right] .
\label{Efarsum2}
\end{eqnarray}
The two terms $\vec{\widetilde{E}}_{1}$ and
$\vec{\widetilde{E}}_{2}$ represent two spherical waves
respectively centered at ${z} = L/2$ and ${z}=-L/2$, that is at
the edges of the straight section. Analysis of Eq.
(\ref{Efarsum2}) shows that both contributions to the total field
are peaked at an angle of order $1/\gamma$. While, as has been
seen before, the total field in dimensional units explicitly
depends on the straights section length $L$, the two expressions
$\vec{\widetilde{E}}_{1}$ and $\vec{\widetilde{E}}_{2}$ exhibit
dependence on $L$ through phase factors only. This fact will have
interesting consequences, as we will discuss later. The two
spherical waves represented by $\vec{\widetilde{E}}_{1}$ and
$\vec{\widetilde{E}}_{2}$ may be thought as originating from two
separate virtual sources located at the edges of the straight
section. One may then describe the system with the help of two
separate virtual sources, and interpret the field at any distance
as the superposition of the contributions from two edges. This
viewpoint is completely equivalent to that considered before
involving a single virtual source in the straight line center. We
are presenting here a different description of the same
phenomenon. As we have seen before, we could not specify,
analytically, the single virtual source in the center of the
straight line. In contrast to this it is possible to specify the
two virtual sources at the edges of the setup. In order to so so
we take advantage of a slightly modified version of Eq.
(\ref{virfie}) that accounts for an arbitrary position of the
source ${z}_{s(1,2)}$:

\begin{eqnarray}
\vec{ \widetilde{E}}\left( {z}_{s(1,2)},\vec{r}_{\bot} \right)&=&
\frac{i \omega {z}_o}{2 \pi c}  \int d\vec{\theta}\cr
&&\times\exp{\left[-\frac{i \omega { \theta}^2}{2
c}\left(z_o+{z}_{s(1,2)}\right)\right]}{
\vec{\widetilde{E}}}_{1,2}\left({z}_o,\vec{\theta}\right)
\exp\left[\frac {i \omega}{c} \vec{r}_{\bot}\cdot
\vec{\theta}\right] ~ , \cr &&\label{virfiesh2}
\end{eqnarray}
Separately substituting $\vec{\widetilde{E}}_{1}$ and
$\vec{\widetilde{E}}_{2}$ into Eq. (\ref{virfiesh2}), and with the
help of polar coordinates, we find the following expressions for
the field at the virtual source positions ${z}_{s1} = L/2$ and
${z}_{s2}=-L/2$\footnote{\label{fifo} It should be noted here that
the virtual sources are singular in the point ${r}_\bot=0$, due to
the behavior of the modified Bessel function $K_1$. In footnote 10
we associated the singularity of the undulator field with the
resonance approximation. Similarly we can state here that paraxial
approximation is valid, in the far zone, for angles of observation
much smaller than unity. As a result, features near the downstream
edge of the straight section cannot be resolved in paraxial
approximation because they depend on far field data at large
angles. Analysis of Eq. (\ref{virpm05}) shows that the intensity
associated to each source exhibits a weak, logarithmic singularity
of the flux in the near zone around $r_\bot=0$. This corresponds
to a logarithmic divergence of the flux in the far zone at large
angles. These divergences, in the near as well as in the far zone
are outside the region of applicability of our approximation,
because we discuss the far-zone field within angles much smaller
than unity as well as near-zone field for transverse displacements
$r_\bot \gg \lambdabar$, where we have no singularity at all.}:

\begin{eqnarray}
\vec{\widetilde{E}}\left(\pm \frac{L}{2},\vec{r}_\bot\right) = \mp
\frac{2 i e \omega }{c^2\gamma} \exp\left[\pm \frac{i\omega L }{4
c \gamma^2}\right] \frac{\vec{r}_\bot}{r} K_1\left(\frac{\omega
r_\bot}{c \gamma} \right)~,\label{virpm05}
\end{eqnarray}
where $K_1(\cdot)$ is the modified Bessel function of the first
order. Analysis of Eq. (\ref{virpm05}) shows a typical scale
related to the sources dimension of order $c \gamma/\omega$ in
dimensional units, corresponding to $1/\sqrt{\hat{\phi}}$ in
normalized units. This is in agreement with the fact that  both
source contributions to the far field are peaked at an angle of
order $\sqrt{\hat{\phi}}$. Note that this remark is only
qualitative. The peak angle in the far zone does not correspond
univocally to the width of the field distribution, nor the typical
scale $1/\sqrt{\hat{\phi}}$ of the virtual sources can be
univocally associated to the width of the sources. Also note that
since the far zone field in dimensional units exhibit dependence
on $L$ only through phase factors only has its counterpart in the
fact that the field at the virtual sources, written in dimensional
units, exhibit dependence on $L$ only through phase factors as
well.

Application of the Fresnel formula allows to calculate the field
at any distance ${z}_o$ in free space. Of course, Eq.
(\ref{virpm05}) can also be used as input to any Fourier code to
calculate the field evolution in the presence of whatever optical
beamline. However, we restrict ourselves to the free-space case.
Taking advantage, once more, of polar coordinates and using the
definition $\vec{\hat{\theta}}=\vec{\hat{r}}_\bot/\hat{z}_o$ we
obtain the following scaling law for the intensity in normalized
units:

\begin{eqnarray}
\hat{I}\left(\hat{z}_o,\vec{\hat{\theta}}\right) &=&
\mathrm{const} \cdot \left|
\left\{-\frac{\vec{\hat{\theta}}}{\hat{\theta}} \frac{2 i
\sqrt{\hat{\phi}} \exp\left[i\hat{\phi}/4\right] }{\hat{z}_o-1/2}
\exp\left[\frac{i\hat{\theta}^2 \hat{z}_o^2
}{2\left(\hat{z}_o-1/2\right)}\right]\right.\right.\cr &&
\left.\left. \times \int_0^{\infty} d\hat{r}'_\bot \hat{r}'_\bot
K_1\left(\sqrt{\hat{\phi}}\hat{r}'_\bot\right) J_1
\left(\frac{\hat{\theta}\hat{r}'_\bot
\hat{z}_o}{\hat{z}_o-1/2}\right)
\exp\left[\frac{i\hat{r}_\bot^{'2}
}{2\left(\hat{z}_o-1/2\right)}\right] \right\} \right.\cr &&\left.
+ \left\{\frac{\vec{\hat{\theta}}}{\hat{\theta}} \frac{2 i
\sqrt{\hat{\phi}} \exp\left[-i\hat{\phi}/4\right] }{\hat{z}_o+1/2}
\exp\left[\frac{i\hat{\theta}^2 \hat{z}_o^2
}{2\left(\hat{z}_o+1/2\right)}\right]\right.\right.\cr &&
\left.\left. \times \int_0^{\infty} d\hat{r}'_\bot \hat{r}'_\bot
K_1\left(\sqrt{\hat{\phi}}\hat{r}'_\bot\right) J_1
\left(\frac{\hat{\theta}\hat{r}'_\bot
\hat{z}_o}{\hat{z}_o+1/2}\right)
\exp\left[\frac{i\hat{r}_\bot^{'2}
}{2\left(\hat{z}_o+1/2\right)}\right] \right\} \right|^2~.\cr
&&\label{fipr}
\end{eqnarray}
In the limit for $\hat{z}_o \gg 1$ Eq. (\ref{fipr}) gives back the
square modulus of Eq. (\ref{Efarsum2}) in normalized units.
Similarly, in the limit for $\hat{\phi} \ll 1$, and using the fact
that $K_1(\sqrt{\hat{\phi}} \hat{r}_\bot) \simeq 1/(\hat{r}_\bot
\sqrt{\hat{\phi}})$ one recovers the square modulus of Eq.
(\ref{vir4e}). In general, the integrals in Eq. (\ref{fipr})
cannot be calculated analytically, but they can be integrated
numerically.

\begin{figure}
\begin{center}
\includegraphics*[width=140mm]{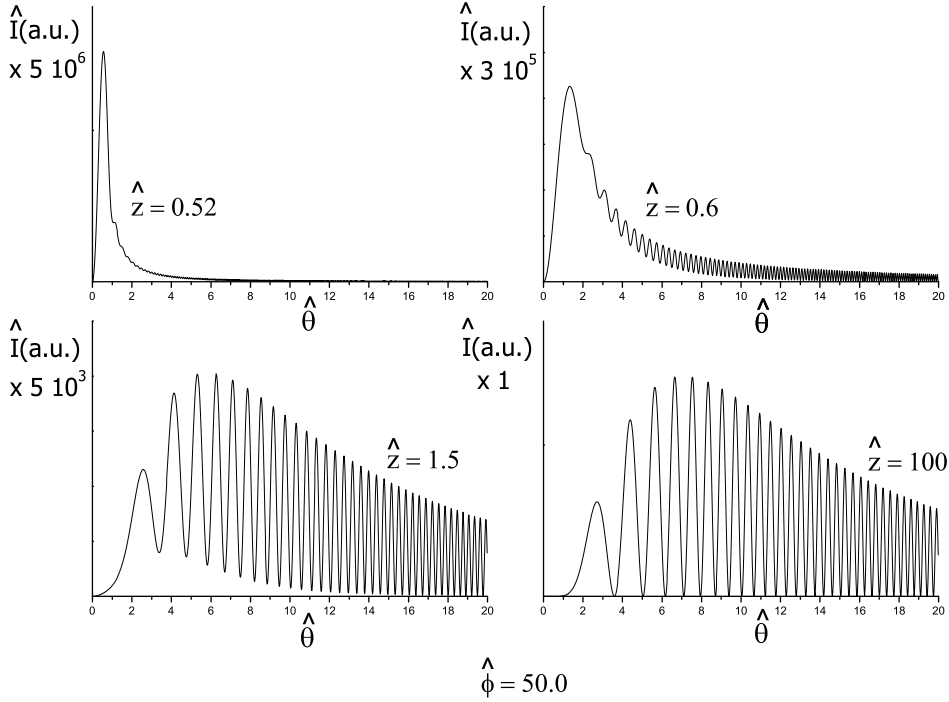}% Here is how to import EPS art
\caption{\label{evolve} Intensity profile for edge radiation for
$\hat{\phi}=50$. These profiles are shown as a function of angles
at different observation distances $\hat{z}_o=0.52$,
$\hat{z}_o=0.6$, $\hat{z}_o=1.5$ and $\hat{z}_o=100.0$.}
\end{center}
\end{figure}
First we checked that we are able to recover, posing $\hat{z}_o
=0$, the intensity profile for the single virtual source, already
calculated numerically and shown in Fig. \ref{Intvirphi} and Fig.
\ref{phil}. Then we propagated the field at non-virtual positions,
for $\hat{z}_o > 1/2$. In Fig. \ref{evolve} we plotted, in
particular, results for the propagation in case $\hat{\phi}=50$.
Radiation profiles are shown as a function of angles
$\hat{\theta}$ at different observation distances
$\hat{z}_o=0.52$, $\hat{z}_o=0.6$, $\hat{z}_o=1.5$ and
$\hat{z}_o=100.0$.

From a technical viewpoint, it is easier to deal with two sources
than with one, because the expression for the virtual two sources
is analytical, whereas that for a single one is not. Moreover, as
said before, the two-sources picture gives new physical insight
for the asymptotic limit $\hat{\phi} \gg 1$.

Qualitatively, we can deal with two limiting cases of the theory,
the first for $\hat{\phi} \ll 1$ and the second for $\hat{\phi}
\gg 1$.

Let us first discuss the case $\hat{\phi} \ll 1 $. The field at
any observation distance is given by Eq. (\ref{vir4e}). There are
only two observation zones of interest.

\begin{itemize}
\item \textbf{Far zone.} In the limit for $\hat{z}_o \gg 1$ one has the far field Eq. (\ref{lllim}).
\item \textbf{Near zone.}  When $\hat{z}_o \lesssim 1$
instead, one has the near field Eq. (\ref{vir4e}).
\end{itemize}
As it can be seen from Eq. (\ref{vir4e}), the total field is
given, both in the near and in the far zone, by the interference
of the two virtual sources located at the straight section edges.
Eq. (\ref{fipr}) shows that the transverse dimension of these
virtual sources is given by $\gamma \lambdabar$ in dimensional
units. This is the typical scale in $r'_\bot$ after which the
integrands in $d \hat{r}'_\bot$ in Eq. (\ref{fipr}) are suppressed
by the function $K_1$. Thus, the sources at the edges of the
straight section have a dimension that is independent of $L$. In
the center of the setup instead, the virtual source has a
dimension of order $\sqrt{\lambdabar L}$ as it can be seen Eq.
(\ref{vir4ei}). When $\hat{\phi}\ll 1$ the source in the center of
the setup is much smaller than those at the edge. This looks
paradoxical. The explanation is that the two contributions due to
edge sources interfere in the center of the setup. In particular,
when $\hat{\phi}\ll 1$ they nearly compensate, as they have
opposite sign. As a result of this interference, the single
virtual source in the center of the setup (and its far-zone
counterpart) has a dimension dependent on $L$ (in non-normalized
units)  while at the edges (and in their far-zone counterpart) the
dependence on $L$ is limited to phase factors only. Due to the
fact that edges contributions nearly compensate for $\hat{\phi}\ll
1$ one may say that the single-source picture is particularly
natural in the case $\hat{\phi} \ll 1$.

Let us now discuss the case $\hat{\phi} \gg 1$. In this situation
the two-sources picture becomes more natural. Let us define with
${d}_{1,2} = {z}_o \mp L/2$ the distances of the observer from the
edges. As seen before, the transverse dimension of the sources at
the edges of the straight section is $r'_\bot \sim
\gamma\lambdabar$. Moreover, we see from Eq. (\ref{ABcontre}) and
Eq. (\ref{phiabe}) that when $\hat{\phi} \gg 1$ the formation
length is $L_f = \gamma^2 \lambdabar$, much shorter than the
system dimension $L$. As a result, one can recognize four regions
of interest, that are more naturally discussed in the two-source
picture.

\begin{itemize}
\item \textbf{Two-edge radiation. Far zone.} When $d_{1,2} \gg L$ we are
summing far field contributions from the two edge sources. This
case is well represented in Fig. \ref{evolve} for $\hat{z}_o =
100$, where interference effects between the two edges
contribution are well visible.
\item \textbf{Two-edge radiation. Near zone.}  When $d_{1,2} \sim L$ the observer is located far
away with respect to the formation length of the sources $L'_f$.
Both contributions from the sources are important, but that from
the nearest source begins to become the main one, as $d_1$ and
$d_2$ become sensibly different. This case is well represented in
Fig. \ref{evolve} for $\hat{z}_o =1.5$.
\item \textbf{Single-edge radiation. Far zone.} When
$\gamma^2\lambdabar \ll d_1 \ll L$ the contribution due to the
near edge is dominant, while the far edge contribution is
negligible. Such tendency is clearly depicted in Fig. \ref{evolve}
for $\hat{z}_o =0.6$. Interference tends to disappear as the near
edge becomes the dominant one, while the intensity distribution
tends to approximate

\begin{eqnarray}
{I}\left({\theta}\right) = \mathrm{const} \times
\left|\vec{\widetilde{E}}_2 \left(z_o,\theta\right)\right|^2 =
\mathrm{const} \times \frac{4 e^2}{c^2 z_o^2} \frac{\gamma^4
{\theta}^2} {\left( \gamma^2 \theta^2+1\right)^2} ~,\label{Isefar}
\end{eqnarray}
where $\vec{\widetilde{E}}_2$ is the single edge far field limit
in Eq. (\ref{Efarsum2}).

\item \textbf{Single-edge radiation. Near zone.} When $0 < d_1 \lesssim \gamma^2
\lambdabar$ we have the near-field contribution from a single
edge. As ${d}_1$ becomes smaller and smaller the intensity
distribution tends to reproduce the singular behavior from a
single virtual source, i.e. the square modulus of Eq.
(\ref{virpm05}):

\begin{eqnarray}
{I}\left(\frac{L}{2},\vec{r}_\bot\right) = \mathrm{const} \times
\left|\vec{\widetilde{E}} \left(\frac{L}{2},
\vec{r}_\bot\right)\right|^2 = \mathrm{const} \times \frac{4  e^2
}{\gamma^2\lambdabar^2 c^2} \left|K_1\left(\frac{r_\bot}{ \gamma
\lambdabar} \right)\right|^2~.\label{Iviri}
\end{eqnarray}
This tendency can be seen in Fig. \ref{evolve} for $\hat{z}_o
=0.52$.
\end{itemize}

It should be noted that the straight section contribution has been
calculated  within the applicability region of the paraxial
approximation, but independently of all assumptions on the
switchers. If switchers or other setup parts are present, their
contribution must be separately calculated and added to the
contribution due to the straight section. For instance, if an
observer is placed after a switcher located downstream of the
straight section, the field due to the straight section has
physical meaning (i.e. the theoretical intensity can be compared
with experiment) only if the zero-length switchers approximation
can be applied. Otherwise it is just a mathematically convenient
term, a partial result to be summed with the switcher
contribution. If, instead, the observer is placed on a transverse
plane placed before the switcher, at the downstream edge of the
straight section, the switcher (or any other part of the setup
following the straight section)  will not influence the field with
ultra-relativistic accuracy.  Eq. (\ref{virpm05}) (or Eq.
(\ref{Iviri})) can then be compared with experimental results. As
has been discussed in footnote \ref{fifo}, the singularity in the
field at $r_\bot=0$ is fundamentally related with our ignorance
about the structure of the electron.

It should be noted that our Eq. (\ref{virpm05}) is identical to
the frequency-domain expression for the transverse component of
the field originating from an ultra-relativistic electron moving
with constant velocity (see \cite{JACK}). In fact, our treatment
for the straight-section contribution in the case of single-edge
radiation in the near zone gives, within an accuracy of
$1/\gamma^2$, the same result found in \cite{JACK} for
bremsstrahlung in a collision between an ultra-relativistic
electron and an atomic nucleus by means of the method of virtual
quanta. In the bremsstrahlung case the electron harmonic is
switched off within a length $d_s \ll \lambdabar$ even in the
model for an electron on a uniform motion, because non-negligible
contributions to the field are generated by the part of the
trajectory before the nucleus only, while the trajectory after the
nucleus gives negligible contribution within an accuracy
$1/\gamma^2$. Exact calculation, retaining all parts of the
Lienard-Wiechert field show that the field from the relativistic
particle is equivalent to two virtual pulses of radiation. One
moves longitudinally, i.e. in the same direction as the electron.
A second moves transversely, i.e. perpendicularly with respect to
the electron direction. As already said, the component of the
frequency spectrum due to the longitudinal pulse is given by our
Eq. (\ref{virpm05}). At any observation position, the component
due to the transverse pulse can be neglected with an accuracy
$1/\gamma^2$. In other words, the weak transverse pulse is a
correction to the main longitudinal pulse. It may be interpreted
as an evanescent wave decaying in the longitudinal direction,
since it propagates transversely, i.e. at large angle $\pi/2$ with
respect to the $z$ axis.  Dropping this term gives back our
result, that was found in the paraxial approximation.

Our previous result in Eq. (\ref{virpm05}) constitutes a novel
finding in theory of edge radiation dealing with the situation of
a particle in straight motion. The case of a particle in straight
motion is of fundamental importance, because of two reasons.
First, our result can be applied to any setup where the
zero-length switchers approximation applied. Second, straight
sections are basic building blocks for any magnetic setup, and our
expression for the near field from a straight section can be used,
as part of the total field, also in the case when the zero-length
switchers approximation fails.

\subsection{\label{sub:zerle} Supplementary remarks on the
zero-length switcher approximation}

For the sake of completeness it remains to qualitatively discuss
the conditions when the field due to switchers is negligible,
because in this case one can work in the "zero-length switchers
approximation". We consider the case when switchers are present in
the form of bending magnets. In particular, here we report results
from an analysis of the problem in the limit $L \gg \gamma^2
\lambdabar$. In this case the formation length associated with the
straight section is given by $\gamma^2 \lambdabar$, while the
formation length associated with the bending magnet is given by
$(\lambdabar \rho^2)^{1/3}$, $\rho$ being the radius of the bend.
The ratio between the latter and the former is indicated with
$\epsilon^2 = (\lambda_c/\lambda)^{2/3} \ll 1$, where $\lambda_c =
4 \pi \rho/(3 \gamma^3)$ is the critical wavelength for SR from
the bend and we assumed $\lambda \gg \lambda_c$, as we are
interested in edge radiation. The product between the small
quantity $\epsilon$ and the formation length for the straight
section is indicated with $d_c = \epsilon \gamma^2 \lambdabar$,
and turns out to constitute an extra characteristic-length for our
system. Let us indicate with $d$ the distance between a transverse
plane of observation and the downstream edge of the straight
section. At very short distances, for $d \ll d_c$, the straight
section contribution dominates the bending magnet one for
transverse displacements from the $z$-axis $r_\bot$ of order
$r_\bot > \epsilon \gamma \lambdabar$. Within the area $r_\bot <
\epsilon \gamma \lambdabar$ the two contributions are comparable.
Note that in this area the contribution from the straight section
is much larger than  in the area for $\epsilon \gamma \lambdabar <
r_\bot < \gamma \lambdabar$ due to singular behavior of the field
contribution from the straight section at $r_\bot \longrightarrow
0$. When $d \sim d_c$ the zero-length switcher approximation
cannot be used, and contributions from the bend should be
calculated explicitly. When $\gamma^2 \lambdabar \gtrsim d \gg
d_c$ we have room for direct application of our theory for the
straight section within the range $0 < r_\bot \ll (\lambdabar^2
\rho)^{1/3}$, because the characteristic size of the radiation is
$\epsilon (\lambdabar^2 \rho)^{1/3}$. Finally, for $d \gg \gamma^2
\lambdabar$, we have room for direct application of our theory for
the straight section within the angular range $0 < \theta \ll
(\lambdabar/ \rho)^{1/3}$, because the characteristic angle of the
radiation is $\epsilon (\lambdabar/ \rho)^{1/3}$.

\section{\label{sec:five} A more sophisticated application. Transition undulator radiation}

\begin{figure}
\begin{center}
\includegraphics*[width=140mm]{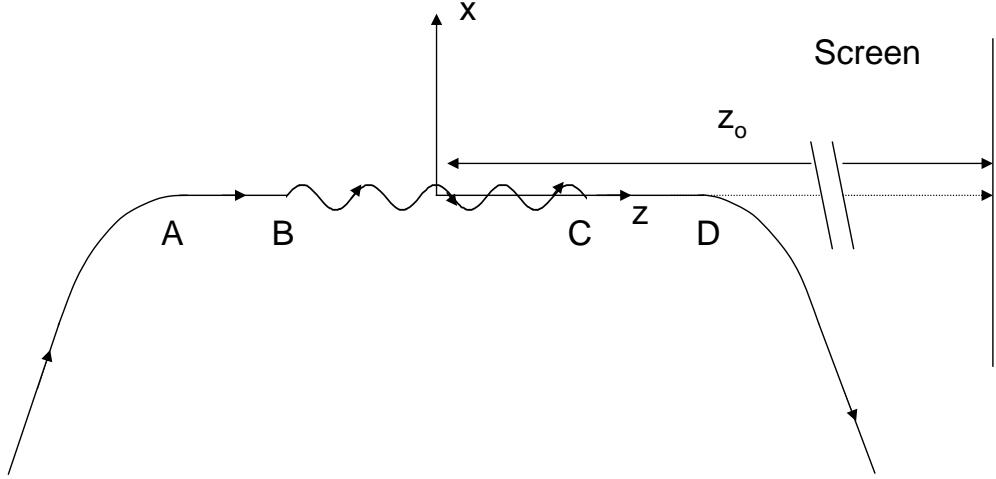}% Here is how to import EPS art
\caption{\label{geom} Edge radiation geometry. The beam enters the
system through a bending magnet, passes a straight section, an
undulator and another straight section before leaving the setup
through another bend. Radiation is collected at a distance $z_o$
from the origin of the reference system, located in the middle of
the undulator. }
\end{center}
\end{figure}
In this Section we apply the method described in Section
\ref{sec:meth} to the more complicated case of an undulator setup.

Instead of the setup in Fig. \ref{geome}, we now consider the
system depicted in Fig. \ref{geom} and we consider a single
particle moving along the system. The electron enters the setup
via a bending magnet, passes through a straight section (segment
$AB$), an undulator (segment $BC$), and another straight section
(segment $CD$). Finally, it exit the setup via another bend.
Radiation is collected at a distance $z_o$ from the center of the
reference system, located in the middle of the undulator. The
passage of the electron through the setup results in collimated
emission of radiation in the low photon energy range, a mechanism
analogous to transition radiation. This kind of radiation is known
in literature as TUR \cite{KIM1,BOS4,KINC,CAST,BOS6,ROY2}, even
though in this paper we prefer to denote it as edge radiation from
undulator setup.

In our case of study the trajectory and, therefore, the space
integration in Eq. (\ref{generalfin}) can be split in five parts:
the two bends, that will be indicated with $b_1$ and $b_2$, the
two straight sections $AB$ and $CD$ and the undulator $BC$. One
may write

\begin{eqnarray}
\vec{\widetilde{{E}}}_\bot(z_o, \vec{r}_{\bot o},\omega) &=&
\vec{\widetilde{{E}}}_{b1}(z_o, \vec{r}_{\bot o},\omega)+
\vec{\widetilde{{E}}}_{AB}(z_o, \vec{r}_{\bot o},\omega)+
\vec{\widetilde{{E}}}_{BC}(z_o, \vec{r}_{\bot o},\omega)\cr &&+
\vec{\widetilde{{E}}}_{CD}(z_o, \vec{r}_{\bot o},\omega)+
\vec{\widetilde{{E}}}_{b2}(z_o, \vec{r}_{\bot o},\omega)~,
\label{split}
\end{eqnarray}
with obvious meaning of notation.

We will denote the length of the segment $AD$ with
$L_\mathrm{tot}$, while we will indicate the length of the
straight section $AB$ with $L_1$, the length of the straight
section $CD$ with $L_2$ and the length of the undulator with
$L_w$. It follows

\begin{equation}
L_\mathrm{tot} = L_1+L_w+L_2~. \label{ffield}
\end{equation}
This means that point $A$ is located at longitudinal coordinate
$z_A= -L_1 - L_w/2$, while $B$, $C$ and $D$ are located
respectively at $z_B= -L_w/2$, $z_C = L_w/2$ and $z_D =  L_w/2
+L_2$.

First, with the help of Eq. (\ref{generalfin}) we will derive an
expression for the field in the far zone. The intensity
distribution in this case, results to be in agreement with that
given in \cite{BOS4}. Then, we will calculate the field
distribution at the virtual source with the help of Eq.
(\ref{virfie}). Finally, Eq. (\ref{fieldpropback}) will allow us
to find an expression for the field both in the near and in the
far zone. We will the turn to analyze the case when a undulator is
present, and describe the field from the setup as superposition of
three laser-like beam from straight sections and undulator. In the
following we will ignore the presence of the bending magnets, i.e.
the radiation switchers. As before, we will treat them as if they
had zero length. As a result, Eq. (\ref{split}) can be simplified
to

\begin{eqnarray} \vec{\widetilde{{E}}}_\bot(z_o,
\vec{r}_{\bot o},\omega) &=& \vec{\widetilde{{E}}}_{AB}(z_o,
\vec{r}_{\bot o},\omega)+ \vec{\widetilde{{E}}}_{BC}(z_o,
\vec{r}_{\bot o},\omega)+ \vec{\widetilde{{E}}}_{CD}(z_o,
\vec{r}_{\bot o},\omega)~. \label{split2}
\end{eqnarray}
We already addressed the question of the applicability region of
the "zero-length switcher approximation" in previous Section
\ref{sec:four}.

\subsection{\label{sub:far} Far field from the undulator setup}

Let us describe the far field from the undulator setup in Fig.
\ref{geom} by separately characterizing the field contributions
along undulator, straight sections and bends and finally adding
them together.

\subsubsection{\label{sub:undu} Field contribution calculated along
the undulator}

We first consider the contribution $\vec{\widetilde{{E}}}_{BC}$
from the undulator, following \cite{OUR1}. Assuming a planar
undulator we write the following expression for the transverse
velocity of an electron:

\begin{equation}
\vec{v}_\bot(z') = - {c K\over{\gamma}} \sin{\left(k_w z'\right)}
\vec{x}~, \label{vuz}
\end{equation}
that is Eq. (\ref{vuzo}). Here $K$ is the undulator parameter
defined in Eq. (\ref{Kpara}). Moreover, $k_w=2\pi/\lambda_w$,
where $\lambda_w$ is the undulator period, so that the undulator
length is $L = N_w \lambda_w$.  The transverse position of the
electron is therefore

\begin{equation}
\vec{r'}_\bot(z') = \frac{K}{\gamma k_w} \cos{\left(k_w z'\right)}
\vec{x}~. \label{erz}
\end{equation}
An expression for the curvilinear abscissa $s$ as a function of
the longitudinal position $z'$ is given by

\begin{equation}
s(z') =  {\frac{\beta}{\beta_{av}}}z' - {K^2\over{8\gamma^2 k_w}}
\sin\left(2 k_w z'\right)~,\label{zsss}
\end{equation}
where $\beta_{av}$ is the time-averaged velocity along the $z$
direction, that can be expressed as:

\begin{equation}
\beta_{av} = \beta \left(1-{K^2\over{4\gamma^2}}\right)~.
\label{bbar}
\end{equation}
We can now substitute Eq. (\ref{erz}) and Eq. (\ref{vuz}) in Eq.
(\ref{generalfin}). Such substitution leads to a general
expression, valid for any observer distance $z_o$.  It is possible
to obtain, similarly to many SR textbooks, a simplified expression
valid in the limit for large values of $z_o$. Since we are
interested in the contribution of the undulator device to the
total field at the observer position, we will integrate Eq.
(\ref{generalfin}) only along the undulator. Then all terms in
$(z_o-z')^{-1}$ in the phase factor of Eq. (\ref{generalfin}) can
be expanded around $z_o$. In the far field limit we can retain
first order terms in $z'$. Dropping negligible terms we obtain

\begin{eqnarray}
\vec{\widetilde{{E}}}_{BC}(z_o, \vec{r_{\bot o}},\omega) &=&
\frac{i \omega e}{c^2 z_o} \int_{z_B}^{z_C} dz' {\exp{\left[i
\Phi_{BC}\right]}} \left\{\left[{K\over{\gamma}} \sin\left(k_w
z'\right)+\theta_x\right]{\vec{x}} +\theta_y{\vec{y}}\right\},
\label{unduraddd}
\end{eqnarray}
where

\begin{eqnarray}
\Phi_{BC} &=& {\omega} \left\{{\theta_x^2+\theta_y^2\over{2c}}
z_o+ \frac{z'}{2  c
}\left(\frac{1}{\gamma_z^2}+\theta_x^2+\theta_y^2\right) -
\frac{K\theta_x}{\gamma k_w c}\cos(k_w z') \right.\cr&&\left.-
\frac{K^2}{8\gamma^2 k_w c} \sin(2 k_w z') \right\}
~.\label{phitundu}
\end{eqnarray}
The longitudinal Lorentz factor $\gamma_z$ in Eq. (\ref{phitundu})
is defined by

\begin{equation}
\gamma_z = \frac{\gamma}{\sqrt{1+{K^2}/{2}}} \label{gz}
\end{equation}
and is always smaller than $\gamma$, because the average
longitudinal velocity of the electron inside the undulator is
smaller than that along the straight sections.

In this paper we will be interested up to frequencies much lower
than the resonance frequency, i.e.  $\omega \ll 2 \gamma_z^2 c
k_w$\footnote{Although we are outside of the applicability range
of the resonance approximation, we still assume that the number of
undulator period $N_w$ is large, i.e. $N_w \gg 1$ to calculate the
undulator contribution. As a result, the accuracy related with the
undulator contribution is $1/(2\pi N_w)$, and not $1/\gamma^2$ as
in the straight section case. It follows that results regarding
the undulator contribution to the field are not as fundamental as
those regarding the straight section, because the extra
approximation $N_w \gg 1$ has been used, that is specific for the
undulator device. }. Then, the contribution due to the term in
$\sin(k_w z')$ in Eq. (\ref{unduraddd}) can always be neglected
when compared with the maximal field magnitude of the terms in
$\theta_{x,y}$. Similarly, in Eq. (\ref{phitundu}), phase terms in
$\cos(k_w z')$ and $\sin(2k_w z')$ can can also be neglected. As a
result, Eq. (\ref{unduraddd}) can be simplified as

\begin{eqnarray}
\vec{\widetilde{{E}}}_{BC}(z_o, \vec{r_{\bot o}},\omega) &=&
\frac{i \omega e}{c^2 z_o} \int_{z_B}^{z_C} dz' {\exp{\left[i
\Phi_{BC}\right]}} \left(\theta_x {\vec{x}}
+\theta_y{\vec{y}}\right) \label{undurad}
\end{eqnarray}
while

\begin{eqnarray}
\Phi_{BC} &=& {\omega} \left[{\theta_x^2+\theta_y^2\over{2c}} z_o+
\frac{z'}{2  c
}\left(\frac{1}{\gamma_z^2}+\theta_x^2+\theta_y^2\right) \right]~.
~\label{phitundu4}
\end{eqnarray}
Finally, note that within these approximations, the curvilinear
abscissa $s$ in Eq. (\ref{zsss}) can be simplified to

\begin{equation}
s(z') =  {\frac{\beta}{\beta_{av}}}z' ~.\label{zsssus}
\end{equation}

\subsubsection{\label{sub:str} Field contribution calculated along
the straight sections}

Let us now calculate the curvilinear abscissa along segments $AB$
and $CD$. Eq. (\ref{zsssus}) implies that the curvilinear abscissa
at the undulator edges $z'=\pm L_w/2$ is given by $ \pm \beta
L_w/(2\beta_{av})$. Moreover the curvilinear abscissa must be of
the form $s(z') = z' + \mathrm{const.}$, because segments $AB$ and
$CD$ are straight lines. It follows that

\begin{equation}
s(z') =  z'+ \frac{L_w}{2}
\left(1-\frac{\beta}{\beta_{av}}\right)\simeq z'+
\frac{L_w}{4\gamma^2} -
\frac{L_w}{4\gamma_z^2}~~~~~\mathrm{for}~z_A<z'<z_B
~\label{zssssa}
\end{equation}
and

\begin{equation}
s(z') =  z'- \frac{L_w}{2}
\left(1-\frac{\beta}{\beta_{av}}\right)\simeq z'
-\frac{L_w}{4\gamma^2}+
\frac{L_w}{4\gamma_z^2}~~~~~\mathrm{for}~z_C<z'<z_D
~,\label{zssssb}
\end{equation}
where we have used the fact that

\begin{equation}
\frac{1}{\beta_{av}} \simeq 1+ \frac{1}{2\gamma_z^2}
~.\label{bava}
\end{equation}
With the help of Eq. (\ref{generalfin}) we write the contribution
from the straight line $AB$ as

\begin{equation}
\vec{\widetilde{E}}_{AB}={i \omega e\over{c^2 z_o}}
\int_{z_A}^{z_B} dz' \exp{\left[i\Phi_{AB}\right]} \left(\theta_x
{\vec{x}}+\theta_y {\vec{y}}\right) \label{ABcontr}
\end{equation}
where $\Phi_{AB}$ in Eq. (\ref{ABcontr}) is given by

\begin{equation}
\Phi_{AB} = \omega \left[ {\theta_x^2+\theta_y^2\over{2c}} z_o  -
\frac{L_w}{4c\gamma_z^2}+ \frac{L_w}{4c\gamma^2}+
{z'\over{2c}}\left({1\over{\gamma^2}}+\theta_x^2+\theta_y^2\right)\right]~,
\label{phiab}
\end{equation}
$\theta_x = x/z_o$ and $\theta_y = y/z_o$ being the observation
angles in the horizontal and vertical direction. Note that we also
used $v \simeq c$ when calculating the term
${L_w}/{(4c\gamma_z^2)}$ in Eq. (\ref{phiab}). The contribution
from the straight line $CD$ is similar to that from the straight
line $AB$ and reads

\begin{equation}
\vec{\widetilde{E}}_{CD}={i \omega e\over{c^2 z_o}}
\int_{z_C}^{z_D} dz' \exp{\left[i\Phi_{CD}\right]} \left(\theta_x
{\vec{x}}+\theta_y {\vec{y}}\right) \label{CDcontr}
\end{equation}
where $\Phi_{CD}$ in Eq. (\ref{CDcontr}) is given by

\begin{equation}
\Phi_{CD} = \omega \left[ {\theta_x^2+\theta_y^2\over{2c}} z_o
+\frac{L_w}{4c\gamma_z^2}-\frac{L_w}{4c\gamma^2}+
{z'\over{2c}}\left(\frac{1}{\gamma^2}+\theta_x^2+\theta_y^2\right)\right]~.
\label{phicd}
\end{equation}
In general, the phases $\Phi_{CD}$ and $\Phi_{AB}$ start
exhibiting oscillatory behavior when $\omega z'/(2 c \gamma^2)
\sim 1$, which gives a maximal integration range in the
longitudinal direction. Similarly as before, in general one has
that the formation lengths $L_\mathrm{fs1}$ and $L_\mathrm{fs2}$
for the straight sections $AB$ and $CD$ can be written as

\begin{equation}
L_\mathrm{fs(1,2)} \sim \min\left[ \lambda
\gamma^2,L_{(1,2)}\right] ~. \label{lstf}
\end{equation}
Depending on the wavelength of interest then, $L_\mathrm{fs(1,2)}
\sim \lambda \gamma^2$ or $L_\mathrm{fs(1,2)} \sim L_{(1,2)}$.

\subsubsection{\label{sub:long} Total field and energy spectrum of radiation}

Consider Eq. (\ref{split2}). The contributions for segment $AB$
and segment $CD$ are given by Eq. (\ref{ABcontr}) and Eq.
(\ref{CDcontr}). One obtains

\begin{eqnarray}
\vec{\widetilde{E}}_{AB}&&=-\frac{ 2    e
\left(\theta_x\vec{x}+\theta_y \vec{y}
\right)}{\left(1/\gamma^2+\theta_x^2+\theta_y^2\right) c z_o}
\exp\left[i \omega {\theta_x^2+\theta_y^2\over{2c}} z_o
\right]\exp\left[-\frac{i \omega L_w}{4 c \gamma_z^2} +\frac{i
\omega L_w}{4 c \gamma^2}\right]\cr && \times \left\{\exp\left[- i
\frac{\omega (L_1+ L_w/2)}{2 c}
\left(\frac{1}{\gamma^2}+\theta_x^2+\theta_y^2\right)
\right]\right.\cr &&\left. -\exp\left[- i \frac{\omega L_w }{4
c}\left(\frac{1}{\gamma^2}+\theta_x^2+\theta_y^2\right)
\right]\right\}\cr &&
 \label{ABcontrint4}
\end{eqnarray}
that can also be written as

\begin{eqnarray}
\vec{\widetilde{E}}_{AB}&&= \frac{i \omega e L_1}{c^2 z_o}
\vec{\theta} \mathrm{sinc}\left[\frac{\omega L_1}{4
c}\left(\frac{1}{\gamma^2}+\theta^2\right)\right] \exp\left[\frac
{i \omega \theta^2 z_o}{2 c} \right]\cr &&\times
\exp\left[-\frac{i \omega L_w
}{4c}\left(\frac{1}{\gamma_z^2}+\theta^2\right)\right]\exp\left[-\frac{i
\omega L_1
}{4c}\left(\frac{1}{\gamma^2}+\theta^2\right)\right]\label{ABfar}
\end{eqnarray}

Similarly,

\begin{eqnarray}
\vec{\widetilde{E}}_{CD}&&=-\frac{ 2    e
\left(\theta_x\vec{x}+\theta_y \vec{y}
\right)}{\left(1/\gamma^2+\theta_x^2+\theta_y^2\right) c z_o}
\exp\left[i \omega {\theta_x^2+\theta_y^2\over{2c}} z_o \right]
\exp\left[\frac{i \omega L_w}{4 c \gamma_z^2}-\frac{i \omega
L_w}{4 c \gamma^2} \right] \cr && \times \left\{-\exp\left[ i
\frac{\omega (L_2+L_w/2) }{2 c}
\left(\frac{1}{\gamma^2}+\theta_x^2+\theta_y^2\right)
\right]\right.\cr &&\left. +\exp\left[ i \frac{\omega L_w }{4
c}\left(\frac{1}{\gamma^2}+\theta_x^2+\theta_y^2\right)
\right]\right\}~.\cr &&
 \label{CDcontrint4}
\end{eqnarray}
that can also be written as

\begin{eqnarray}
\vec{\widetilde{E}}_{CD}&&= \frac{i \omega e L_2}{c^2 z_o}
\vec{\theta} \mathrm{sinc}\left[\frac{\omega L_2}{4
c}\left(\frac{1}{\gamma^2}+\theta^2\right)\right] \exp\left[\frac
{i \omega \theta^2 z_o}{2 c} \right]\cr &&\times \exp\left[\frac{i
\omega L_w
}{4c}\left(\frac{1}{\gamma_z^2}+\theta^2\right)\right]\exp\left[\frac{i
\omega L_2
}{4c}\left(\frac{1}{\gamma^2}+\theta^2\right)\right]\label{CDfar}
\end{eqnarray}

Finally, the contribution for the segment $BC$ is written as in
Eq. (\ref{undurad}). Calculations yield:

\begin{eqnarray}
\vec{\widetilde{E}}_{{BC}}&=&-\frac{ 2    e
\left(\theta_x\vec{x}+\theta_y \vec{y}
\right)}{\left(1/\gamma_z^2+\theta_x^2+\theta_y^2\right) c z_o}
\exp\left[i \omega \frac{\theta_x^2+\theta_y^2}{2c} z_o \right]
\cr &&\times \left\{ {-\exp\left[\frac{i\omega L_w}{4c}
\left(\frac{1}{\gamma_z^2}+\theta_x^2+\theta_y^2\right)\right]}
\right.\cr &&\left.+{\exp\left[-\frac{i\omega L_w}{4c}
\left(\frac{1}{\gamma_z^2}+\theta_x^2+\theta_y^2\right)\right]}\right\}
~.\cr && \label{linead4b}
\end{eqnarray}
that can be written as

\begin{eqnarray}
\vec{\widetilde{E}}_{BC}&&= \frac{i \omega e L_w}{c^2 z_o}
\vec{\theta} \mathrm{sinc}~ \left[\frac{\omega L_w}{4
c}\left(\frac{1}{\gamma_z^2}+\theta^2\right)\right]
\exp\left[\frac {i \omega \theta^2 z_o}{2 c} \right]\label{BCfar}
\end{eqnarray}
The total field produced by the setup is obtained by summing up
Eq. (\ref{ABcontrint4}), Eq. (\ref{CDcontrint4}) and Eq.
(\ref{BCfar}). Note that the same definition for the observation
angle $\vec{\theta} = \vec{r}_\bot/z_o$ is used in these
equations. This means that the observation angle is measured
starting from the center of the undulator, located at $z=0$. The
energy density of radiation as a function of angles and
frequencies can be written substituting the resultant total field
in Eq. (\ref{endene}). We obtain

\begin{eqnarray}
&&\frac{d W}{d\omega d \Omega} = \frac{ e^2}{ \pi^2 c} \frac{
\gamma^4 \theta^2}{\left(1+\gamma^2\theta^2\right)^2 } \Bigg|
-\exp\left[- i \frac{\omega L_w }{4
c\gamma^2}\left(1+\frac{K^2}{2}+\gamma^2\theta^2\right) \right]\cr
&&+\exp\left[- i \frac{\omega L_1}{2 c \gamma^2}
\left(1+\gamma^2\theta^2\right) - i \frac{\omega  L_w}{4 c
\gamma^2} \left(1+\frac{K^2}{2}+\gamma^2\theta^2\right) \right]
\cr && +\frac{1/\gamma^2+\theta^2}{1/\gamma_z^2+\theta^2}\left\{
{-\exp\left[\frac{i\omega L_w}{4c\gamma^2}
\left({1}+\frac{K^2}{2}+\gamma^2\theta^2\right)\right]} \right.\cr
&&\left.+{\exp\left[-\frac{i\omega L_w}{4c\gamma^2}
\left(1+\frac{K^2}{2}+\gamma^2\theta^2\right)\right]}\right\}
+\exp\left[ i \frac{\omega L_w }{4
c\gamma^2}\left(1+\frac{K^2}{2}+\gamma^2\theta^2\right) \right]\cr
&& -\exp\left[ i \frac{\omega L_2 }{2 c\gamma^2}
\left(1+\gamma^2\theta^2\right)+\frac{i \omega L_w}{4 c
\gamma^2}\left(1+\frac{K^2}{2}+\gamma^2\theta^2\right)
\right]\Bigg|^2~, \label{enden2}
\end{eqnarray}
that is equivalent to the analogous expression in \cite{BOS4}.

\subsection{\label{sub:virt} Virtual source characterization and
field propagation}

Expressions in Eq. (\ref{ABfar}), Eq. (\ref{CDfar}) and Eq.
(\ref{BCfar}) can be interpreted as far field radiation from
separate virtual sources. Note that the same definition for the
observation angle $\vec{\theta} = \vec{r}_\bot/z_o$ is used in
these equations. Let us find the locations of the virtual sources.
We will see that these locations corresponds to points along the
longitudinal axis where wavefronts are plane, i.e. where maximal
simplification arises.

When $F(0)$ is real, Eq. (\ref{fieldpropback3tris}) describes the
far field in terms of a spherical wave centered at $z=0$. In the
more generic case of a spherical wave centered at some position
$z_s$ Eq. (\ref{fieldpropback3tris}) should be substituted by Eq.
(\ref{fieldpropback3}) with $z = z_s$, as the typical phase phase
factor of a spherical wave centered in $z_s$ is $\exp[i \omega
\theta^2 (z_o+z_s)/(2c)]$. In this case $F(z_s)$ in Eq.
(\ref{fieldpropback3}) is a real function. Accordingly, Eq.
(\ref{virfie}) should be modified to Eq. (\ref{virfiemody}). Once
we substitute Eq. (\ref{ABfar}), Eq. (\ref{CDfar}) or Eq.
(\ref{BCfar}) in Eq. (\ref{virfiemody}), there are particular
values of $z_s$ ($z_{s1}$ for the segment $AB$, $z_{s2}$ for $BC$
and $z_{s3}$ for $CD$) such that the phase in $\theta^2$ is
cancelled and that ${ \widetilde{E}}( z_s,\vec{r}_{\bot} )$ gives
a plane wave. These are:

\begin{equation}
{z}_{s1} = -\frac{{L}_w }{2  }-\frac{{L}_1}{2} ~,\label{co1}
\end{equation}
\begin{equation}
{z}_{s2} = 0 \label{co2}
\end{equation}
and

\begin{equation}
{z}_{s3} = \frac{{L}_w }{2}+\frac{{L}_2}{2} ~.\label{co3}
\end{equation}
Substituting Eq. (\ref{ABfar}), Eq. (\ref{CDfar}) and Eq.
(\ref{BCfar}) into Eq. (\ref{virfiemody}) we obtain the following
plane wavefronts describing the three virtual sources at positions
$z_{s1}$, $z_{s2}$ and $z_{s3}$:

\begin{eqnarray}
\vec{\widetilde{E}}_{AB}\left(z_{s1},\vec{r}_{\bot}\right) &=& -
\frac{ \omega^2 e L_1}{2\pi c^3 } \exp\left[-\frac{i\omega
}{4c}\left(\frac{L_w}{\gamma_z^2}+\frac{L_1}{\gamma^2}\right)\right]\cr
&& \times \int d\vec{\theta} ~\vec{\theta}~
\mathrm{sinc}\left[\frac{\omega L_1}{4 c}
\left(\theta^2+\frac{1}{\gamma^2}\right) \right]\exp\left[\frac{i
\omega}{c}\vec{r}_{\bot}\cdot \vec{\theta} \right]
~,\label{vir1eaf}
\end{eqnarray}
\begin{eqnarray}
\vec{\widetilde{E}}_{BC}(z_{s2},\vec{r}_{\bot}) &=& -\frac{
\omega^2 e L_w}{2\pi c^3 } \int d\vec{\theta} ~\vec{\theta}~
\mathrm{sinc}\left[\frac{\omega L_w}{4 c}
\left(\theta^2+\frac{1}{\gamma_z^2}\right)
\right]\exp\left[\frac{i \omega}{c}\vec{r}_{\bot}\cdot
\vec{\theta} \right] ~\cr && \label{vir1ebf}
\end{eqnarray}
and

\begin{eqnarray}
\vec{\widetilde{E}}_{CD}\left(z_{s3},\vec{r}_{\bot}\right) &=& -
\frac{ \omega^2 e L_2}{2\pi c^3 }
\exp\left[\frac{i\omega}{4c}\left(\frac{L_w}{\gamma_z^2}+\frac{L_2}{\gamma^2}\right)\right]\cr
&& \times \int d\vec{\theta} ~\vec{\theta}~
\mathrm{sinc}\left[\frac{\omega L_1}{4 c}
\left(\theta^2+\frac{1}{\gamma^2}\right) \right]\exp\left[\frac{i
\omega}{c}\vec{r}_{\bot}\cdot \vec{\theta} \right]
~.\label{vir1edf}
\end{eqnarray}
Note that $L_1$, $L_2$ and $L_w$ can assume different values.
$\gamma$ and $\gamma_z$ are also different. It may therefore seem
convenient to introduce different normalized quantities, referring
to the undulator and the straight lines. However, in the end we
are in summing up contribution all contributions from different
sources, so that it is important to keep a common definition of
vertical displacement (or observation angle). Therefore we
prescribe the same normalization for all quantities:

\begin{equation}
\vec{\hat{\theta}} = \sqrt{\frac{\omega L_\mathrm{tot}}{c}}
\vec{\theta}~, \label{thnoe}
\end{equation}
\begin{equation}
\hat{\phi} = \frac{\omega L_\mathrm{tot}}{\gamma^2 c}~,
\label{phie}
\end{equation}
and

\begin{equation}
\vec{\hat{r}}_{\bot} = \sqrt{\frac{\omega }{L_\mathrm{tot}c}}
\vec{r}_{\bot}~. \label{rsnoe}
\end{equation}
Then, we introduce parameters $\hat{L}_1 = L_1/L_\mathrm{tot}$,
$\hat{L}_2 = L_2/L_\mathrm{tot}$, $\hat{L}_w =  L_w/
L_\mathrm{tot}$ and $\hat{\phi}_w = \gamma^2/\gamma_z^2
\hat{\phi}$.  Finally, we define $\hat{z}_{s}=z_s/L_\mathrm{tot}$.
Eq. (\ref{vir1eaf}), Eq. (\ref{vir1ebf}) and Eq. (\ref{vir1edf})
can then be written as

\begin{eqnarray}
\vec{\hat{E}}_{AB}\left(\hat{z}_1,\vec{\hat{r}}_{\bot}\right)
&=-&\hat{L}_1
\exp\left[-\frac{i}{4}\left(\hat{L}_w\hat{\phi}_w+\hat{L}_1\hat{\phi}\right)\right]
\cr && \times \int d\vec{\hat{\theta}} ~\vec{\hat{\theta}}~
\mathrm{sinc}\left[\frac{\hat{L}_1}{4}
\left(\hat{\theta}^2+\hat{\phi}\right)
\right]\exp\left[i\vec{\hat{r}}_{\bot}\cdot \vec{\hat{\theta}}
\right] ~,\label{vir1ean}
\end{eqnarray}
\begin{eqnarray}
\vec{\hat{E}}_{BC}(\hat{z}_2,\vec{\hat{r}}_{\bot})
&=&-\hat{L}_w\int d\vec{\hat{\theta}} ~\vec{\hat{\theta}}~
\mathrm{sinc}\left[\frac{\hat{L}_w}{4}
\left(\hat{\theta}^2+\hat{\phi}_w\right) \right]\exp\left[i
\vec{\hat{r}}_{\bot}\cdot \vec{\hat{\theta}} \right]
~\label{vir1ebn}
\end{eqnarray}
and

\begin{eqnarray}
\vec{\hat{E}}_{CD}\left(\hat{z}_3,\vec{\hat{r}}_{\bot}\right)
&=&-{\hat{L}_2}
\exp\left[\frac{i}{4}\left(\hat{L}_w\hat{\phi}_w+\hat{L}_2\hat{\phi}\right)\right]\cr
&&\times \int d\vec{\hat{\theta}} ~\vec{\hat{\theta}}~
\mathrm{sinc}\left[\frac{\hat{L}_2}{4}
\left(\hat{\theta}^2+\hat{\phi}\right)
\right]\exp\left[i\vec{\hat{r}}_{\bot}\cdot \vec{\hat{\theta}}
\right] ~\label{vir1edn}
\end{eqnarray}
where we  defined

\begin{equation}
\vec{\widetilde{E}} =  \frac{\omega e}{c^2} \vec{\hat{E}}~.
\label{etildehat}
\end{equation}
Besides factors in the $\mathrm{sinc}$ functions, the integrals in
Eq. (\ref{vir1ean}), Eq. (\ref{vir1ebn}) and Eq. (\ref{vir1edn})
are mathematically identical to Eq. (\ref{vir1e}). One may check
that, in the limit for $\hat{\phi} \ll 1$ and $\hat{\phi}_w \ll 1$
one obtains the same results as for edge radiation from a single
straight section.

Besides this obvious limit, there is a second region of interest
in the parameter space that can be dealt with analytically and
corresponds to $\hat{\phi} \ll 1$ and $\hat{\phi}_w \gg 1$. In
this particular limit, the contribution from the undulator, Eq.
(\ref{vir1ebn}) can be neglected, because the
$\mathrm{sinc}(\cdot)$ is strongly suppressed. In this case one
has the following virtual sources:

%[[[ CHECK OUT SIGNS AND FACTOR 4 -REMEMBER NEXT WEEK !!!!! ]]]

\begin{eqnarray}
\vec{\hat{E}}(\hat{z}_1,\vec{\hat{r}}_\bot) = \frac{4
\vec{\hat{r}}_\bot}{\hat{L}_1} \exp\left[-\frac{i \hat{L}_w
\hat{\phi}_w }{4}\right]
\mathrm{sinc}\left(\left|\vec{\hat{r}}_\bot\right|^2/\hat{L}_1\right)
~,\label{virt1}
\end{eqnarray}
\begin{eqnarray}
\vec{\hat{E}}(\hat{z}_2,\vec{\hat{r}}_\bot) \simeq
0~,\label{virt2}
\end{eqnarray}
\begin{eqnarray}
\vec{\hat{E}}(\hat{z}_3,\vec{\hat{r}}_\bot) = \frac{4
\vec{\hat{r}}_\bot}{\hat{L}_2} \exp\left[\frac{i \hat{L}_w
\hat{\phi}_w }{4}\right]
\mathrm{sinc}\left(\left|\vec{\hat{r}}_\bot\right|^2/\hat{L}_2\right)
~,\label{virt3}
\end{eqnarray}
Note that the previous three equations describe virtual sources
characterized by plane wavefronts. Application of the Fresnel
propagation formula, Eq. (\ref{fieldpropback}) allows one to
reconstruct the field both in the near and in the far region.
Results can directly be obtained from Eq. (\ref{vir4e}) accounting
for the fact that we want to calculate contributions at the same
observation point, which requires substitution of
$z_o/L_{(1,w,2)}$ with $(z_o-z_{s(1,2,3)})/{L_{(1,w,2)}}$. From
Eq. (\ref{vir4e}) we obtain the following results for the two
surviving contributions:

\begin{eqnarray}
\vec{\hat{E}}_{AB}(\hat{z}_o,\vec{\hat{r}}_\bot) &=& -\frac{4
\vec{\hat{r}}_\bot}{\hat{r}_\bot^2}\exp\left[-\frac{i \hat{L}_w
\hat{\phi}_w }{4}\right]\exp\left[i
\frac{\hat{r}_\bot^2}{2(\hat{z}_o-\hat{z}_{s1})}\right]\cr &&
\times \left[ \exp\left(-\frac{i \hat{L}_1\hat{r}_\bot^2 }{2
(\hat{z}_o-\hat{z}_{s1}) (\hat{L}_1+2\hat{z}_o-2\hat{z}_{s1})}
\right)\right.\cr&&\left.-\exp\left(\frac{i
\hat{L}_1\hat{r}_\bot^2 }{2 (\hat{z}_o-\hat{z}_{s1})
(-\hat{L}_1+2\hat{z}_o-2\hat{z}_{s1})} \right) \right]
~.\label{vir4A}
\end{eqnarray}
\begin{eqnarray}
\vec{\hat{E}}_{CD}(\hat{z}_o,\vec{\hat{r}}_\bot) &=& -\frac{4
\vec{\hat{r}}_\bot}{\hat{r}_\bot^2}\exp\left[\frac{i \hat{L}_w
\hat{\phi}_w }{4}\right]\exp\left[i
\frac{\hat{r}_\bot^2}{2(\hat{z}_o-\hat{z}_{s3})}\right] \cr
&&\times \left[ \exp\left(-\frac{i \hat{L}_2\hat{r}_\bot^2 }{2
(\hat{z}_o-\hat{z}_{s3}) (\hat{L}_2+2\hat{z}_o-2\hat{z}_{s3})}
\right)\right.\cr&&\left.-\exp\left(\frac{i
\hat{L}_2\hat{r}_\bot^2 }{2 (\hat{z}_o-\hat{z}_{s3})
(-\hat{L}_2+2\hat{z}_o-2\hat{z}_{s3})} \right) \right]
~.\label{vir4D}
\end{eqnarray}
Eq. (\ref{vir4A}) and Eq. (\ref{vir4D}) solve the field
propagation problem for both the near and the far field in the
limit for $\hat{\phi} \ll 1$ and $\hat{\phi}_w \gg 1$. The
directivity diagram is obtained by summing up Eq. (\ref{vir4A})
and Eq. (\ref{vir4D}) and taking square modulus of the sum.

In the particular case $L_1=L_2=L_w=L_{\mathrm{tot}}/3$ we obtain
simplified expressions:

\begin{eqnarray}
\vec{\hat{E}}_{AB}(\hat{z}_o,\vec{\hat{r}}_\bot) &=& -\frac{4
\vec{\hat{r}}_\bot}{\hat{r}_\bot^2}\exp\left[i
\frac{\hat{r}_\bot^2}{2(\hat{z}_o+1/3)}\right] \left[
\exp\left(-\frac{i \hat{r}_\bot^2 }{6 (\hat{z}_o+1/3)
(1+2\hat{z}_o)} \right)\right.\cr&&\left.-\exp\left(\frac{i
\hat{r}_\bot^2 }{6 (\hat{z}_o+1/3) (2\hat{z}_o+1/3)} \right)
\right] \exp\left[-\frac{i  \hat{\phi}_w
}{12}\right]~.\label{vir4As}
\end{eqnarray}
\begin{eqnarray}
\vec{\hat{E}}_{CD}(\hat{z}_o,\vec{\hat{r}}_\bot) &=& -\frac{4
\vec{\hat{r}}_\bot}{\hat{r}_\bot^2}\exp\left[i
\frac{\hat{r}_\bot^2}{2(\hat{z}_o-1/3)}\right] \left[
\exp\left(-\frac{i \hat{r}_\bot^2 }{6 (\hat{z}_o-1/3)
(2\hat{z}_o-1/3)} \right)\right.\cr&&\left.-\exp\left(\frac{i
\hat{r}_\bot^2 }{6 (\hat{z}_o-1/3) (2\hat{z}_o-1)} \right)
\right]\exp\left[\frac{i  \hat{\phi}_w }{12}\right]
~.\label{vir4Ds}
\end{eqnarray}
The corresponding directivity diagram is given by

\begin{eqnarray}
\hat{I} &\sim& \frac{1}{\hat{\theta}^2}\Bigg |\exp\left[i
\frac{\hat{z}_o^2\hat{\theta}^2}{2(\hat{z}_o+1/3)}\right] \left[
\exp\left(-\frac{i \hat{z}_o^2 \hat{\theta}^2 }{6 (\hat{z}_o+1/3)
(1+2\hat{z}_o)} \right)\right.\cr&&\left.-\exp\left(\frac{i
\hat{z}_o^2\hat{\theta}^2 }{6 (\hat{z}_o+1/3) (2\hat{z}_o+1/3)}
\right) \right] \exp\left[-\frac{i  \hat{\phi}_w }{12}\right] \cr
&& +\exp\left[i
\frac{\hat{z}_o^2\hat{\theta}^2}{2(\hat{z}_o-1/3)}\right] \left[
\exp\left(-\frac{i \hat{z}_o^2\hat{\theta}^2 }{6 (\hat{z}_o-1/3)
(2\hat{z}_o-1/3)} \right)\right. \cr&& \left.-\exp\left(\frac{i
\hat{z}_o^2\hat{\theta}^2 }{6 (\hat{z}_o-1/3) (2\hat{z}_o-1)}
\right)\right] \exp\left[\frac{i  \hat{\phi}_w
}{12}\right]\Bigg|^2\label{dirlas}
\end{eqnarray}

\begin{figure}
\begin{center}
\includegraphics*[width=140mm]{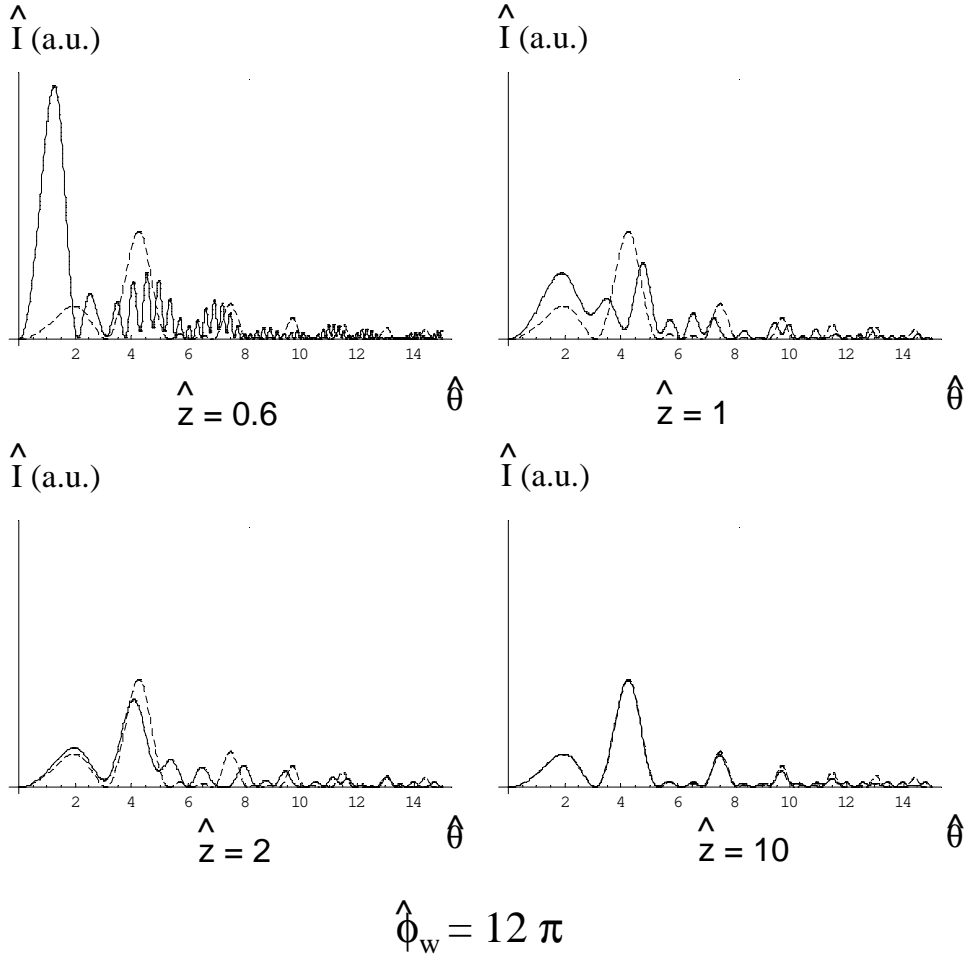}% Here is how to import EPS art
\caption{\label{Anydtur} Evolution of the intensity profile for
TUR for $\hat{\phi}_w = 12\pi $ . These profiles are shown as a
function of angles at different observation distances
$\hat{z}_o=0.6$, $\hat{z}_o=1.0$, $\hat{z}_o=2.0$ and
$\hat{z}_o=10.0$ (solid lines). The dashed line always refers to
the far-zone intensity. }
\end{center}
\end{figure}
\begin{figure}
\begin{center}
\includegraphics*[width=140mm]{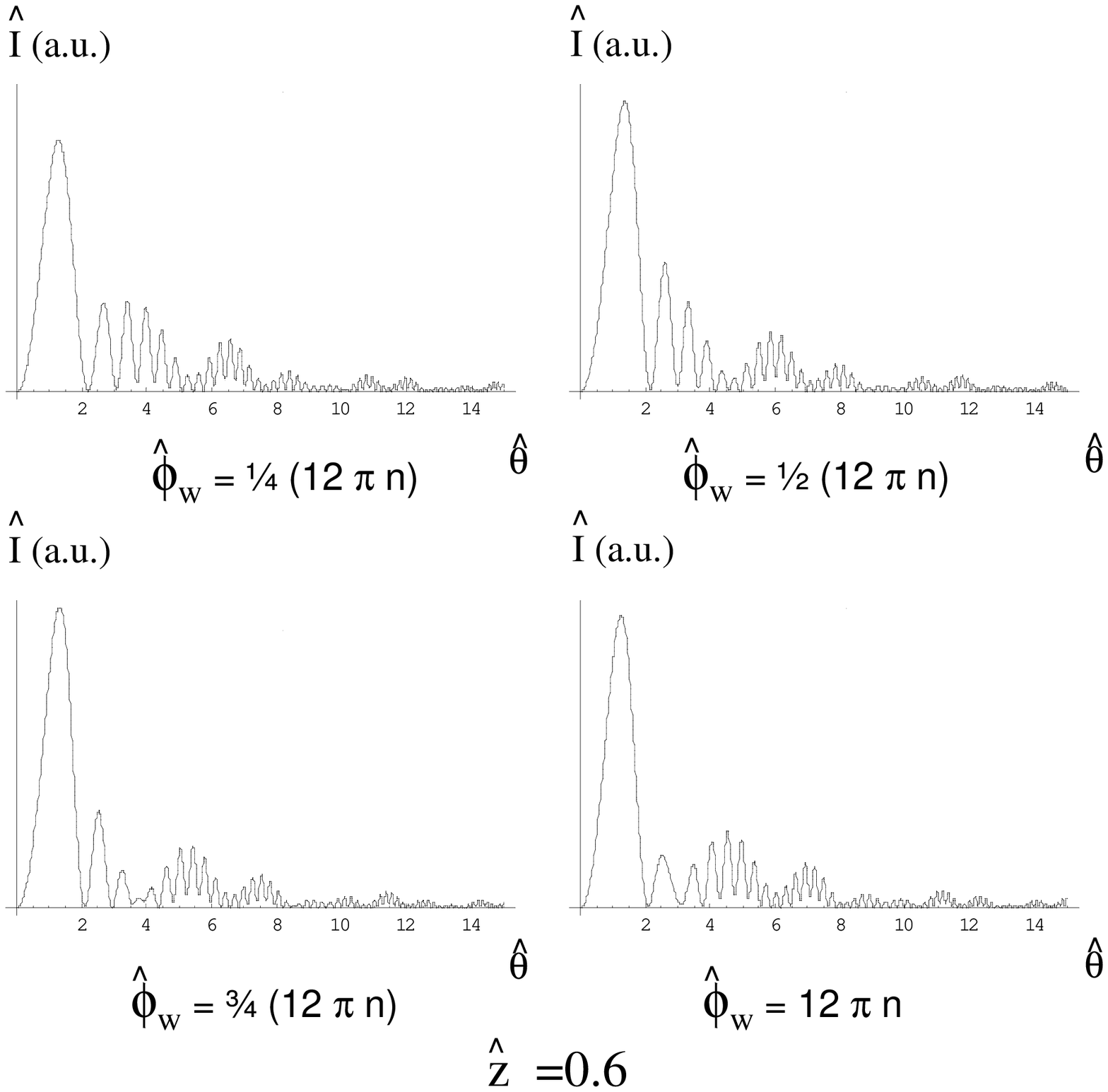}% Here is how to import EPS art
\caption{\label{Anyphiz06} Intensity profiles for TUR at
$\hat{z}_0 =0.6$. at different values $\hat{\phi}_w = (1/4) (12
\pi)$, $\hat{\phi}_w = (1/2) (12 \pi)$, $\hat{\phi}_w = (3/4) (12
\pi)$, $\hat{\phi}_w =  (12 \pi)$. }
\end{center}
\end{figure}
\begin{figure}
\begin{center}
\includegraphics*[width=140mm]{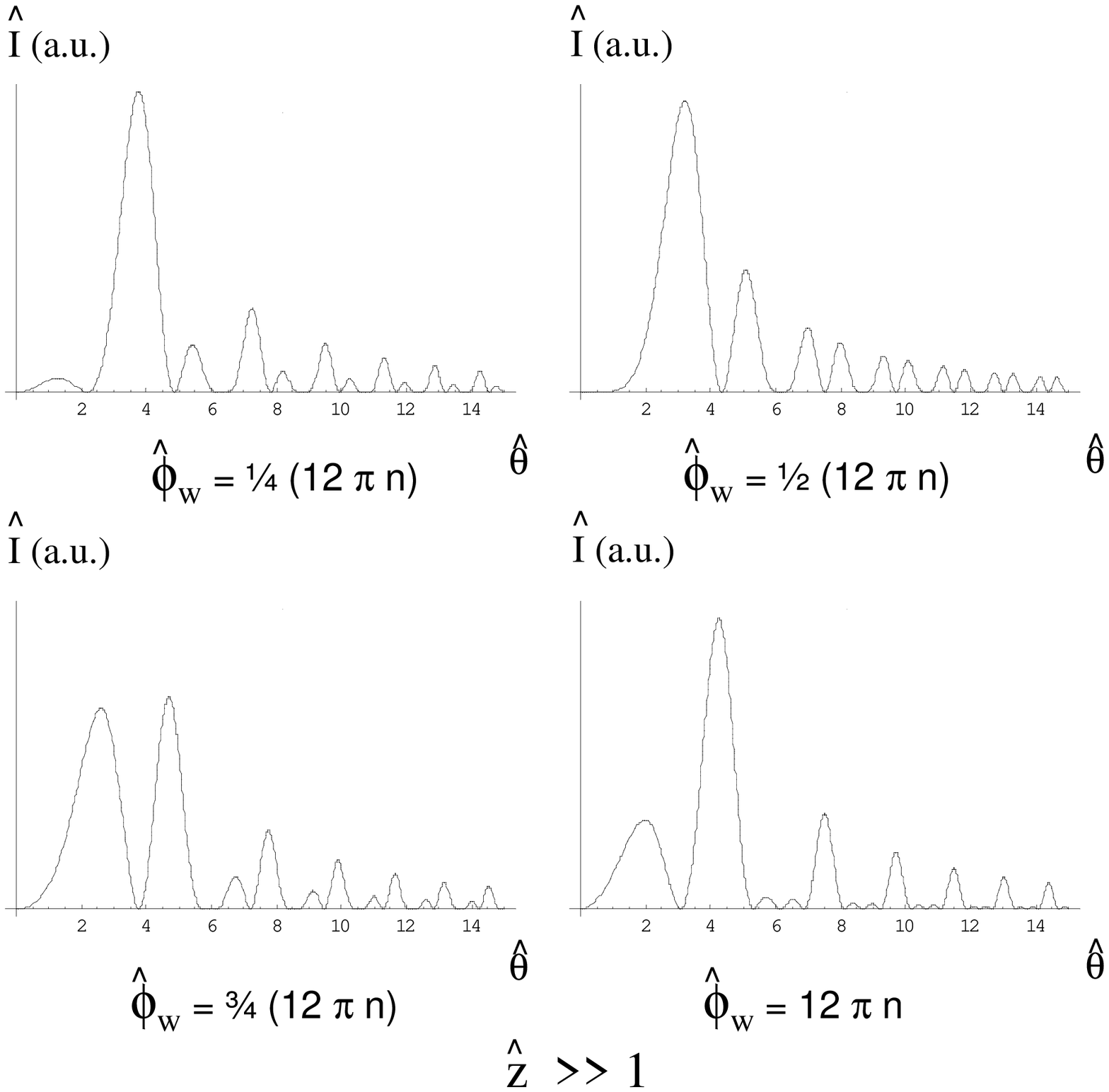}% Here is how to import EPS art
\caption{\label{Anyphizinf} Intensity profiles for TUR in the
asymptotic limit for $\hat{z}_0 \gg 1$. at different values
$\hat{\phi}_w = (1/4) (12 \pi)$, $\hat{\phi}_w = (1/2) (12 \pi)$,
$\hat{\phi}_w = (3/4) (12 \pi)$, $\hat{\phi}_w = (12 \pi)$ }
\end{center}
\end{figure}

Even though it refers to a particular case where
$L_1=L_2=L_w=L_{\mathrm{tot}}/3$, Eq. (\ref{dirlas}) still depends
on the parameter $\hat{\phi}_w \gg 1$. Its dependence on
$\hat{\phi}_w$ is periodic, with period $12 \pi$. For the sake of
exemplification, in Fig. \ref{Anydtur} we plot the intensity
profile for $\hat{\phi}_w = 12\pi $  at different distances
$\hat{z}_o$ and we compare these profiles with the far field
asymptotic behavior. In Fig. \ref{Anyphiz06} and Fig.
\ref{Anyphizinf} we plot, instead, the intensity profile for
different values of $\hat{\phi}_w$ at $\hat{z}_o=0.6$ and in the
asymptotic case for $\hat{z}_o \gg 1$ respectively.


\section{\label{sec:crit} A critical re-examination of conventional edge radiation theory}

In this Section we compare our findings with treatments of edge
radiation and TUR, that can be found in literature. Conventional
understanding is based on the zero-length switcher approximation.
This allows direct comparison of results with the contributions
calculated in the previous Section \ref{sec:four} and Section
\ref{sec:five}. Since there is a lot of literature dealing with
edge radiation and TUR we will limit ourselves to a few
significative works only. In particular, we will refer to
\cite{BOS4}, \cite{BOS3}, \cite{BOS7} and \cite{CHUN}, even though
other works report the same results. For example, the same
understanding can be found in the very recent\footnote{The year of
writing is 2006.} review \cite{WILL}.

We found agreement with literature as concerns calculations in the
far-field zone.

As we have discussed before, the far-field asymptotic is valid at
observation positions $z_o \gg L$, independently of $\hat{\phi}=
L/(\gamma^2\lambdabar)$. It should therefore be stressed that in
the case $\hat{\phi} \gg 1$ one may be in the near zone even
though the observer is located many formation lengths $\gamma^2
\lambdabar$ from the downstream edge (for example, when $z_o =
L$). In the far-zone case Eq. (\ref{ABcontrint4e}) gives the
electric field in the far region. The correspondent directivity
diagram, Eq. (\ref{normI1}), is plotted in Fig. \ref{Dirfar2}, for
different values of the main parameter of the theory,
$\hat{\phi}$. This result is well-known and in perfect agreement
with literature. For example one may see that, aside for
proportionality factors due to different choice of units, there is
agreement between the radiation energy density in Eq.
(\ref{enden2e}) and the photon flux per unit solid angle in Eq.
(18) of \cite{BOS4} (also compare, qualitatively, Fig. 4 of that
reference and our Fig. \ref{Dirfar2}). Similarly, the field in Eq.
(\ref{enden2e}) is in agreement with Eq. (17) of \cite{BOS3} or
Eq. (31) of \cite{BOS7} (in the limit for a large observation
distance) accounting for the fact that the origin of the reference
system is chosen at the exit of the straight section by the author
of \cite{BOS3} (also, the electron charge $e$ is negative in that
reference). Note that in the far zone, the field exhibits a
spherical wavefront (independently of the choice of coordinate
system).

Differences between our results and results in literature arise,
however, in the near zone.

\subsection{Near-field edge radiation}

\paragraph*{Case for $\mathbf{\hat{\phi} \gg 1}$.} Let us pose $d=z_o-L/2$. In the case when $0 \lesssim d \ll
\gamma^2 \lambdabar$, we have seen that the intensity profile for
edge radiation reproduces better and better (as $d$ decreases) the
single edge source, i.e. Eq. (\ref{Iviri}). According to our
understanding then, it follows that the photon flux depends on
$\gamma$, does not exhibit oscillatory dependence on the
transverse position $r$, it has a characteristic scale $\gamma
\lambdabar$, and depends on the frequency as $\omega^2 (\Delta
\omega/\omega) K_1^2[\omega r_\bot/(c\gamma)]$. In contrast to
this, the flux according to Eq. (14) in \cite{BOS4}\footnote{This
equation coincides with Eq. (20) in \cite{CHUN} and Eq. (26) in
the review \cite{WILL}. It is a currently accepted expression
describing the photon flux in the case under study.} reads, in our
notation (number of photons per unit surface per unit time within
the bandwidth $\Delta \omega/\omega$) :

\begin{eqnarray}
\frac{d F}{d S} \simeq  \alpha \frac{\Delta \omega}{\omega}
\frac{I_b}{e\pi^2} \frac{1}{r_\bot^2} \sin^2\left[\frac{\omega
r_\bot^2}{4 c d}\right]~,\label{wron}
\end{eqnarray}
where $\alpha = e^2/(\hbar c)$ is the fine structure constant and
$I_b$ is the beam current. This suggests that for $d \ll \gamma^2
\lambda$ the photon flux does not depend on $\gamma$, exhibits
oscillatory dependence on the transverse position $r_\bot$  on the
scale $\sqrt{\lambda d}$, and depends on the frequency as $(\Delta
\omega/\omega) \sin^2[\omega r_\bot^2/(4 c d)]$.   The field from
which Eq. (\ref{wron}) is derived is given explicitly in several
papers (e.g. Eq. (24) in \cite{BOS7}, Eq. (4) in \cite{BOS2}). In
our units, it reads:

\begin{eqnarray}
\vec{\widetilde{E}}({z}_o,\vec{r}_\bot) &&=- \frac{4 i e}{c}
\frac{ \vec{{r}}_\bot}{{r}_\bot^2}\exp\left[-i \frac{\omega
{r}_\bot^2}{4 c d }\right] \sin\left[\frac{\omega {r}_\bot^2}{4 c
d }\right]~.\label{wron2}
\end{eqnarray}
We disagree with the conclusions in Eq. (\ref{wron}) and Eq.
(\ref{wron2}). In particular the expression for the field in Eq.
(\ref{wron2}) is in contrast with our Eq. (\ref{virpm05}).

The region of parameters for $d \sim \gamma^2 \lambdabar$ is not
discussed in literature. As we have seen, this corresponds to a
region where the near-field contribution from the downstream edge
is present.

When $\gamma^2 \lambdabar \ll d \ll L$ the contribution of the
downstream edge is dominant, and the intensity distribution tends
to approximate $\mathrm{const} \times
\hat{\theta}^2/(\hat{\theta}^2+\hat{\phi})^2$, that is the single
edge far field limit. This case is practically described in
literature when discussing "an electron exiting a bending magnet
at the upstream end of a straight section" of infinite length
(cited from \cite{BOS4}. Compare, for example with Eq. (8) in the
same reference). Although this situation is \textit{per se} not
physical, in the sense that an infinite straight section cannot be
built, it has a practical realization (for $\hat{\phi} \gg 1$) in
the before-mentioned observation region $\gamma^2 \lambdabar \ll d
\ll L$.

The region of parameters for $d \sim L$ is not discussed in
literature, because the near-field region is understood for $d
\lesssim \gamma^2\lambdabar$ only. However, as we have seen, the
case $d\sim L$ corresponds to a region where interference begins
to become important. In our understanding, this is a transition
region between the near and the far zone ($d \gg L$).

\paragraph*{Case for $\mathbf{\hat{\phi} \ll 1}$.} In this situation we can compare results in literature with our
Eq. (\ref{vir4e}). In particular, Eq. (31) of \cite{BOS7} (or,
equivalently, Eq. (17) of \cite{BOS3} for the Fourier transform of
the field  $\vec{\bar{E}}$) is said to be applicable at for
observation regions where $d \ll \gamma^2 \lambdabar$ (using our
notations) and for $L \ll \gamma^2 \lambdabar$, i.e. when
$\hat{\phi} \ll 1$. Accounting for the fact that the origin of the
reference system in \cite{BOS7} is at the downstream edge of the
straight section, we can write Eq. (31) of \cite{BOS7} in our
dimensional units as:

\begin{eqnarray}
\vec{\widetilde{E}}({z}_o,\vec{r}_\bot) &&= \frac{2 e}{c} \frac{
\vec{{r}}_\bot}{{r}_\bot^2}\exp\left[i \frac{\omega {r}_\bot^2}{2
c{z}_o}\right]\exp\left[i \frac{\omega {r}_\bot^2}{2
c(L-2{z}_o)}\right] \cr&&\times\left[ \exp\left(-\frac{i \omega
{r}_\bot^2 }{2  c {z}_o (1+2{z}_o/L)} \right)-\exp\left(\frac{i
\omega {r}_\bot^2 }{2  c {z}_o (-1+2{z}_o/L)} \right) \right]
.\label{vir4eloro}
\end{eqnarray}
Thus, Eq. (31) in \cite{BOS7} appears to differ of the phase
factor $\exp \{i \omega {r}_\bot^2/[2 c(L-2{z}_o)]\}$ with respect
to our Eq. (\ref{vir4e}), directly calculated by propagating the
field at the virtual source position. We disagree with the result
in Eq. (\ref{vir4eloro}) (i.e. Eq. (31) in \cite{BOS7}). The
intensity pattern, however, is correct in both cases\footnote{Note
that the intensity profile in Eq. (27) of the review \cite{WILL}
is the square modulus of Eq. (31) in \cite{BOS4}.}.

It is of some interest to understand the origin of the difference
between our Eq. (\ref{vir4e}) and Eq. (31) in \cite{BOS7}. In this
regard, in \cite{BOS7} it is reported that when $L\gg d $ (and $L
\ll \gamma^2 \lambda$) , the interference between upstream and
downstream edge is negligible. In this limit, Eq. (31) in
\cite{BOS7} gives back Eq. (24) in \cite{BOS7} that is Eq.
(\ref{wron2}). On the one hand Eq. (24) in \cite{BOS7} is said to
be valid for an infinitely long straight section followed by a
downstream edge. On the other hand, Eq. (31) in \cite{BOS7} (and
thus its limiting case for $L\gg d$ as well) is reported to be
valid for $d,L \ll \gamma^2 \lambda$. Since the limiting case of
Eq. (31) in \cite{BOS7} for $L\gg d$ coincides with Eq. (24) in
\cite{BOS7} one concludes that an infinitely long straight section
is identified \textit{de facto} with the case $L\gg d$. We hold
this to be a misconception, in disagreement with our
understanding. The length of the straight section is naturally
measured by the parameter $\hat{\phi} = L/(\gamma^2 \lambdabar)$.
Such length is independent of the position of the observer. Only
when $\hat{\phi} \gg 1$ it is possible to have radiation from a
single edge. In our understanding thus, the condition $L \gg d$ is
still necessary, but not sufficient to have single-edge radiation.
Describing the situation for $\hat{\phi} \ll 1$ we have seen from
Eq. (\ref{vir4e}) that the total field is always given as the
result of the interference of the two virtual sources located at
the straight section edges and one never has radiation from single
edge. In contrast to this radiation from a single edge (in Eq.
(24) of \cite{BOS7}) is explicitly reported to hold  for $L \ll
\gamma^2 \lambda$ in the limit for $L\gg d$.

We conclude that the origin of the difference between our results
and those in literature may be ascribed to a different
understanding of the parameters of the theory. While we hold
$\hat{\phi}$ to be the main parameter of our treatment, result in
literature seem to neglect its existence. By doing this,
conventional treatments neglect the fact that $d$ and $L$ should
always be compared with the third main length scale of the
problem, that is $\gamma^2 \lambdabar$.

\subsection{Transition Undulator Radiation}

As we have seen, TUR can be addressed as a more complicated
edge-radiation setup. In this case we have contributions from
three parts, two straight lines and the undulator. As we have
seen, the undulator contribution is similar to a straight line
contribution, the only difference being a different average
longitudinal velocity of the electron. Then, the far-zone region
can be identified by distances $z_o \gg L_{tot}$. In the far zone,
well-accepted expressions for the TUR emission are reported in
literature \cite{KIM1,KINC,CAST}, that are equivalent to the
following equation for the radiation energy density as a function
of angle and frequency:

\begin{eqnarray}
\frac{d W}{d \omega d\Omega} &=& \frac{e^2}{\pi^2 c}
\left[\frac{2\gamma^2 \theta K^2
}{(1+K^2/2+\gamma^2\theta^2)(1+\gamma^2\theta^2)}\right]^2
 \sin^2\left[\frac{\pi L_w}{2\gamma^2 \lambda}
\left(1+\frac{K^2}{2}+\gamma^2\theta^2\right)\right]. \cr &&
\label{wron3}
\end{eqnarray}
We disagree with this expression. In our understanding, there
cannot be any range of parameters in the setup in Fig. \ref{geom}
where Eq. (\ref{wron3}) is valid.

In order to prove this it is sufficient to compare Eq.
(\ref{wron3}) with Eq. (\ref{enden2}), that is equivalent to Eq.
(20) of reference \cite{BOS4}\footnote{We believe that there is a
small misprint in Eq. (8) of \cite{BOS4}. The symbol "$\theta$"
after the sign "$\approx$" should be replaced by the
fine-structure constant "$\alpha$".}. In particular, since Eq.
(\ref{wron3}) does not depend on the straight section lengths
$L_1$ or $L_2$, we conclude that the only regions where such
comparison makes sense can be for $L_1=L_2=0$, or $L_1 =L_2 = L_w$
\footnote{Referring to Fig. \ref{geom}, note that when $L_2 \gg
\gamma^2 \lambdabar$ the contribution from the straight section
edge in $D$ is always larger than those from the undulator edges
in $B$ and $C$ at any distance in the near and in the far zone.
This is the case because outside the undulator the longitudinal
velocity is nearer to $c$ than inside ($\gamma_z^2 < \gamma^2$).
Thus, for long straight sections $L_1 = L_2$, with $L_2 \gg
\gamma^2 \lambdabar$, the straight section contribution will be
dominant compared to the undulator contribution.}.

When $L_1=L_2=0$ Eq. (\ref{enden2}) reduces to:

\begin{eqnarray}
\frac{d W}{d\omega d \Omega} = \frac{ e^2}{ \pi^2 c}\left[ \frac{2
\gamma^2\theta}{1+K^2/2+\gamma^2\theta^2 }\right]^2
\sin^2\left[\frac{\pi L_w}{2 \gamma^2 \lambda }
\left({1}+\frac{K^2}{2}+\gamma^2\theta^2\right)\right] ~,
\label{enden2zero}
\end{eqnarray}
that is obviously different from Eq. (\ref{wron3}.)

When $L_1=L_2=L_w$ Eq. (\ref{enden2}) becomes:

\begin{eqnarray}
\frac{d W}{d\omega d \Omega} &=& \frac{ e^2}{ \pi^2 c}
\left[\frac{2 \gamma^2 \theta
K^2}{(1+K^2/2+\gamma^2\theta^2)\left(1+\gamma^2\theta^2\right)}
\right]^2 \cr \times &&\Bigg| \frac{1}{2}~ {\sin\left[\frac{\pi
L_w}{2\gamma^2 \lambda}
\left({1}+\frac{K^2}{2}+\gamma^2\theta^2\right)\right]} \cr && -
\frac{(1+K^2/2+\gamma^2\theta^2)}{K^2}\sin \left[ \frac{ \omega
L_w}{4 c \gamma^2}\left(3+\frac{K^2}{2}+3\gamma^2\theta^2\right)
\right]\Bigg|^2~, \label{enden2Lw}
\end{eqnarray}
that is also different from Eq. (\ref{wron3})\footnote{For
example, in the limit for $K \ll 1$ Eq. (\ref{wron3}) tends to
zero, while Eq. (\ref{enden2Lw}) gives back Eq. (\ref{enden2e})
with $L=3 L_w$ as it should be.}. We conclude that Eq.
(\ref{wron3}) has no physical meaning.

Finally, we put attention on the fact that there is a general
need, in the FEL community, to extend the current theory of TUR to
cover the near zone. Some attempt in this direction has been
reported in \cite{ADA1}. The author of \cite{ADA1} discusses a
possible use of coherent TUR to produce visible-to-infrared light
synchronized with X-rays from an X-ray free-electron laser. For
simplicity we will ignore the delicate issue of technical
realization of the scheme proposed in \cite{ADA1} for the LCLS
facility, and we will restrict our discussion of a issue
pertaining fundamental questions of electrodynamics.

In our view, the extension of the theory of TUR in Appendix A of
\cite{ADA1} to the near zone also includes some misconceptions.
The treatment begins with Eq. (A1), that is an expression for the
vector potential $\vec{A}$ at a given observation position and at
a given frequency of interest. We report this expression here, for
the readers' commodity:

\begin{eqnarray}
\vec{A}(\omega) \sim \int_{-T/2}^{T/2} \exp\left[i \omega \left(t+
R(t)/c\right)\right]
\frac{d}{dt}\left[\frac{\vec{n}\times\left(\vec{n}\times\vec{\beta}
\right)}{1-\vec{\beta}\cdot\vec{n}}\right]~.\label{A1}
\end{eqnarray}
Here $T = L/c$ is the time that a photon takes to travel the
undulator length, $R(t)$ is the distance from the electron
position at retarded time $t$ to the observer, the unit vector
$\vec{n}$ is the direction from the retarded position of the
electron to the observer and $\vec{\beta}(t)$ is the electron
velocity in units of the speed of light.  Eq. (\ref{A1}) is a
modification of the well-known Eq. (14.62) of \cite{JACK}. In
fact, integration limits in Eq. (\ref{A1}) have been changed from
$\pm \infty$ to $\pm T/2$ "under the assumption that the electrons
go on straight paths before and after the undulator" (cited from
\cite{ADA1}). As a first remark, as discussed above, the previous
assumption is not justified in our view, as it does not make sense
to discuss generically about infinite straight lines preceding and
following the undulator. However, we believe that there is another
problem with the use made of Eq. (\ref{A1}). The author of
\cite{ADA1} starts with Eq. (\ref{A1}) and modifies it accounting
for the fact that, in the near zone, $\vec{n}$ is not constant.
The problem with this is that the derivation of Eq. (\ref{A1})
relies \textit{a priori} on the assumption that $\vec{n} =
\mathrm{const}$. In particular, the classical result

\begin{equation}
{\vec{n}\times[(\vec{n}-\vec{\beta})\times{\dot{\vec{\beta}}}]
\over{(1-\vec{n}\cdot\vec{\beta})
\left|\vec{r}_o-\vec{r'}(t)\right|}} ={1\over{r_o}}
{d\over{dt}}\left[
{\vec{n}\times(\vec{n}\times{\vec{\beta}})\over{1-\vec{n}\cdot\vec{\beta}}}\right]~,
\label{classdiff}
\end{equation}
is used, where $r_o = |\vec{r}_o|$ is the distance of the observer
from the origin of the coordinate system. Eq. (\ref{classdiff})
assumes $\vec{n} = \mathrm{const}$.

Our conclusion is that Eq. (\ref{A1}) cannot be manipulated
assuming that $\vec{n}$ is not constant, as has been done in
\cite{ADA1}. When $\vec{n}$ is not constant, as in this case, Eq.
(\ref{classdiff}) should be modified to the expression presented
before in Eq. (\ref{luccioparti}) that may be found in the
interesting but perhaps little-known reference \cite{LUCC}, dating
back more than twenty years.

However, in this paper we have seen that it is possible to develop
a more convenient theory of near-field Synchrotron Radiation
Theory (including the case of TUR) in the space-frequency domain,
where Fourier Optics techniques can be taken advantage of.

\section{\label{sec:conc} Conclusions}

In this paper we presented a general connection between Fourier
Optics and classical relativistic electrodynamics. Consistent use
of such connection  resulted in the formulation of a modern theory
of Synchrotron Radiation in terms of laser-beam optics.

In particular, we developed a theory of near-field SR in the
space-frequency domain based of Fourier Optics techniques. These
techniques can be taken advantage of without limitations for SR
setups, because the paraxial approximation can always be applied
in the case of electrons in ultra-relativistic motion. We
restricted ourselves to the analysis of  single-particle
radiation.   As we demonstrated in Section \ref{sec:meth}, and
discussed in Section \ref{sec:dis}  the use of paraxial
approximation allows reconstruction of the field in the near-zone
from the knowledge of far-field data only. The solvability  of the
inverse problem  for the field starting from far-field data allows
characterization of any synchrotron radiation setup by means of
virtual sources. In cases of interest these sources exhibit a
plane wavefront, and can be pictured as waists of laser-like
beams. These laser-like beams help developing our theory in close
relation with laser-beam optics. In particular, usual Fourier
Optics can be used to describe the field at any distance, thus
providing a tool for designing and analyzing SR setups.

We gave several examples of applications of our theory in Section
\ref{sec:exa}, Section \ref{sec:four} and Section \ref{sec:five}.
These examples provide both physical insight and understanding of
several non-intuitive phenomena. In Section \ref{sec:exa} we
treated the case of undulator radiation. In the case of perfect
resonance we found an analytic expression for the virtual source
and we propagated such expression both in the near and in the far
zone. The applications in the following Section \ref{sec:four} and
Section \ref{sec:five} actually constitute the first comprehensive
theory of edge radiation and transition undulator radiation.

In Section \ref{sec:four} we dealt with the fundamental case of an
electron moving on a straight section. Such kind of system can be
considered, in all generality, as a fundamental building block for
a number of more complicated setups. We demonstrated that two
equivalent pictures in terms of laser-like sources can be
presented. The first takes advantage of a single virtual source in
the middle of the straight section. The second relies on two
virtual sources at the edges. The virtual sources at the edges can
always be described analytically, independently of the region of
parameters considered. The main parameter of the theory is
identified to be the ratio between the straight section length $L$
and $\gamma^2\lambdabar$. When this ratio is small the
single-source picture is found to be natural, and two asymptotic
regions of observation are found, the near and the far zone. When
the ratio is large, the two-source picture is more natural and one
is left with four asymptotic regions of  observation: far zone for
two interfering sources, near zone for two interfering sources,
far zone for a single edge and near zone for a single edge.
Results in Section \ref{sec:four} can be directly used to describe
an edge-radiation setup in the case upstream and downstream bends
can be considered to have zero length. These bends act like
switchers, switching the radiation harmonic (and the beam current
harmonic) on and off. Thus, our results for the straight section
can be applied to any situation with zero-length switchers. Our
expressions are of importance also in the case when switchers
cannot be considered having zero-length. In fact in general, as
said before, straight sections can be considered as basic building
blocks for any magnetic setup.

To demonstrate this last point, in Section \ref{sec:five} we used
results of Section \ref{sec:four} as building blocks for
describing the case of TUR in the zero-length switcher
approximation. In particular, we dealt with a setup composed by an
undulator preceded and followed by two straight sections. In the
zero-length switchers approximation, such setup describes TUR
emission. Although more sophisticated with respect to the
straight-section motion, the example in Section \ref{sec:five}
does not present any conceptual difficulty in addition to those
encountered in the treatment of the straight-section motion.

Finally, in Section \ref{sec:crit}, we compared our findings with
current understanding of edge radiation and TUR. We found that
several misconceptions exist in the conventional understanding of
edge radiation in the near zone, as well as in the treatment of
TUR. We extensively discussed these misconceptions and proposed
our understanding of the correct description of these phenomena.

We find it interesting to remark that, although classical
relativistic electrodynamics is a relatively old branch of
physics, there are many novel and valuable results pertaining this
field that have been obtained only recently. Undoubtedly, many
more will follow. These recent developments illustrate a recurring
fact in the history of Physics. Namely, subjects appearing
well-understood, and perhaps even a little old-fashioned may still
have surprises in store for us.

\section{\label{sec:graz} Acknowledgements}

The authors are grateful to Martin Dohlus, Hermann Franz, Petr
Ilinski  and Helmut Mais (DESY) for many useful discussions and to
Massimo Altarelli, Edgar Weckert and Jochen Schneider (DESY) for
their interest in this work.


\begin{thebibliography}{99}


\bibitem{OUR1} G. Geloni, E. Saldin, E. Schneidmiller and M.
Yurkov, "Paraxial Green's functions in SR theory", DESY 05-032,
ISSN 0418-9833 (2005)
\bibitem{OUR2} G. Geloni, E. Saldin, E. Schneidmiller and M.
Yurkov, "Understanding transverse coherence properties of X-ray
beams in third generation SR source", DESY 05-109, ISSN 0418-9833
(2005)
\bibitem{OUR3} G. Geloni, E. Saldin, E. Schneidmiller and M.
Yurkov, "Statistical Optics approach to the design of beamlines
for Synchotron Radiation", DESY 06-037, ISSN 0418-9833 (2006)
\bibitem{GOOD} J. W. Goodman, "Introduction to Fourier Optics", Mc Graw-Hill Book
Company(1968)
\bibitem{ZEMA} ZEMAX code, at http://www.zemax.com/
%\bibitem{SMOL} N.V. Smolyakov, Sov. Phys. Tech. Phys. 31, 741
%(1986)
%\bibitem{BASH} Y.A. Bashmakov, Rev. Sci. Instrum. 63, 343 (1991)
\bibitem{BOSF} R.A. Bosch et al. Rev. Sci. Instrum. 67, 3346 (1995)
\bibitem{CHUN} R.A. Bosch and O.V. Chubar, "Long wavelength edge
radiation in an electron storage ring", Proc. SRI'97, tenth US
National Conference, AIP Conf. Proc. 417, edited by E. Fontes
(AIP, Woodbury, NY), pp. 35-41 (1997)
%\bibitem{MATH} Y.-L. Mathis et al. Phys. Rev. Lett. 80, 6 (1998)
\bibitem{BOS4} R.A. Bosch, Il Nuovo Cimento, 20, 4 p. 483 (1998)
\bibitem{BOS7} R.A. Bosch,  Nucl. Instr. Meth. Phys. Res. A, 431
(1999) 320
\bibitem{PROY} P. Roy, J-B. Brubach et al., Nucl. Instr. Meth.
Phys. Res. A, 426 (2001)
\bibitem{BOS2} R.A. Bosch, Nucl. Instr. Meth. Phys. Res. A, 482,
789 (2002)
\bibitem{BOS3} R.A. Bosch, Phys. Rev. ST AB 5, 020701 (2002)
\bibitem{WILL} G. P. Williams, Rep. Prog. Phys. 69 (2006), 301
\bibitem{KIM1} K.-J. Kim, Phys. Rev. Lett. 76, 8 (1996)
\bibitem{KINC} B.M. Kincaid, Il Nuovo Cimento Soc. Ital. Fis, 20D,
495 (1998) and LBL-38245 (1996)
\bibitem{CAST} M. Castellano, Nucl. Instr and Meth. in Phys. Res.
A 391 (1997) 375
\bibitem{BOS6} R.A. Bosch, Nucl. Instr. Meth. Phys. Res. A, 386
(1997) 525
\bibitem{ROY2} P. Roy et al. Phys. Rev. Lett. 84, 3 (2000)
\bibitem{REI1} S. Reiche and Z. Huang, in Proceedings of the 2004 FEL Conference, R. Bakker et al. Editors, Trieste, pp. 193-195 (2004)
\bibitem{ADA1} B. Adams, Rev. Sci. Instrum. 76, 2 (2005)
\bibitem{JACK} J. Jackson, "Classical Electrodynamics", 3rd ed., Wiley, New York (1999)
\bibitem{CHU2} O. Chubar and P. Elleaume, in Proc. of the 6th
European Particle Accelerator Confernce EPAC-98, 1177-1179 (1998)
\bibitem{LUCC} Y. Hirai, A. Luccio and Li-hua Yu, J. Appl. Phys. 55(1),
25 (1984)
%\bibitem{CHUB} O. Chubar, P.Elleaume et al., Nucl. Instr. Meth.
%Phys. Res. A, 435, 495 (1999)
\bibitem{HAR2} G. Geloni, E. Saldin, E. Schneidmiller and M.
Yurkov, "Exact solution for the Second Harmonic Generation in
XFELs", DESY 05-137, ISSN 0418-9833 (2005), submitted to Optics
Communications
\bibitem{PHAS} J. Bahrdt, in Proceedings of the 2005 FEL Conference, R. Reitmeyer Editor, Stanford, pp. 694-701 (2005)
%\bibitem{EMMA} P. Emma et al. Phys. Rev. Lett. 92, 074801 (2004)
%\bibitem{STUP} G. Stupakov, Y. Ding and Z. Huang, "Calculation of
%the beam field in the LCLS bunch length monitor", SLAC-PUB-11890
%(2006)
\end{thebibliography}
\end{document}